\documentclass[showpacs,showkeys,preprintnumbers,pre,amsmath,amssymb,superscriptaddress,twocolumn]{revtex4}
\pdfoutput=1

\usepackage{array}
\usepackage{amssymb}
\usepackage{amsmath}
\usepackage{graphicx}
\usepackage{color}
\usepackage[colorlinks=true,linkcolor=blue,urlcolor=blue,citecolor=blue]{hyperref}

\usepackage{amssymb}
\usepackage{enumerate}
\usepackage{amsthm,amsfonts,amsmath}
\usepackage{bm}
\usepackage{graphicx}
\usepackage{graphics,epsf,psfrag,pifont,dsfont,bbold}
\usepackage{siunitx}
\usepackage{color}
\usepackage{xspace}
\usepackage{accents}

\usepackage{xspace}
\usepackage{graphicx,color}
\usepackage{amsmath,amssymb,amsfonts,mathrsfs}
\usepackage{url}
\usepackage[normalem]{ulem}
\usepackage{booktabs}
\usepackage{float}

\usepackage[toc,page]{appendix}
\usepackage[capitalize]{cleveref}
\usepackage{mathtools}
\usepackage[normalem]{ulem}

\newcommand{\be}{\begin{equation}}
\newcommand{\ee}{\end{equation}}
\newcommand{\ben}{\begin{equation*}}
\newcommand{\een}{\end{equation*}}

\newcommand{\delF}[1]{\textcolor{blue}{}}

\definecolor{forestgreen}{rgb}{0.13, 0.55, 0.13}

\begin{document}

\title{Structure and Isotropy of Lattice Pressure Tensors for Multi-range Potentials}
\author{Matteo Lulli}
\email{lulli@sustech.edu.cn}
\affiliation{Department of Mechanics and Aerospace Engineering, Southern University of Science and Technology, Shenzhen, Guangdong 518055, China}
\author{Luca Biferale}
\email{biferale@roma2.infn.it}
\affiliation{Department of Physics \& INFN, University of Rome ``Tor Vergata'', Via della Ricerca Scientifica 1, 00133, Rome, Italy.}
\author{Giacomo Falcucci}
\email{giacomo.falcucci@uniroma2.it}
\affiliation{Department of Enterprise Engineering ``Mario Lucertini'', University of Rome ``Tor Vergata", Via del Politecnico 1, 00133 Rome, Italy; John A. Paulson School of Engineering and Applied Physics, {\it Harvard University},  33 Oxford Street, 02138 Cambridge, Massachusetts, USA.}
\author{Mauro Sbragaglia}
\email{sbragaglia@roma2.infn.it}
\affiliation{Department of Physics \& INFN, University of Rome ``Tor Vergata'', Via della Ricerca Scientifica 1, 00133, Rome, Italy.}
\author{Xiaowen Shan}
\email{shanxw@sustech.edu.cn}
\affiliation{Department of Mechanics and Aerospace Engineering, Southern University of Science and Technology, Shenzhen, Guangdong 518055, China}
\pacs{47.11.-j, 05.20.Jj, 68.05.-n}
\keywords{Lattice Boltzmann Methods, Pressure Tensor, Non-Ideal Interfaces}
%%%%%%%%%%%%%%%%%%%%%%%%%%%%%%%%%%%%%%%%%%%%%%%%%%%%%%%%%%%%%%%%%%%%%%%%%%%%%%%%%%%%%%%%%%%%%%%%%%%%%%%%%
% Abstract
%%%%%%%%%%%%%%%%%%%%%%%%%%%%%%%%%%%%%%%%%%%%%%%%%%%%%%%%%%%%%%%%%%%%%%%%%%%%%%%%%%%%%%%%%%%%%%%%%%%%%%%%%
\begin{abstract}
We systematically analyze the tensorial structure of the lattice pressure tensors for a class of multi-phase lattice Boltzmann models (LBM) with multi-range interactions. Due to lattice discrete effects, we show that the built-in isotropy properties of the lattice interaction forces are not necessarily mirrored in the corresponding lattice pressure tensor. This finding opens a different perspective for constructing forcing schemes, achieving the desired isotropy in the lattice pressure tensors via a suitable choice of multi-range potentials. As an immediate application, the obtained LBM forcing schemes are tested via numerical simulations of non-ideal equilibrium interfaces and are shown to yield weaker and less spatially extended spurious currents with respect to forcing schemes obtained by forcing isotropy requirements only. From a general perspective, the proposed analysis yields an approach for implementing forcing symmetries, never explored so far in the framework of the Shan-Chen method for LBM. We argue this will be beneficial for future studies of non-ideal interfaces.
\end{abstract}

\maketitle

%%%%%%%%%%%%%%%%%%%%%%%%%%%%%%%%%%%%%%%%%%%%%%%%%%%%%%%%%%%%%%%
\section{Introduction}\label{sec:intro}
%%%%%%%%%%%%%%%%%%%%%%%%%%%%%%%%%%%%%%%%%%%%%%%%%%%%%%%%%%%%%%%
The study of multi-phase fluids pertains a vast spectrum of scientific disciplines, from theoretical physics to biology and engineering~\cite{brennen2005,baumgarten2006,helmersson2006,pierson1999}. The investigation of multi-phase flows poses a challenge that lies at the heart of fluid dynamics, as proven by the multitude of analytical and numerical approaches encompassed by the vast scientific literature on the subject~\cite{brennen2005,crowe2011multiphase,tryggvason2011,succi2018lattice}. Among these, the lattice Boltzmann method (LBM)~\cite{succi2018lattice} stands out for its remarkable capability in handling multi-phase flows. The first pioneering applications of LBM for the simulations of multi-phase flows started to appear around 30 years ago~\cite{Gunstensen91,Grunau93,ShanChen93,ShanChen94,Swift95,Swift96}. Since then, various studies have been reported in the literature, witnessing the versatility and robustness of the methodology in simulating multi-phase flows with an ample spectrum of applications across widely separated time and space scales~\cite{Huang15,kruger2017lattice,succi2018lattice}. Among all the facets of the LBM methodology for multi-phase flows, the so-called ``Shan-Chen'' (SC) method~\cite{ShanChen93,ShanChen94,ShanDoolen95,ShanDoolen96} has undoubtedly marked a major contribution to the field and its applications have experienced an increasing success in the recent years~\cite{YuanSchaefer06,Sbragaglia07,Falcucci07,HyvaluomaHarting08,ZhangTian08,Huang11,jansen2011bijels,Frijters12,SegaSbragaglia13,Chen14,Belardinelli15,Liu16,Xue18,milan2018lattice,chiappini2019,Frometal19}. In a nutshell, the method hinges on the evolution of a lattice Boltzmann dynamics equipped with multi-range interaction forces directly computed on the lattice nodes. The resulting dynamics reproduces multi-phase flows whose non-ideal interfaces emerge from the underlying mesoscale interactions without the need of being tracked in time during the evolution. The early SC implementations feature a limited set of interaction links, typically coinciding with the links characterizing the LBM dynamics. In the recent years, however, some extensions have been proposed including multi-range potentials, i.e. SC forces with an arbitrary range of interactions~\cite{Shan06,Sbragaglia07,Falcucci07,Falcucci10}. The use of multi-range potentials was first introduced by Shan~\cite{Shan06} to construct forcing schemes with the desired isotropy properties: the higher the degree of isotropy, the larger the number of weights characterizing the lattice force. Shortly after, Sbragaglia {\it et al.}~\cite{Sbragaglia07} showed that the methodology could be used to separately control both bulk properties and surface tension in the context of multi-phase flows. Falcucci {\it et al.}~\cite{Falcucci07} studied the consequences on the surface tension of employing the multi-range pseudopotential, and in \cite{falcucci2008lattice} the methodology was used to deliver configurations with multi-droplets and inhibited coalescence. In~\cite{Falcucci10}, the \textit{gamut} of multi-range interactions was mapped, boosting the density ratio between the coexisting phases, reducing the spurious current magnitude and yielding enhanced numerical stability. The multi-range approach has also allowed to model multi-component yield-stress fluids, e.g.  emulsions, along with their complex flowing behavior~\cite{Benzi2009, Sbragaglia2012, Benzi2013} by introducing competing self-interactions giving rise to an effective disjoining pressure between the surfaces of two droplets. Colosqui {\it et al.}~\cite{colosqui2012mesoscopic} proposed a dynamic optimization strategy to set proper speeds of sound for the liquid and vapor phases, thus allowing to reach high density (up to $1:1000$) and compressibility (up to $25000:1$) ratios.
More recently, in~\cite{li2013achieving}, an alternative approach was proposed to tune the surface tension without affecting the mechanical stability of the interface. Extended forcing schemes have also proved instrumental for implementing the thermodynamic consistency of the Shan-Chen model, as it was first detailed in~\cite{SbragagliaShan10} and further developed in~\cite{Khajepor2015}, with both works based on the lattice pressure tensor first detailed in~\cite{Shan08}. The multi-range pseudopotential approach has been applied to complex non-ideal phenomena of technical interest as well, for example in the simulation of flow-induced cavitation in orifices~\cite{falcucci2013direct}, providing robust evidence of cavitation inception. As apparent from the available literature, the multi-range approach has been key in shedding light on pivotal multi-phase applications, both from a scientific and technological point of view. However, several interesting phenomena connected to non-ideal interfaces have never been charted, yet, such as the curvature dependencies of the surface tension~\cite{Tolman1949, RowlinsonWidom82, Blokhuis1992, Blokhuis2006}, or others that still endure as open questions, such as nucleation~\cite{Menzl2016, Lohse2016, Aasen2020}: in such cases, the multi-range \textit{may} provide a valuable tool for both fundamental investigations and engineering applications.

To mark a further step towards these interesting and promising perspectives, in this work we aim to systematically focus on the pressure tensor, whose precise knowledge is crucial for an accurate characterization of all interface properties (i.e. bulk densities, surface tension, etc.)~\cite{RowlinsonWidom82}. The SC method is based on lattice forces, hence the pressure tensor needs to be constructed once the latter are assigned. Over the years, various attempts have been made to compute the pressure tensor for the SC method. While a pioneering analysis on the SC pressure tensor was already presented in the seminal paper by Shan \& Chen~\cite{ShanChen94}, it is only in the last 15 years that the topic has attracted considerable interest. Sbragaglia {\it et al.}~\cite{Sbragaglia07} presented an analysis to compute the ``continuum'' pressure tensor for multi-range potentials. Instead of invoking a continuum approximation, Shan~\cite{Shan08} presented a systematic analysis to construct the ``lattice'' pressure tensors: the crucial advantage of the lattice formulation of the pressure tensor is that it solves the mechanical equilibrium condition of zero divergence directly on the lattice; hence, it can be used as a starting point to retrieve more accurate interfacial predictions. Based on this lattice formulation, Sbragaglia \& Shan~\cite{SbragagliaShan10} drew some guidelines on the suitable choice of the pseudo-potentials to achieve thermodynamic consistency. The lattice formulation for the pressure tensor has also been extended to multicomponent fluids~\cite{SbragagliaBelardinelli13}. In a recent paper, From {\it et al.}~\cite{Frometal19} studied the lattice pressure tensor on higher order lattices truncating the expansion at second order derivatives of the pseudo-potentials and analyzed the corresponding mechanical equilibrium conditions for a flat interface, verifying the thermodynamic consistency along the lines of the analysis proposed in~\cite{SbragagliaShan10}. These results have been later applied in~\cite{From2020} for the calculation of the diffusion constants and contact angles in multi-component systems. In this paper, we delve deeper in detail with the analysis of the tensorial structures of lattice pressure tensors for multi-range potentials. Given the forcing schemes with some prescribed isotropy properties, it will be shown that such isotropy properties are not exactly mirrored in the lattice pressure tensors introduced in~\cite{Shan08}, i.e. the lattice pressure tensor possesses anisotropic contributions that are absent in the forcing. The desired isotropy can be retrieved by proper adjustments of the multi-range potentials, resulting in new forcing schemes where both forces and lattice-based pressure tensors possess the desired isotropy properties. We stress that the present results are not concerned with the details of the forcing implementation in the LBM. Rather, for a given forcing scheme, the results focus on the determination of the interactions (i.e. the weights) in order to impose a higher degree of isotropy for the lattice pressure tensor.

Numerical tests will be conducted to highlight the improvements introduced by the new forcing schemes. In the present work, we choose to focus on the spurious currents developed near a curved interface. We isolate the role of the new pressure tensor isotropy conditions by proposing 4 new sets of 5 weights (24 forcing directions) and comparing them to the 6-th, 8-th, 10-th and 12-th order forcing isotropy schemes already proposed in the literature~\cite{Shan06, Sbragaglia07}. The comparison is made by ``mimicking'' with the new schemes the previous ones, i.e. by setting the same equation of state, flat interface profile and surface tension. All new schemes yield weaker and less extended spurious currents. The meaning of this result is two-fold: on one hand, there is a clear computational advantage brought in by the ability to obtain with 5 weights weaker spurious currents than by using 10 weights; on the other hand, the results have a clear theoretical importance since they show the existence of a new ``dimension'', that of the lattice pressure tensor, that can be used to implement the symmetries of the forcing in a so far unexplored way.

The paper is organized as follows: in Sec.~\ref{sec:LB} we review some basic concepts and definitions of the LBM while in Sec.~\ref{sec:F} we give some technical details on the analysis of the forcing isotropy. In Sec.~\ref{sec:PT} we review the essential features of the lattice pressure tensor and in Sec.~\ref{sec:THEORY} we present a systematic analysis of the structure of the pressure tensor for multi-range potentials, highlighting the anisotropic contributions and proposing new strategies to cure them. In Sec.~\ref{sec:NUM} we present results of numerical simulations to test the improvements brought by the new forcing schemes. Conclusions will follow in Sec.~\ref{sec:CONCLUSIONS}. The source code for the simulations can be found on the github repository \href{https://github.com/lullimat/idea.deploy}{https://github.com/lullimat/idea.deploy}~\cite{sympy, scipy, numpy0, numpy1, scikit-learn, matplotlib, ipython, pycuda_opencl}, where a Jupyter notebook~\cite{ipython} is available to reproduce the results reported in this paper.
%%%%%%%%%%%%%%%%%%%%%%%%%%%%%%%%%%%%%%%%%%%%%%%%%%%%%%%%%%%%%%%
\section{Lattice Boltzmann}\label{sec:LB}
%%%%%%%%%%%%%%%%%%%%%%%%%%%%%%%%%%%%%%%%%%%%%%%%%%%%%%%%%%%%%%%
A brief overview of the method is here provided. Extensive details can be found elsewhere~\cite{kruger2017lattice,succi2018lattice}. The lattice Boltzmann method (LBM)~\cite{Benzi1992,Chen1998, Wolf2004} is based on a discrete version of the Boltzmann transport equation in which the single-particle probability density function $f(\textbf{x}, \boldsymbol{\xi}, t)$ is defined on the the nodes $\{\textbf{x}\}$ of a $d$-dimensional lattice, at discrete times $t$. The velocities $\{\boldsymbol{\xi}_i\}$, with $i=0,\ldots,N_p$, are discretized as well~\cite{Shan06b,kruger2017lattice,succi2018lattice}, so that for each of them the probability density function only depends on space and time $f_i(\textbf{x}, t) = f(\textbf{x},\boldsymbol{\xi}_i,t)$. The latter are commonly referred to as populations. The discretized velocities are chosen as vectors connecting different points on the lattice (similarly to what is shown in Fig.~\ref{fig:stencil} with the force vectors) and feature a set of weights $\{w_i\}$, such that $\sum_{i=0}^{N_{p}}w_{i}=1$: these are chosen in order to recover the isotropic $n$-rank tensors from the sum of the velocity tensor products, i.e. $\xi^{\mu_1}_i \cdots \xi^{\mu_n}_i$, up to a given maximum order. As an example, the second order isotropic tensor can be written as
\begin{equation}
\sum_{i=0}^{N_{p}}w_i \xi_{i}^{\alpha}\xi_{i}^{\beta}=c_{s}^{2}\delta^{\alpha\beta},
\end{equation}
where the prefactor $c_s^2$ is the square of the lattice sound speed, which is specific to the given set of velocities $\{\boldsymbol{\xi}_i\}$. Greek indices run over the vector components. In the next Section we are going to analyze in detail a similar construction applied to the inter-particles forces.\\
The moments of the discretized distribution function are computed directly by summing the populations. For the first two moments, i.e. the mass density $n$ and the momentum density $n\textbf{u}$, one has
\begin{equation}\label{eq:0and1M}
  n\left(\mathbf{x},t\right)=\sum_{i=0}^{N_{p}}f_{i}\left(\mathbf{x},t\right),\;\;\; n\left(\mathbf{x},t\right)\textbf{u}\left(\mathbf{x},t\right)=\sum_{i=0}^{N_{p}}\boldsymbol{\xi}_{i}f_{i}\left(\mathbf{x},t\right).
\end{equation}
The Boltzmann equation can be discretized over a unitary time lapse $\Delta t = 1$ as
\begin{equation}
  f_{i}\left(\mathbf{x}+\boldsymbol{\xi}_{i},t+1\right)-f_{i}\left(\mathbf{x},t\right)=\Omega_{i}\left(\mathbf{x},t\right),
\end{equation}
which is typically understood as describing two different processes: collision on the right-hand side, conserving mass and momentum, i.e. $\sum_i \Omega_i = \sum_i \xi^\alpha_i \Omega_i = 0$, and streaming on the left-hand side. The collision operator acts locally and it is responsible for the local relaxation of the momenta of the probability distribution, while the streaming operator is responsible for the space-time propagation of the relaxed populations along the lattice. In this work we employ the single-time relaxation BGK collision operator
\begin{equation}
\Omega_{i}^{\left(\text{BGK}\right)}\left(\mathbf{x},t\right)=-\frac{1}{\tau}\left[f_{i}\left(\mathbf{x},t\right)-f_{i}^{\left(\text{eq}\right)}\left(\mathbf{x},t\right)\right],
\end{equation}
which relaxes the populations towards a local equilibrium distribution $f_{i}^{\left(\text{eq}\right)}\left(\mathbf{x},t\right)$ at a characteristic rate given by the inverse of the relaxation time $\tau$. The local equilibrium is chosen as the second order expansion of the Maxwellian distribution
\begin{equation}
  f_{i}^{\left(\text{eq}\right)}=w_{i}n\left[1+\frac{\xi_{i}^{\alpha}u_{\alpha}^{\left(\text{eq}\right)}}{c_{s}^{2}}+\frac{(\xi_{i}^{\alpha}u_{\alpha}^{\left(\text{eq}\right)})^{2}}{c_{s}^{4}}-\frac{u_{\alpha}^{\left(\text{eq}\right)}u^{\left(\text{eq}\right)}_{\alpha}}{2c_{s}^{2}}\right],
\end{equation}
where we use the summation over repeated indices and omit the space-time dependence. In the previous expression one substitutes $u_{\alpha}^{\left(\text{eq}\right)}\left(\mathbf{x},t\right)$ with the fluid velocity computed from the local populations as described in~\eqref{eq:0and1M}. By means of the Chapman-Enskog expansion~\cite{Benzi1992,Chen1998,Wolf2004,kruger2017lattice,succi2018lattice}, it can be shown that the discretized transport equations converge to a conservation equation for the density $n$ and to the Navier-Stokes equation with a kinematic viscosity given by $\nu = c_s^2 (\tau - 1/2)$, and ideal gas equation of state given by $p = nc_s^2$. In order to implement the inter-particles forcing, we adopted the scheme proposed by Guo~\cite{Guo2002,kruger2017lattice}, according to which one modifies the equilibrium fluid velocity and the collision term as follows
\begin{equation}
u^{\left(\text{eq}\right)}_{\alpha}\left(\mathbf{x},t\right)=\frac{1}{n\left(\mathbf{x},t\right)}\sum_{i=0}^{N_{p}}\xi_{i}^{\alpha}f_{i}\left(\mathbf{x},t\right)+\frac{1}{2n\left(\mathbf{x},t\right)}F^{\alpha}\left(\mathbf{x},t\right),
\end{equation}
\begin{equation}
  \begin{split}\Omega_{i} & =\Omega_{i}^{\left(\text{BGK}\right)}\\
 & +\left(1-\frac{1}{2\tau}\right)w_{i}\left[\frac{1}{c_{s}^{2}}\xi_{i}^{\alpha}+\frac{1}{c_{s}^{4}}\left(\xi_{i}^{\alpha}\xi_{i}^{\beta}-c_{s}^{2}\delta^{\alpha\beta}\right)u_{\beta}^{\left(\text{eq}\right)}\right]F^{\alpha}
\end{split}
\end{equation}
which essentially represents a particular case of a multiple relaxation time approach~\cite{kruger2017lattice} with collisional matrix proportional to the identity matrix. With this scheme, we are able to implement the inter-particles forces described in the next Section, which modify the equation of state allowing for the coexistence of a liquid and a gas phase for suitable choices of the forcing parameters. Now that the LBM implementation of the forcing has been described, we will focus on the properties of the forcing itself, so that all the symmetry features of LBM, i.e. Galilean invariance, remain untouched by the following considerations.
%%%%%%%%%%%%%%%%%%%%%%%%%%%%%%%%%%%%%%%%%%%%%%%%%%%%%%%%%%%%%%%
%%%%%%%%%%%%%%%%%%%%%%%%%%%%%%%%%%%%%%%%%%%%%%%%%%%%%%%%%%%%%%%
\section{Lattice Force Isotropy}\label{sec:F}
%%%%%%%%%%%%%%%%%%%%%%%%%%%%%%%%%%%%%%%%%%%%%%%%%%%%%%%%%%%%%%%
%%%%%%%%%%%%%%%%%%%%%%%%%%%%%%%%%%%%%%%%%%%%%%%%%%%%%%%%%%%%%%
%%%%%%%%%%%%%%%%%%%%%%%%%%%%%%%%%%%%%%%%%%%%%%%%%%%%%%%%%%%%%
In this Section, we review the SC multi-phase forcing scheme and analyze its isotropy properties. The SC scheme~\cite{ShanChen94} is based on the definition of a body force resulting from the inter-particles interactions at each lattice point involving only a limited number of neighbors. The component $\mu$ of this local force is defined as
\begin{equation}\label{eq:sc_f}
F^{\mu}\left(\mathbf{x}\right)=-Gc_{s}^{2}\psi\left(\mathbf{x}\right)\sum_{\mathbf{e}_{a}\in\mathcal{G}}W\left(|\mathbf{e}_{a}|^{2}\right)\psi\left(\mathbf{x}+\mathbf{e}_{a}\right)e_{a}^{\mu},
\end{equation}
where $G$ is a (self-)coupling constant and the function $\psi(\mathbf{x}, t)=\psi(n(\mathbf{x}, t))$ is the so-called pseudo-potential, which is a generic function of the local density, hence implicitly depending on time and position. With $\mathbf{e}_a$ we indicate the stencil vectors which connect any given point $\mathbf{x}$ to its neighbors in a finite set $\mathcal{G}$, and finally with $W$ (distinguishing them from the weights $w_i$ of the lattice velocities) we indicate a set of weights which only depend on the squared length of the stencil vectors, i.e. $W(|\mathbf{e}_a|^2)$.\\
Given the discrete nature of this definition, one should look at the isotropy properties of the continuum limit of the forcing. This can be done by considering the Taylor expansion of the lattice force
\begin{equation}
\begin{split}F^{\mu}\left(\mathbf{x}\right)\simeq & -Gc_{s}^{2}\psi\left(\mathbf{x}\right)\left[\partial_{\alpha}\psi\left(\mathbf{x}\right)\sum_{\mathbf{e}_{a}\in\mathcal{G}}W\left(|\mathbf{e}_{a}|^{2}\right)e_{a}^{\alpha}e_{a}^{\mu}\right.\\
+\frac{1}{3!} & \left.\partial_{\alpha}\partial_{\beta}\partial_{\gamma}\psi\left(\mathbf{x}\right)\sum_{\mathbf{e}_{a}\in\mathcal{G}}W\left(|\mathbf{e}_{a}|^{2}\right)e_{a}^{\alpha}e_{a}^{\beta}e_{a}^{\gamma}e_{a}^{\mu}+\ldots\right],
\end{split}
\end{equation}
\begin{figure}[!t]
%\begin{center}
%[scale=0.415]
%[scale=0.545]
\includegraphics[scale=0.89]{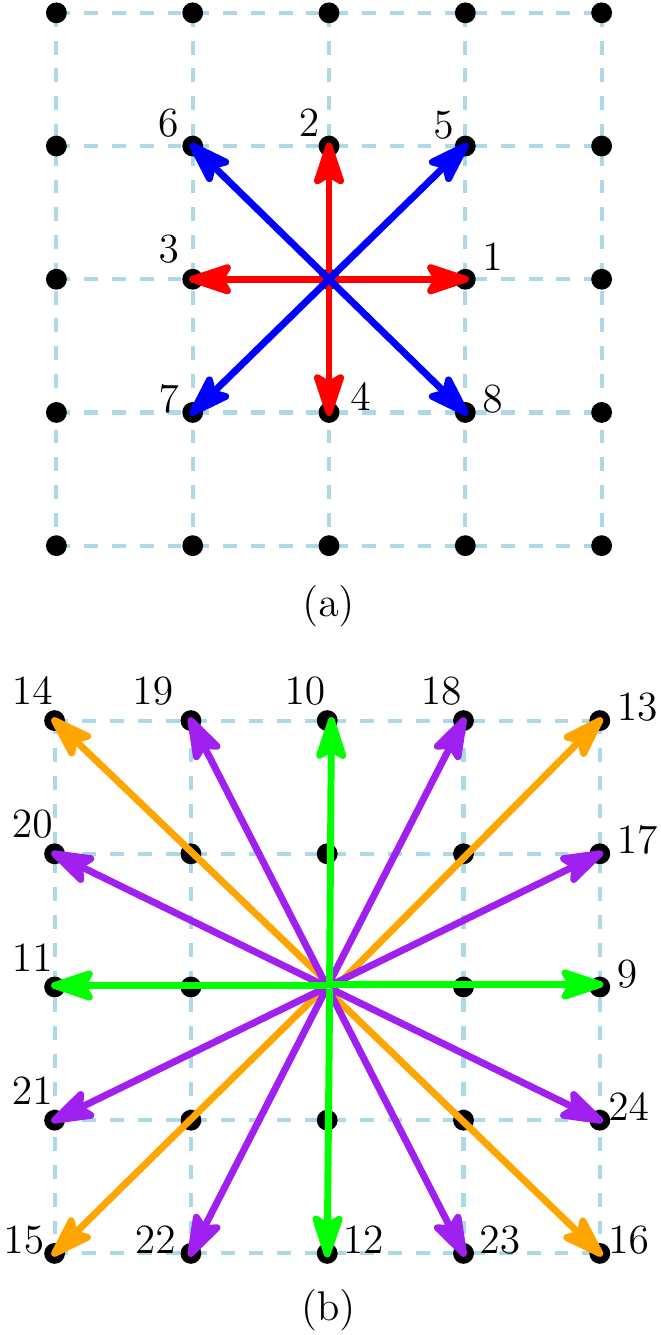}
%\end{center}
\caption{Set of basis vectors $\{\mathbf{e}_a\}$ (with $a=1,\ldots,24$) used to construct the forcing schemes presented in Sec.~\ref{sec:F}. Panels (a) and (b) show the numbers referred to lattice vectors, with the color coding for the different squared lengths $\{|\mathbf{e}_{a}|^{2} = 1,2,4,5,8\}$. Using these vectors, one can define $4$-th, $6$-th and $8$-th order isotropic forcing schemes~\cite{Shan06,Sbragaglia07,Falcucci07}, labeled as $\boldsymbol{E}^{(4)}$, $\boldsymbol{E}^{(6)}$ and $\boldsymbol{E}^{(8)}$, respectively.}\label{fig:stencil}
\end{figure}where one can notice the summations involving the products of an even number of basis vectors. We will now analyze in detail the isotropy properties of these quantities, which in turn determine the isotropy of the forcing. As a first step, we collect the $\mathbf{e}_a$ vectors in groups, according to their squared lengths, i.e. $\mathcal{G}_{\ell} = \{\mathbf{e}_a : |\mathbf{e}_a|^2 = \ell\}$ (although $\ell$ is not a unique label for $\ell \geq 25$ in 2D~\footnote{Note that the square length might not be a unique label when the former is large enough: in two dimensions, for example, this happens for $\ell = 25$ which can be obtained starting from either $\mathbf{e}_{\ell = 25} = (5, 0)$ or from $\mathbf{e}_{\ell = 25} = (4, 3)$, which, however, are not related by a spatial parity or coordinates permutations operations. Since we will present the details only for the stencil featuring vectors such that $\ell = |\textbf{e}_a|^2 \leq 8$ (cf. Fig.~\ref{fig:stencil}), we will keep on using the simplified notation $\mathcal{G}_\ell$.}). Typical requirements are that each group $\mathcal{G}_\ell$ contains vectors that are related either by spatial parity or coordinates permutations combined with alternate sign changes. In the following we will be using only vectors such that $\ell = |\textbf{e}_a|^2 \leq 8$ (cf. Fig.~\ref{fig:stencil}). Such stencil can be employed to define $4$-th, $6$-th or $8$-th order isotropy multi-range forcing that we denote~\cite{Shan06,Sbragaglia07,Falcucci07,Shan08} as $\boldsymbol{E}^{(4)}$, $\boldsymbol{E}^{(6)}$ and $\boldsymbol{E}^{(8)}$, respectively. The symmetry requirement for vectors belonging to the same group are  enough to ensure that the sum of the product of an odd number of stencil vectors will add up to zero, i.e.
\begin{equation}
\sum_{\mathbf{e}_{a}\in\mathcal{G}_{\ell}} e_{a}^{\mu_{1}}e_{a}^{\mu_{2}}\cdots e_{a}^{\mu_{2n+1}}=0.
\end{equation}
Hence, we introduce the $2n$-indices quantities defined by the relation
\begin{equation}\label{eq:Eall}
  \begin{split}E^{\mu_{1}\ldots\mu_{2n}} & =\sum_{\mathbf{e}_{a}\in\mathcal{G}}W\left(|\mathbf{e}_{a}|^{2}\right)e_{a}^{\mu_{1}}e_{a}^{\mu_{2}}\cdots e_{a}^{\mu_{2n}}\\
 & =E_{\text{iso}}^{\mu_{1}\ldots\mu_{2n}}+E_{\text{aniso}}^{\mu_{1}\ldots\mu_{2n}},
\end{split}
\end{equation}
where $E_{\text{iso}}^{\mu_{1}\ldots\mu_{2n}}$ and $E_{\text{aniso}}^{\mu_{1}\ldots\mu_{2n}}$ indicate the isotropic and anisotropic contributions respectively. Notice that the previous decomposition holds for $2n > 2$ since for $2n = 2$ all the contributions are proportional to the Kronecker delta. The main idea~\cite{Wolfram1986, Shan06, Sbragaglia07} is to choose the weights $\{W(|\textbf{e}_a|^2)\}$ so that only the isotropic contributions survive
\begin{equation}\label{eq:iso_cond_short}
E^{\mu_{1}\ldots\mu_{2n}} = E_{\text{iso}}^{\mu_{1}\ldots\mu_{2n}} = e_{2n}\,\Delta^{\mu_1\ldots\mu_{2n}},
\end{equation}
where the isotropy constants $e_{2n}$ multiply the fully isotropic $2n$-rank tensor $\Delta^{\mu_1\ldots\mu_{2n}}$~\cite{Wolfram1986, Shan06, Sbragaglia07}. Generalizing, in two dimensions, the approach of~\cite{Wolfram1986}, the anisotropic contributions can be written as
\begin{equation}\label{eq:E_aniso}
E_{\text{aniso}}^{\mu_{1}\ldots\mu_{2n}}=\sum_{k=0}^{M\left(n\right)/2}I_{2n,k}\left[\delta^{\mu_{1}\ldots\mu_{2k}}\delta^{\mu_{2k+1}\ldots\mu_{2n}}+\text{perms}\right],
\end{equation}
where $\delta^{\mu_1\ldots\mu_{2n}}$ is the higher rank Kronecker delta, which is not isotropic and equals 1 only if all indices take the same value, and the upper limit for $2k$ is $M(n) = n - (2 + n\,\mbox{mod}\,2)$ with $n \geq 2$; finally, ``perms'' stands for all the possible independent indices permutations, whose number is $(2n)!/(2n - 2k)!(2k)!$. The isotropy coefficients $e_{2n}$, multiplying $\Delta^{\mu_1\ldots\mu_{2n}}$, and the anisotropy ones $I_{2n,k}$, multiplying terms proportional to $\delta^{\mu_1\ldots\mu_{2n}}$, can be generally written as combinations of the weights
\begin{equation}
  \begin{split} & e_{2n}=\sum_{\ell}\mathcal{A}^{\left(2n\right)}\left(\ell\right)W\left(\ell\right),\\
 & I_{2n,k}=\sum_{\ell}\mathcal{B}_{2n-2k}^{\left(2n\right)}\left(\ell\right)W\left(\ell\right),
\end{split}
\end{equation}
where the coefficients $\mathcal{A}^{\left(2n\right)}$ and $\mathcal{B}_{2n-2k}^{\left(2n\right)}$ depend on $\ell=|\textbf{e}_a|^2$. More details are reported in the Appendix~\ref{app:IsoDetails} and~\ref{app:IsoCondCompare}. In order to obtain the weights for the 6-th order isotropic forcing $\boldsymbol{E}^{(6)}$~\cite{Shan06,Sbragaglia07,Falcucci07} one sets $I_{4,0}=0$ and $I_{6,0}=0$ which are linear combinations of $\{W(1), W(2), W(4)\}$. For the 8-th order isotropic forcing, $\boldsymbol{E}^{(8)}$, one has to consider, alongside $I_{4,0}=0$ and $I_{6,0}=0$, the conditions $I_{8,0}=0$ and $I_{8,1}=0$, which are now combinations of $\{W(1), W(2), W(4), W(5), W(8)\}$. Similar arguments hold for higher order isotropy.

We wish to stress that the $2n$-order isotropy can only be achieved for the tensorial structure in Eq.~\eqref{eq:Eall} and not for the same structure computed for each group separately, because the isotropy conditions can only be satisfied by using linear combinations of the weights. However, the restriction to a single group of Eq.~\eqref{eq:Eall} plays a crucial role in the identification of the anisotropic terms of the Taylor expansion of the pressure tensor (see Section~\ref{sec:THEORY} and Appendix~\ref{app:IsoDetails} and~\ref{app:IsoCondCompare} for details).
%%%%%%%%%%%%%%%%%%%%%%%%%%%%%%%%%%%%%%%%%%%%%%%%%%%%%%%%%%%%%%%
\section{Lattice Pressure Tensor}\label{sec:PT}
%%%%%%%%%%%%%%%%%%%%%%%%%%%%%%%%%%%%%%%%%%%%%%%%%%%%%%%%%%%%%%%
\begin{figure}[!t]
\includegraphics[scale=0.9]{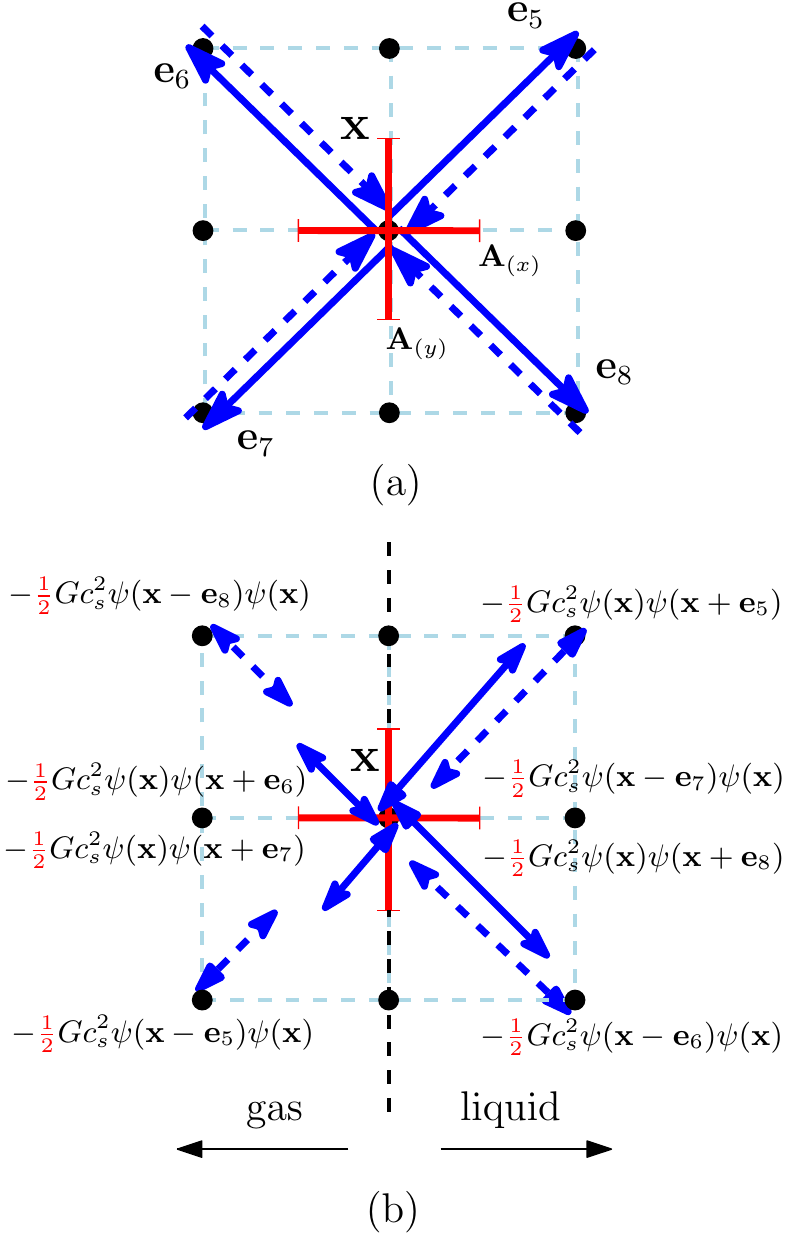}
\caption{ Panel (a): Forcing directions $\textbf{e}_a$ belonging to the $\mathcal{G}_2$ group and centered at the point $\textbf{x}$. Vectors starting at $\textbf{x}$ are reported in solid lines, while those ending in $\textbf{x}$ are dashed. Unit area elements $\textbf{A}_{(k)}$ are reported in red. Panel (b): different contributions to the average force $\bar{F}_a$ [see Eq.~\eqref{eq:aveF5}] at a gas-liquid flat interface: double arrows stand for the magnitude of the contribution specified in the adjacent expression. Notice that the contributions on the gas side are smaller in magnitude due to smaller pseudo-potential (i.e. smaller density) values. The average force $\bar{F}_{a}$ is the same for all directions.}\label{fig:force_area_G2}
\end{figure}

Let us now review the definition of the lattice formulation of the pressure tensor for the SC model~\cite{Shan08}: this will be the starting point for the study of its isotropy properties. All details will be specified for the forcing stencils reported in Fig.~\ref{fig:stencil}, i.e. using five weights $\{W(1), W(2), W(4), W(5), W(8)\}$ in two dimensions. The procedure described in~\cite{Shan08} allows us to define the interaction pressure tensor, $P_{\text{int}}^{\mu\nu}$ directly on the lattice. The total lattice pressure tensor is given by summing the latter to the kinetic pressure tensor which for LBM simply amounts to the ideal gas isotropic contribution $P_{\text{kin}}^{\mu\nu}(\textbf{x}) = n(\textbf{x}) c_s^2 \delta^{\mu\nu}$, hence $P_{\text{tot}}^{\mu\nu} = P_{\text{kin}}^{\mu\nu} + P_{\text{int}}^{\mu\nu}$. Given this distinction, we will use the notation $P^{\mu\nu}$ for the interaction part in the rest of the paper, adding the ideal contribution when needed. We report a detailed review for the definition of the lattice pressure tensor in Appendix~\ref{app:LPT_def} and briefly report here the main points. Following~\cite{Shan08,SbragagliaBelardinelli13} we write, in tensorial form on the lattice, the total force crossing a given unit area element as the pressure flux through the same element, which for each group $\mathcal{G}_\ell$ reads
\begin{equation}\label{eq:discrete_p}
F_{\ell,\left(k\right)}^{\mu}\left(\mathbf{x}\right)=\sum_{\mathbf{e}_{a}\in\mathcal{G}_{\ell}}F_{a,\left(k\right)}^{\mu}\left(\mathbf{x}\right)=-\sum_{\mathbf{e}_{a}\in\mathcal{G}_{\ell}}P_{a}^{\mu\alpha}\left(\mathbf{x}\right)\;A_{\left(k\right)}^{\alpha},
\end{equation}
where $\mathbf{A}_{(y)} = \mathbf{e}_1$ and $\mathbf{A}_{(x)} = \mathbf{e}_2$ are the unit areas (characterized by their normal vectors), with $\mathbf{e}_1$ and $\mathbf{e}_2$ the coordinate basis vectors (see Fig.~\ref{fig:stencil}(a) and~\ref{fig:force_area_G2}(a)), and $F_{\ell,\left(k\right)}^{\mu}\left(\mathbf{x}\right)$ is the group total force crossing the area element $\mathbf{A}_{(k)}$, while $F_{a,\left(k\right)}^{\mu}\left(\mathbf{x}\right)$ is the specific force contribution along the direction $\textbf{e}_a$ (see Fig.~\ref{fig:stencil}(a) and (b)). It is possible (see Appendix~\ref{app:LPT_def}) to rewrite the latter as
\begin{equation}\label{eq:single_farea}
F_{a,\left(k\right)}^{\mu}\left(\mathbf{x}\right)=\bar{F}_{a}\left(\mathbf{x}\right)\,e_{a}^{\alpha}e_{a}^{\mu}\,A_{\left(k\right)}^{\alpha}=-P_{a}^{\mu\alpha}\left(\mathbf{x}\right)\,A_{\left(k\right)}^{\alpha},
\end{equation}
from which we read the definition of the lattice pressure tensor
\begin{equation}\label{eq:LP_def}
P_{a}^{\mu\nu}\left(\mathbf{x}\right)=-\bar{F}_{a}\left(\mathbf{x}\right)\,e_{a\
}^{\mu}e_{a}^{\nu}.
\end{equation}
We define $\bar{F}_a$ as a weighted average of the norm of the force vectors crossing the largest number of times any of the area elements $\textbf{A}_{(k)}$ (cf. Appendix~\ref{app:LPT_def}). As an example, for $\textbf{e}_5$ the average force is given by
\begin{equation}\label{eq:aveF5}
  \bar{F}_{5}=-Gc_{s}^{2}W\left(2\right)\psi\left(\mathbf{x}\right)\left[\frac{1}{2}\psi\left(\mathbf{x}-\mathbf{e}_{5}\right)+\frac{1}{2}\psi\left(\mathbf{x}+\mathbf{e}_{5}\right)\right].
\end{equation}
We report in Fig.~\ref{fig:force_area_G2}(b) a sketch depicting the above expression for all different contributions of the group $\ell = 2$, to which $\textbf{e}_5$ belongs, at one node of a flat gas-liquid interface.

Now, we write the lattice pressure tensor for each group of vectors. Starting from the single-force directions (see Appendix~\ref{app:LPT_def} for details) we can write the total sum for the groups $\{\mathcal{G}_{1}, \mathcal{G}_{2}, \mathcal{G}_{4},\mathcal{G}_{8}\}$ in a compact form
\begin{equation}\label{eq:p_12}
  P_{\left(1,2\right)}^{\mu\nu}=\frac{Gc_{s}^{2}}{2}\psi\left(\mathbf{x}\right)\sum_{\mathbf{e}_{a}\in\left\{ \mathcal{G}_{1},\mathcal{G}_{2}\right\} }W\left(|\mathbf{e}_{a}|^{2}\right)\psi\left(\mathbf{x}+\mathbf{e}_{a}\right)e_{a}^{\mu}e_{a}^{\nu},
\end{equation}
\begin{equation}\label{eq:p_48}
  \begin{split}P_{\left(4,8\right)}^{\mu\nu}=\frac{Gc_{s}^{2}}{4}\psi\left(\mathbf{x}\right) & \sum_{\mathbf{e}_{a}\in\left\{ \mathcal{G}_{4},\mathcal{G}_{8}\right\} }W\left(|\mathbf{e}_{a}|^{2}\right)\psi\left(\mathbf{x}+\mathbf{e}_{a}\right)e_{a}^{\mu}e_{a}^{\nu}\\
+\frac{Gc_{s}^{2}}{4}\sum_{\mathbf{e}_{a}\in\left\{ \mathcal{G}_{4},\mathcal{G}_{8}\right\} } & W\left(|\mathbf{e}_{a}|^{2}\right)\psi\left(\mathbf{x}+\frac{\mathbf{e}_{a}}{2}\right)\psi\left(\mathbf{x}-\frac{\mathbf{e}_{a}}{2}\right)e_{a}^{\mu}e_{a}^{\nu}.
\end{split}
\end{equation}
Considering the group $\mathcal{G}_5$ and following~\cite{Shan08}, we define two different contributions for the pressure tensor, namely $(5a)$ including the directions starting or ending in $\textbf{x}$ and $(5b)$ for those starting and ending on the neighbors:
\begin{equation}\label{eq:p_5a}
P_{5a}^{\mu\nu}=\frac{Gc_{s}^{2}}{4}W\left(5\right)\psi\left(\mathbf{x}\right)\sum_{\mathbf{e}_{a}\in\mathcal{G}_{5}}\psi\left(\mathbf{x}+\mathbf{e}_{a}\right)\,e_{a}^{\mu}e_{a}^{\nu},
\end{equation}
\begin{equation}\label{eq:p_5b}
\begin{split}P_{5b}^{\mu\nu}= & \frac{Gc_{s}^{2}}{4}W\left(5\right)\left[\psi_{-1,-1}\psi_{1,0}+\psi_{-1,0}\psi_{1,1}\right]\,e_{17}^{\mu}e_{17}^{\nu}\\
+ & \frac{Gc_{s}^{2}}{4}W\left(5\right)\left[\psi_{-1,-1}\psi_{0,1}+\psi_{0,-1}\psi_{1,1}\right]\,e_{18}^{\mu}e_{18}^{\nu}\\
+ & \frac{Gc_{s}^{2}}{4}W\left(5\right)\left[\psi_{0,1}\psi_{1,-1}+\psi_{-1,1}\psi_{0,-1}\right]\,e_{19}^{\mu}e_{19}^{\nu}\\
+ & \frac{Gc_{s}^{2}}{4}W\left(5\right)\left[\psi_{1,0}\psi_{-1,1}+\psi_{1,-1}\psi_{-1,0}\right]\,e_{20}^{\mu}e_{20}^{\nu},
\end{split}
\end{equation}
where we used the short-hand notation $\psi_{a,b} = \psi(\textbf{x} + a\textbf{e}_1 + b\textbf{e}_2)$~\cite{Shan08}.

The interaction lattice pressure tensor for the multi-range SC forcing defined on the stencils in Fig.~\ref{fig:stencil} can be obtained by summing all the different contributions, i.e.
\begin{equation}\label{eq:pt_E8}
  P_{}^{\mu\nu}\left(\mathbf{x}\right) = P_{\left(1,2\right)}^{\mu\nu}\left(\mathbf{x}\right) + P_{\left(4,8\right)}^{\mu\nu}\left(\mathbf{x}\right) + P_{5a}^{\mu\nu}\left(\mathbf{x}\right) + P_{5b}^{\mu\nu}\left(\mathbf{x}\right).
\end{equation}
In the next Section we analyze the isotropy properties of this lattice pressure tensor using a $4$-th order expansion.
%%%%%%%%%%%%%%%%%%%%%%%%%%%%%%%%%%%%%%%%%%%%%%%%%%%%%%%%%%%%%%%
\section{Isotropy Analysis \& Modified Forcing Schemes}\label{sec:THEORY}
%%%%%%%%%%%%%%%%%%%%%%%%%%%%%%%%%%%%%%%%%%%%%%%%%%%%%%%%%%%%%%%
We study now the continuum limit of the lattice pressure tensor by using its Taylor expansion up to second order derivatives and products of first ones. This, in turn, will yield an analysis of the isotropy properties up to the $4$-th order. We do not consider any specific solution for the weights so that we can analyze the role of the anisotropic terms.

Starting from Eqs~\eqref{eq:p_12}, \eqref{eq:p_48}, \eqref{eq:p_5a} and \eqref{eq:p_5b}, and following the procedure detailed in Appendix~\ref{app:PTDetails}, we merge together the Taylor expansions of all the different contributions and obtain the general form for the $4$-th order expansion of the lattice pressure tensor, i.e. involving second order and products of first order derivatives, for the $\boldsymbol{E}^{(4)}$, $\boldsymbol{E}^{(6)}$ and $\boldsymbol{E}^{(8)}$ forcing stencils
\begin{equation}\label{eq:P4E8}
\begin{split}P^{\mu\nu}= & \left(nc_{s}^{2}+\frac{Gc_{s}^{2}e_{2}}{2}\psi^{2}\right)\delta^{\mu\nu}\\
 & +Gc_{s}^{2}\left(\Lambda_{N}\psi\nabla^{2}\psi-\chi_{N}\big|\nabla\psi\big|^{2}\right)\delta^{\mu\nu}\\
 & +Gc_{s}^{2}\left(\Lambda_{T}\psi\partial^{\mu}\partial^{\nu}\psi-\chi_{T}\partial^{\mu}\psi\partial^{\nu}\psi\right)\\
 & +Gc_{s}^{2}\left(\Lambda_{I}\psi\partial_{\alpha}\partial_{\beta}\psi-\chi_{I}\partial_{\alpha}\psi\partial_{\beta}\psi\right)\delta^{\alpha\beta\mu\nu},
\end{split}
\end{equation}
with the constants of the isotropic contributions given by $\Lambda_{N}=W\left(2\right)+12W\left(8\right)+7W\left(5\right)$, $\chi_{N}=W\left(5\right)+4W\left(8\right)$, $\Lambda_{T}=2\left[W\left(2\right)+12W\left(8\right)+6W\left(5\right)\right]$ and $\chi_{T}=4\left[W\left(5\right)+2W\left(8\right)\right]$. Anisotropic contributions of derivatives contracted with $\delta^{\alpha\beta\mu\nu}$, appear. The latter are multiplied by the coefficients
\begin{equation}\label{eq:p4_niso}
\begin{split}\Lambda_{I} & =\frac{1}{2}W\left(1\right)-2W\left(2\right)+6W\left(4\right)-6W\left(5\right)-24W\left(8\right),\\
\chi_{I} & =2W\left(4\right)-W\left(5\right)-8W\left(8\right).
\end{split}
\end{equation}
Equation~\eqref{eq:P4E8} is a general expression for the expansion of the lattice pressure tensor for $\boldsymbol{E}^{(4)}$, $\boldsymbol{E}^{(6)}$ and $\boldsymbol{E}^{(8)}$ in tensorial form, displaying clear information about the isotropy properties of the pressure tensor.

%% $\{I_{2n,k}=0\}$, i.e. Eq.~\eqref{eq:IsoCond},
Now, one should ask whether the coefficients $\Lambda_{I}$ and $\chi_{I}$ automatically vanish when the isotropy conditions for the forcing are satisfied. The answer is negative. Indeed, one can see that, for the present choice of the vectors $\{\textbf{e}_a\}$, the $4$-th order isotropy equation for the forcing, i.e. $I_{4,0}=0$, is given by a combination of the coefficients $\chi_I$ and $\Lambda_I$
\begin{equation}\label{eq:I4_LC}
\begin{split}I_{4,0} & =2W(1)-8W(2)+32W(4)-28W(5)-128W(8)\\
 & =4\left(\Lambda_{I}+\chi_{I}\right)=0.
\end{split}  
\end{equation}
The last result implies that requiring the $4$-th order isotropy for the lattice pressure tensor expansion, i.e. $\chi_I = \Lambda_I = 0$, does imply the $4$-th order isotropy condition for the forcing, but not vice versa. Indeed, all multi-range forcings above the $4$-th order isotropy, i.e. above the single belt, suffer this issue. However, the $4$-th order, or single-belt, stencil $\boldsymbol{E}^{(4)}$ automatically yields an isotropic expression of the continuum limit of the lattice pressure tensor at the 4-th order. This happens because in the single belt case $\chi_I = 0$ trivially, so that $I_{4,0} = 4\Lambda_I = 0$, i.e. 4-th order pressure and forcing isotropy are obtained with the same condition. This is probably the reason why the anisotropy of the pressure tensor went unnoticed so far.

Indeed, the fact that the 4-th order pressure tensor isotropy is implemented by means of two equations, i.e. $\chi_I=0$ and $\Lambda_I = 0$, and not only one as for the forcing case, i.e. $I_{4,0}=0$, implies that, for a fixed number of weights, the solution leading to a higher pressure tensor isotropy must also yield a lower forcing isotropy. However, as we will show in the next Section, this delivers a reduction of the spurious currents, rather than an increase in magnitude and extension as one would have expected~\cite{Shan06,Sbragaglia07,Falcucci07}.

We now wish to understand what are the effects of a higher isotropy order for the pressure tensor. To do so, we will compare forcing schemes with the same values for the isotropy constants $\{e_{2n}\}$ up to a given order, while changing the pressure tensor degree of isotropy. As we discuss in the following, this operative strategy allows to keep the bulk and interface properties, i.e. equation of state, flat interface profile and surface tension, unchanged when comparing the two forcing schemes. This will allow to better highlight the effects induced by the pressure tensor anisotropy.

\subsection{Mechanical Equilibrium Analysis}\label{subsec:MEQ}
%%%%%%%%%%%%%%%%%%%%%%%%%%%%%%%%%%%%%
\begin{table*}[t!]
  \centering
  \begin{ruledtabular}
  \renewcommand{\arraystretch}{1.6}
  \begin{tabular}{llcc|cc|cc|cc}
    & & $\boldsymbol{E}^{(6)}_{P2,F6}$ & $\boldsymbol{E}^{(6)}_{P4,F6}$ & $\boldsymbol{E}^{(8)}_{P2,F8}$ & $\boldsymbol{E}^{(8)}_{P4,F6}$ & $\boldsymbol{E}^{(10)}_{P2,F10}$ & $\boldsymbol{E}^{(10)}_{P4,F6}$ & $\boldsymbol{E}^{(12)}_{P2,F12}$ & $\boldsymbol{E}^{(12)}_{P4,F6}$ \\
    \hline
    & $e_2$ & $1$ & $\mathbf{1}$ & $1$ & $\mathbf{1}$ & $1$ & $\mathbf{1}$ & $1$ & $\mathbf{1}$ \\
    Surface Tension & $e_4$ & $2/5$ & $2/5$ & $4/7$ & $4/7$ & $12/17$ & $12/17$ & $120/143$ & $120/143^{(*)}$\\
    Flat Profile & $\varepsilon$ & $2/17$ & $\mathbf{2/17}$ & $10/31$ & $\mathbf{10/31}$ & $38/89$ & $\mathbf{38/89}$ & $136774/271813$ & $\mathbf{136774/271813}$ \\
    \hline
    Pressure Isotropy Condition & $\Lambda_I$ & $-1/60$ & $0$ & $-8/315$ & $0$ & $-$ & $0$ & $-$ & $0$\\
    & $\chi_I$ & $1/60$ & $\mathbf{0}$ & $8/315$ & $\mathbf{0}$ & $-$ & $\mathbf{0}$ & $-$ & $\mathbf{0}$ \\
    \hline
    Forcing Isotropy Condition & $I_{4,0}$ & $0$ & $\mathbf{0}$ & $0$ & $\mathbf{0}$ & $0$ & $\mathbf{0}$ & $0$ & $\mathbf{0}$ \\
    & $I_{6,0}$ & $0$ & $\mathbf{0}$ & $0$ & $\mathbf{0}$ & $0$ & $\mathbf{0}$ & $0$ & $\mathbf{0}$ \\
  \end{tabular}
  \end{ruledtabular}
  \caption{Values of the isotropy ($e_{2n}$ and $\varepsilon$) constants for different forcing schemes along with the force and pressure isotropy conditions. The non-zero values of $\chi_I$ and $\Lambda_I$ single out the stencils yielding a $2$-nd order isotropy for the pressure tensor. We report in bold the values that are set in Eq.~\eqref{eq:newsys}, used to determine the weights for the new schemes, as reported in Table~\ref{tab:W_vals} (see Section~\ref{subsec:match} for discussion). $^{(*)}$ The actual value is $37800/45013$ differing from $120/143$ by $6\cdot10^{-4}$.}\label{tab:FP_vals}
\end{table*}
%%%%%%%%%%%%%%%%%%%%%%%%%%%%%%%%%%%%%
Let us start by analyzing the mechanical equilibrium condition of a flat interface. Assuming that the density field $n$ depends on $x$ only, we write the normal and tangential component to the interface, i.e. $P^{xx}$ and $P^{yy}$ respectively, as
\begin{equation}\label{eq:p1d}
\begin{split}P^{xx}= & nc_{s}^{2}+\frac{Gc_{s}^{2}e_{2}}{2}\psi^{2}+\frac{Gc_{s}^{2}}{12}\left[\beta\psi\frac{\text{d}^{2}\psi}{\text{d}x^{2}}+\alpha\left(\frac{\text{d}\psi}{\text{d}x}\right)^{2}\right]\\
P^{yy}= & nc_{s}^{2}+\frac{Gc_{s}^{2}e_{2}}{2}\psi^{2}+\frac{Gc_{s}^{2}}{4}\left[\eta\psi\frac{\text{d}^{2}\psi}{\text{d}x^{2}}+\gamma\left(\frac{\text{d}\psi}{\text{d}x}\right)^{2}\right]
\end{split}
\end{equation}
with $P^{xy}=0$. As for the coefficients $\alpha$, $\beta$, $\gamma$ and $\eta$ we use the same notation as in~\cite{Shan08}. These can be expressed using the coefficients of the general expression in Eq.~\eqref{eq:P4E8} as follows: $-\alpha/12=\chi_{N}+\chi_{T}+\chi_{I}$, $\beta/12=\Lambda_{N}+\Lambda_{T}+\Lambda_{I}$, $\eta/4=\Lambda_{N}$ and $-\gamma/4=\chi_{N}$. The mechanical equilibrium condition, i.e. $\partial_\mu P^{\mu\nu} = 0$, implies $P^{xx}(x) = p_0$ for a flat interface, i.e. the pressure normal to the interface must not change from one bulk phase to the other, and through the interface itself. We wish to stress~\cite{SbragagliaBelardinelli13,Shan08} that, as demonstrated in simulations, the lattice pressure tensor in Eq.~\eqref{eq:pt_E8} is observed to be numerically constant at machine precision in the bulk and through the interface, i.e. it exactly implements the mechanical equilibrium condition \emph{on the lattice}. Starting from the mechanical equilibrium condition $P^{xx}(x) = p_0$ and making use of the identity $\frac{\text{d}^{2}\psi}{\text{d}x^{2}}=\frac{1}{2}\frac{\text{d}}{\text{d}\psi}(\frac{\text{d}\psi}{\text{d}x})^{2}$  it is possible to write the following equation for the square of the density profile derivative as a function of the density $n$
\begin{equation}\label{eq:n_prof}
\left(\frac{\text{d}n}{\text{d}x}\right)^{2}=\frac{24\psi^{\varepsilon}}{Gc_{s}^{2}\beta\psi'^{2}}\int_{n_{g}}^{n}\text{d}\bar{n}\frac{\psi'}{\psi^{\varepsilon+1}}\left[p_{0}-\bar{n}c_{s}^{2}-\frac{Gc_{s}^{2}e_{2}}{2}\psi^{2}\right],
\end{equation}
where $\psi' = \text{d}\psi/\text{d}n$, $\varepsilon = -2\alpha/\beta$ and $24/\beta = 8(1 - \varepsilon)/e_2$~\cite{Shan08}\footnote{As an additional remark with respect to~\cite{Shan08}, this result can be obtained from writing the 5 forcing weighs $\{W(|\textbf{e}_a|^2)\}$ as a function of $\varepsilon$ and the 4 isotropy coefficients $e_{2n}$ (cf. Eq.~\eqref{eq:W_E}). In order to fully recover the definitions of~\cite{Shan08} we need to solve with respect to $\varepsilon$ the 4-th order isotropy condition $I_{4,0}=0$ reported in Eq.~\eqref{eq:I_E} and obtain $\varepsilon = (6e_4 - 2e_2)/(6e_4 + e_2)$. We can use this relation and compute the expressions for the constants $\alpha = 3\varepsilon e_{2}/2(\varepsilon - 1) = e_{2} - 3e_{4}$ and $\beta = -3 e_{2}/(\varepsilon - 1) = e_{2} + 6e_{4}$. On the other hand, the two remaining constants do not depend on $\varepsilon$, yielding $\eta = 13e_{4}/12 - e_{6}/4$ and $\gamma = e_{4}/12 - e_{6}/4$ with $\eta - \gamma = e_{4}$ as in~\cite{Shan08}.} (see Appendix~\ref{app:w2e} for details).

Since the density derivative is zero in the bulk phases, we have the integral constraint
\begin{equation}\label{eq:Maxwell}
\int_{n_{g}}^{n_{l}}\text{d}\bar{n}\frac{\psi'}{\psi^{\varepsilon+1}}\left[p_{0}-\bar{n}c_{s}^{2}-\frac{Gc_{s}^{2}e_{2}}{2}\psi^{2}\right]=0 \ ,
\end{equation}
where $n_g$ and $n_l$ are the densities of the bulk gas and liquid phases.
Assuming $G<G_c$ (where $G_c$ is the critical value below which two-phase coexistence is possible), Eq.~\eqref{eq:Maxwell} coupled to the mechanical equilibrium requirement of equal bulk pressures [see Eq.~\eqref{eq:bulk_p}] in both phases
\begin{equation}\label{eq:flat_press}
P^{\mu\nu}_b(n_g) = P^{\mu\nu}_b(n_l) = p_0\delta^{\mu\nu},
\end{equation}
allows to uniquely determine the values of $n_g$ and $n_l$ as functions of the coupling $G$. As one can see, the properties of the stencils of the multi-range forcing explicitly enter in Eq.~\eqref{eq:Maxwell} through the constant $\varepsilon$, which also appears in the definition of the profile derivative in Eq.~\eqref{eq:n_prof}. Hence, by matching the isotropy constant  $e_2$ and $\varepsilon$, we obtain the same equation of state and same density profile for the flat interface.

\subsection{Surface Tension Analysis}
Let us now continue with the surface tension of the flat interface which is given by the integral
\begin{equation}\label{eq:sigma}
\begin{split}\sigma= & \int_{-\infty}^{+\infty}\text{d}x\left[P^{xx}\left(x\right)-P^{yy}\left(x\right)\right]\\
= & -Gc_{s}^{2}\left(\chi_{T}+\Lambda_{T}+\chi_{I}+\Lambda_{I}\right)\int_{-\infty}^{+\infty}\text{d}x\left[\frac{\text{d}\psi\left(x\right)}{\text{d}x}\right]^{2}.
\end{split}
\end{equation}
Assuming the use of a forcing scheme for which the $4$-th order forcing isotropy condition $I_{4,0} = 0$ is fulfilled, given Eq.~\eqref{eq:I4_LC} it automatically follows that $\chi_I + \Lambda_I = 0$, i.e. the surface tension does not depend on the anisotropy coefficients. In other words, the anisotropies of the lattice pressure tensor do not affect the value of $\sigma$, securing its physical meaning (and, thus, its use in practical applications, such as the contact angle calculations~\cite{From2020} and spray formation/break-up~\cite{falcucci2010spray}) for higher order stencils.

In order to complete the comparison with~\cite{Shan08}, we compute the value of the combination $\chi_{T}+\Lambda_{T}$, resulting in (see Appendix~\ref{app:w2e} for details)
\begin{equation}\label{eq:sigma_e4}
\chi_{T}+\Lambda_{T}=\frac{e_{4}}{2},
\end{equation}
which coincides with the result reported in~\cite{Shan08}. Hence, matching $e_4$, in addition to $e_2$ and $\varepsilon$ as discussed in Section~\ref{subsec:MEQ}, eventually yields the same surface tension of any reference multi-range forcing.

\subsection{Macroscopic Matching Strategy}\label{subsec:match}
In order to match the forcing expansion and the bulk and interface properties, we express $\{e_{2n}\}$ as functions of the weights $\{W(|\textbf{e}_a|^2)\}$ (see Appendix~\ref{app:w2e}).
To do so, we employ a new group-wise parametrization of $E^{\mu_{1}\ldots\mu_{2n}}$, extending the 6-th order one presented in~\cite{Wolfram1986} (see Appendix~\ref{app:IsoDetails} for the details).
\begin{table*}[t!]
  \centering
  \begin{ruledtabular}
  \renewcommand{\arraystretch}{1.6}
  \begin{tabular}{lcccccccccc}
    & $W(1)$ & $W(2)$ & $W(4)$ & $W(5)$ & $W(8)$ & $W(9)$ & $W(10)$ & $W(13)$ & $W(16)$ & $W(17)$\\
    \hline
    $\boldsymbol{E}^{(6)}_{P2,F6}$ & $4/15$ & $1/10$ & $1/120$ & $-$ & $-$ & $-$ & $-$ & $-$ & $-$ & $-$\\
    $\boldsymbol{E}^{(6)}_{P4,F6}$ & $19/60$ & $1/15$ & $-1/240$ & $1/120$ & $-1/480$ & $-$ & $-$ & $-$ & $-$ & $-$\\
    \hline
    \hline
    $\boldsymbol{E}^{(8)}_{P2,F8}$ & $4/21$ & $4/45$ & $1/60$ & $2/315$ & $1/5040$ & $-$ & $-$ & $-$ & $-$ & $-$\\
    $\boldsymbol{E}^{(8)}_{P4,F6}$ & $4/15$ & $4/105$ & $-1/420$ & $2/105$ & $-1/336$ & $-$ & $-$ & $-$ & $-$ & $-$ \\
    \hline
    \hline
    $\boldsymbol{E}^{(10)}_{P2,F10}$ & $262/1785$ & $93/1190$ & $7/340$ & $6/595$ & $9/9520$ & $2/5355$ & $1/7140$ & $-$ & $-$ & $-$\\
    $\boldsymbol{E}^{(10)}_{P4,F6}$ & $58/255$ & $4/255$ & $-1/1020$ & $7/255$ & $-1/272$ & $-$ & $-$ & $-$ & $-$ & $-$\\
    \hline
    \hline
    $\boldsymbol{E}^{(12)}_{P2,F12}$ & $68/585$ & $68/1001$ & $1/45$ & $62/5005$ & $1/520$ &  $4/4095$ & $2/4095$ & $2/45045$ & $1/480480$ & $0$\\
    $\boldsymbol{E}^{(12)}_{P4,F6}$ & $254419/1350390$ & $-4474/675195$ & $2237/5401560$ & $96737/2700780$ & $-1575/360104$ & $-$ & $-$ & $-$ & $-$ & $-$\\    
  \end{tabular}
  \end{ruledtabular}
\caption{Values of the weights for different isotropy orders of the lattice pressure tensor. The values for $\boldsymbol{E}^{(6)}_{P2,F6}$, $\boldsymbol{E}^{(8)}_{P2,F8}$, $\boldsymbol{E}^{(10)}_{P2,F10}$ and $\boldsymbol{E}^{(12)}_{P2,F12}$ correspond to those obtained requiring the forcing isotropy only~\cite{Shan06,Sbragaglia07,Falcucci07} at the $6$-th, $8$-th, $10$-th and $12$-th order, respectively, yielding a $2$-nd order isotropy for the pressure tensor. The values for $\boldsymbol{E}^{(6)}_{P4,F6}$, $\boldsymbol{E}^{(8)}_{P4,F6}$, $\boldsymbol{E}^{(10)}_{P4,F6}$ and $\boldsymbol{E}^{(12)}_{P4,F6}$ are obtained by matching the previous schemes as described in Subsection~\ref{subsec:match}.}\label{tab:W_vals}
\end{table*}

Let us now give a schematic description of the procedure we adopted for defining the new forcing schemes. In order to distinguish among the different stencils we introduce a modified notation. We label by $\boldsymbol{E}^{(k)}_{P2Fk}$ any higher order stencil computed on forcing isotropy requirements only~\cite{Shan06,Sbragaglia07,Falcucci07,Falcucci10}. Since such stencils yield a second order isotropic lattice pressure tensor, we use $P2$ in the subscript, whereas with $Fk$ we indicate that the lattice forcing is isotropic at the $k$-th order. We use $\boldsymbol{E}^{(6)}_{P2F6}$, $\boldsymbol{E}^{(8)}_{P2F8}$, $\boldsymbol{E}^{(10)}_{P2F10}$ and $\boldsymbol{E}^{(12)}_{P2F12}$ as target stencils, i.e. we want to ``mimic'' or \emph{match} them via some new sets of weights yielding a $4$-th order isotropy for the lattice pressure tensor. We shall soon motivate that these new schemes will only yield a $6$-th order isotropy for the forcing. Hence, we indicate the new matching schemes as $\boldsymbol{E}^{(6)}_{P4F6}$, $\boldsymbol{E}^{(8)}_{P4F6}$, $\boldsymbol{E}^{(10)}_{P4F6}$ and $\boldsymbol{E}^{(12)}_{P4F6}$, where the superscript now indicates which of the previously introduced stencils is matched~\cite{Shan06,Sbragaglia07,Falcucci07,Falcucci10}. The matching is obtained by imposing a system of linear equations of the weights:
\begin{equation}\label{eq:newsys}
    \begin{cases}
      e_{2}=e_{2}\left(\boldsymbol{E}_{P2Fk}^{\left(k\right)}\right) & \text{(a): equation of state}\\
      \varepsilon=\varepsilon\left(\boldsymbol{E}_{P2Fk}^{\left(k\right)}\right) & \text{(b): bulk densities}\\
      I_{4,0}=0 & \text{(c): 4-th order force}\\
      \chi_{I}=0 & \text{(d): 4-th order pressure}\\
      I_{6,0}=0 & \text{(e): 6-th order force}
    \end{cases}
\end{equation}
where with the symbols $e_{2}\left(\boldsymbol{E}_{P2Fk}^{\left(k\right)}\right)$ and $\varepsilon\left(\boldsymbol{E}_{P2Fk}^{\left(k\right)}\right)$ we indicate the numerical value of these constants for the stencils $\boldsymbol{E}^{(k)}_{P2Fk}$. The expressions for $e_2$, $\varepsilon$, $I_{4,0}$ and $I_{6,0}$ as functions of the weights are reported in Appendix~\ref{app:w2e}, while $\chi_{I}$ is given by Eq.~\eqref{eq:p4_niso}. We detail the computation of $\varepsilon$ for the higher order schemes $\boldsymbol{E}^{(10)}_{P2F10}$ and $\boldsymbol{E}^{(12)}_{P2F12}$ in Appendix~\ref{app:1dPT}, where we explicitly write the coefficients for the flat interface pressure tensor as functions of the weights.
Equations~(\ref{eq:newsys}.a) and (\ref{eq:newsys}.b) are used to match the equation of state and the bulk equilibrium densities and flat profile; Eq.~(\ref{eq:newsys}.c) imposes the $4$-th order isotropy for the forcing, so that Eq.~(\ref{eq:newsys}.d) delivers the $4$-th order isotropy for the pressure tensor [see Eqs.~\eqref{eq:p4_niso} and~\eqref{eq:I4_LC}]; finally,
Eq.~(\ref{eq:newsys}.e) fixes the $6$-th order isotropy for the forcing.\\We wish to stress that it is possible to match any forcing scheme as long as the equations are linearly independent. This fact allows us to match the $12$-th order forcing isotropy stencil $\boldsymbol{E}^{(12)}_{P2F12}$, defined by 10 weights and 56 forcing vectors, by using \emph{only} 5 weights and 24 forcing vectors. Another important property is that this procedure yields the same value of $e_4$, i.e. the surface tension, up to $\boldsymbol{E}^{(10)}_{P2F10}$, while for $\boldsymbol{E}^{(12)}_{P2F12}$ 
a small deviation of the order $10^{-4}$ is found. Indeed, a possible way to fix the value of $e_4$ would be to introduce one more group of vectors, allowing for one further condition. We report in Table~\ref{tab:FP_vals} the values used to define the system of equations and in Table~\ref{tab:W_vals} the five weights for the four new multi-range models $\boldsymbol{E}^{(6)}_{P4,F6}$, $\boldsymbol{E}^{(8)}_{P4,F6}$, $\boldsymbol{E}^{(10)}_{P4,F6}$ and $\boldsymbol{E}^{(12)}_{P4,F6}$, together with those of the usual stencils $\boldsymbol{E}^{(6)}_{P2F6}$, $\boldsymbol{E}^{(8)}_{P2F8}$, $\boldsymbol{E}^{(10)}_{P2F10}$ and $\boldsymbol{E}^{(12)}_{P2F12}$. Such values can be directly inserted into any existing code implementing a two-belt SC forcing scheme. It is interesting to notice that all new schemes have a forcing isotropy which is always smaller or equal to that of the target stencil. Nevertheless, we show in the next Section that, considering the isotropy condition of the pressure tensor, spurious currents decrease in extent and intensity.

%%%%%%%%%%%%%%%%%%%%%%%%%%%%%%%%%%%%%%%%%%%%%%%%%%%%%%%%%%%%%%%
\section{Numerical tests}\label{sec:NUM}
%%%%%%%%%%%%%%%%%%%%%%%%%%%%%%%%%%%%%%%%%%%%%%%%%%%%%%%%%%%%%%%
The following results have been obtained by implementing the methods described in Section~\ref{sec:LB} for a two-dimensional regular square lattice of linear size $L$. We use the $D2Q9$ discrete velocity set $\{\boldsymbol{\xi}_i\}$ with $i=0,\ldots,8$, for which $\boldsymbol{\xi}_0 = \textbf{0}$ and $\boldsymbol{\xi}_a = \textbf{e}_a$ for $a = 1,\ldots,8$ as reported in Fig.~\ref{fig:stencil}(a), and $c_s^2 = 1/3$. In the following, we report the forcing values in the scaled form $Gc_s^2$. Finally, in order to demonstrate the robustness of our findings, we also consider two different functional forms for the pseudo-potential, namely $\psi = \exp(-1/n)$ and $\psi = 1 - \exp(-n)$~\cite{ShanChen93,ShanChen94,Sbragaglia07,Falcucci07,Shan08}. All droplet simulations have been run with a size ratio $L/R=5$, where $R$ is the initial radius value. The initialization is performed by means of the following radial profile
\begin{equation}
  n(r, R) = \frac{1}{2}(n_g + n_l) - \frac{1}{2}(n_l - n_g)\tanh(r - R),
\end{equation}
where the values of $n_g$ and $n_l$ are obtained by solving Eqs.~\eqref{eq:n_prof},~\eqref{eq:Maxwell} and~\eqref{eq:flat_press}.
For the droplet simulations, we set $L\in \{127, 159, 191, 223, 255, 287, 319, 351\}$, while for the flat interfaces the size is fixed to $L_x = 100, L_y = 4$ and the initial profile is given by
\begin{figure}
    \centering
    \includegraphics[scale=0.62]{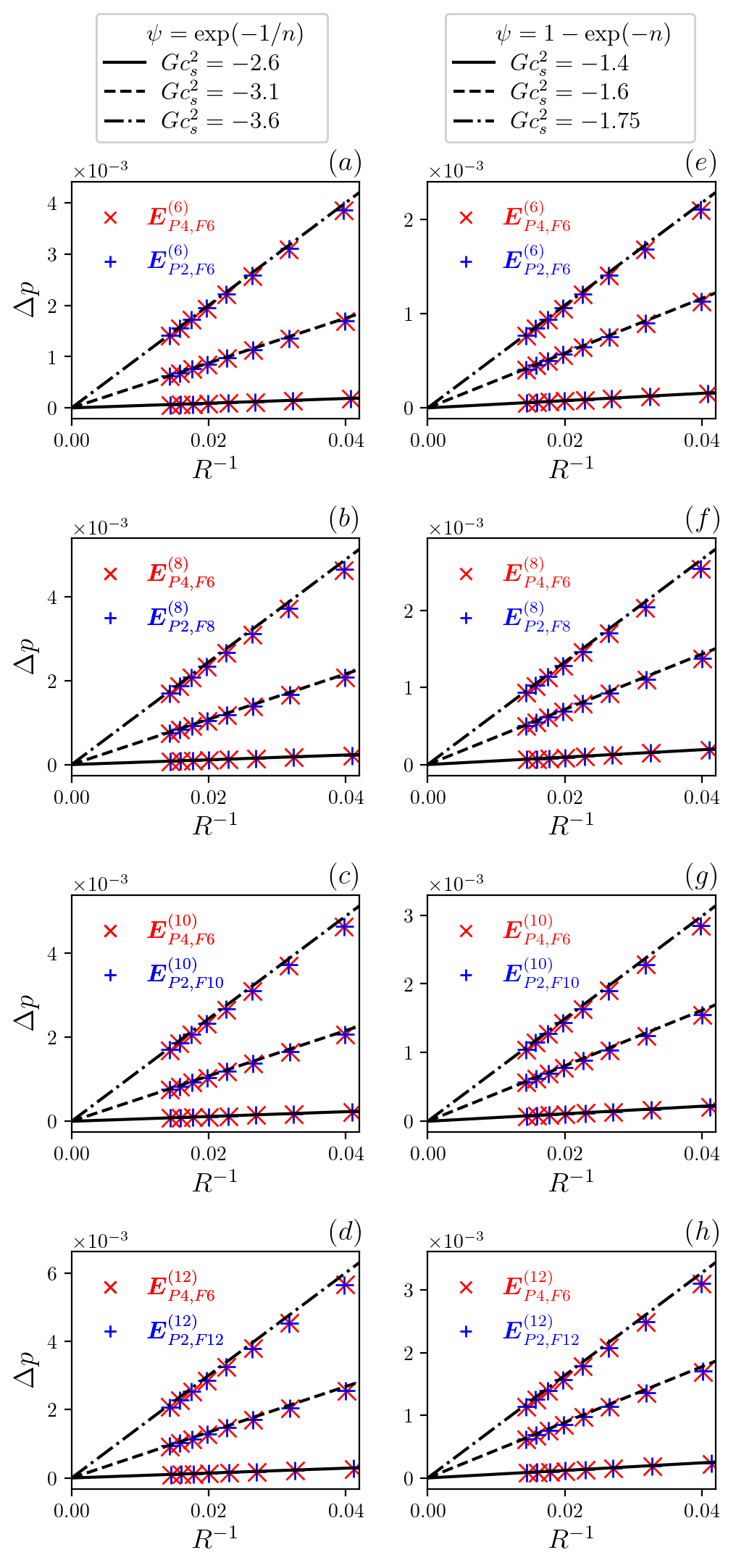}
    \caption{Laplace law comparison for $\psi = \exp(-1/n)$, on the left column, and $\psi = 1 - \exp(-n)$, on the right one, at different values of $Gc_s^2$. The slope of the straight lines corresponds to $\sigma$ obtained by integrating Eqs.~\eqref{eq:n_prof}, \eqref{eq:Maxwell} and \eqref{eq:sigma} (see main text for details). Red `x' points and blue `+' symbols represent the data for the forcing schemes with $4$-th and $2$-nd order lattice pressure tensor isotropy, respectively.}
    \label{fig:res_laplace}
\end{figure}
\begin{equation}
  \begin{split}n\left(x,x_{0},w\right)= & \frac{1}{2}\left(n_{l}+n_{g}\right)\\
- & \frac{1}{2}\left(n_{l}-n_{g}\right)\tanh\left[x-\left(x_{0}-\frac{w}{2}\right)\right]\\
+ & \frac{1}{2}\left(n_{l}-n_{g}\right)\left\{ \tanh\left[x-\left(x_{0}+\frac{w}{2}\right)\right]+1\right\},
\end{split}
\end{equation}
where $x_0$ is the center of the strip and $w$ its width. 
Finally, to fix a convergence criterion, we use the magnitude $\delta u$ of the spatial average of the difference between the components of two velocity fields, $\delta u=L^{-2}\sum_{\textbf{x}}\sum_{\alpha}|u^{\alpha}(\textbf{x},t+\delta t)-u^{\alpha}(\textbf{x},t)|$, at a time distance $\delta t = 2^{14}$: we consider the simulation as converged when $\delta u< 10^{-12}$. All the results have been obtained using $64$-bits floating point variables for all the quantities.
\begin{figure*}[t!]
\includegraphics[scale=0.5]{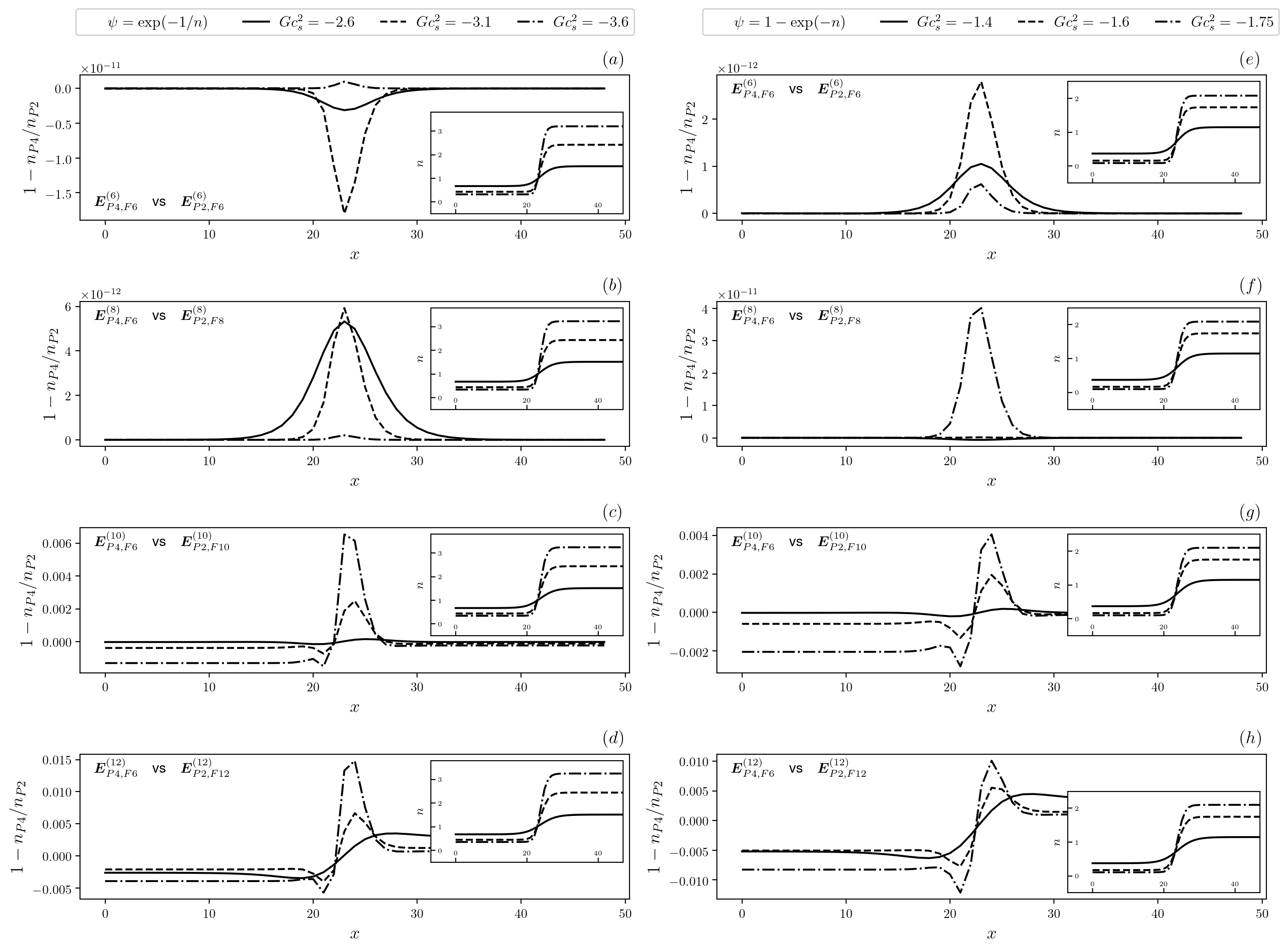}
\caption{\label{fig:res_prof} Comparison of the flat interface profiles for all forcing schemes, at different values of $Gc_s^2$ and for different pseudo-potentials, $\psi = \exp(-1/n)$ and $\psi = 1 - \exp(-n)$. The comparison is carried out using the relative deviation of the flat interface profiles given by $1 - n_{P4}/n_{P2}$ where $n_{P4}$ is the profile for the $4$-th order isotropic pressure tensor (reported in the inset) and $n_{P2}$ is the profile for the $2$-nd order isotropic pressure tensor.}
\end{figure*}
\begin{figure*}[t!]
  \includegraphics[scale=0.366825]{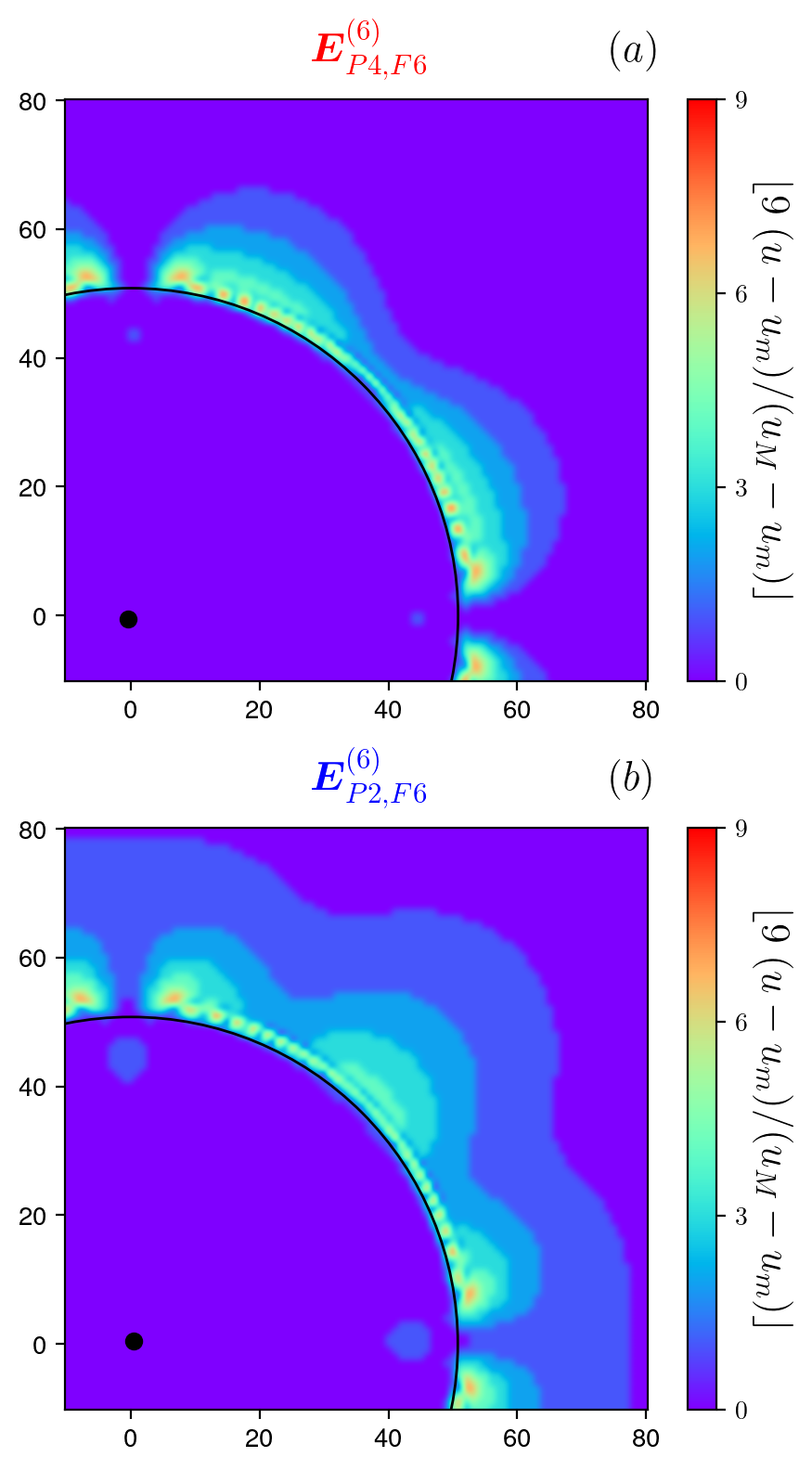}\quad
  \includegraphics[scale=0.366825]{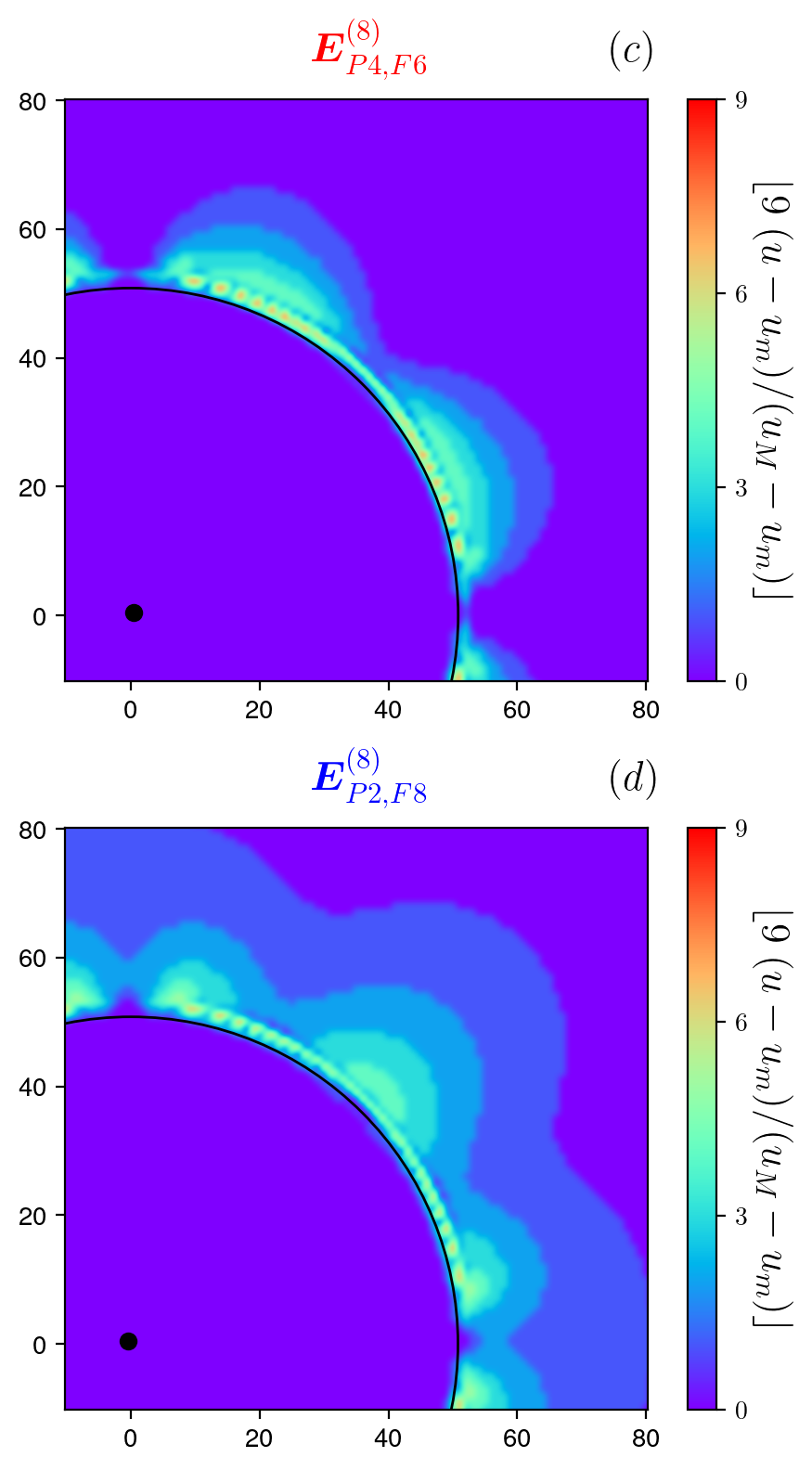}\quad
  \includegraphics[scale=0.366825]{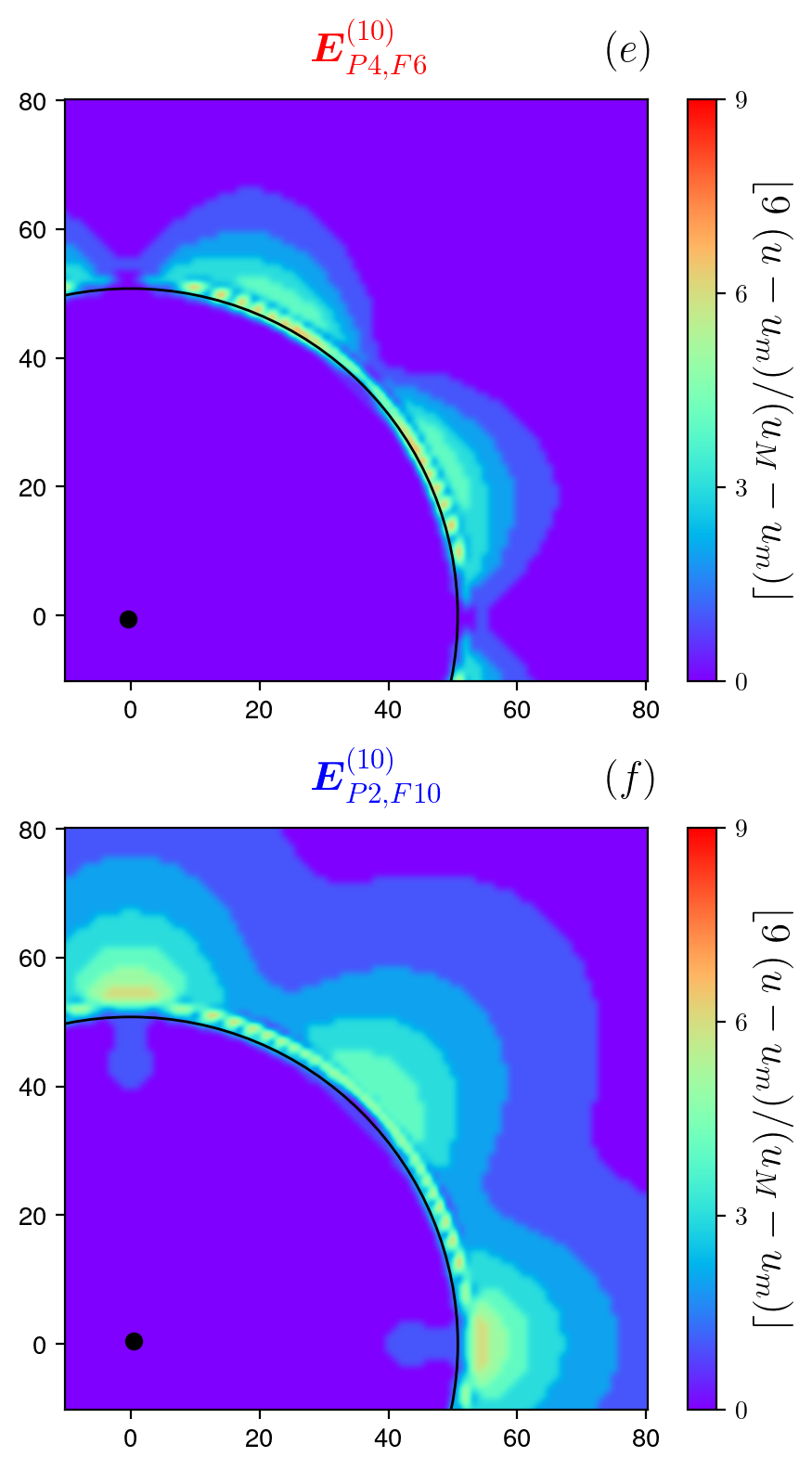}\quad
  \includegraphics[scale=0.366825]{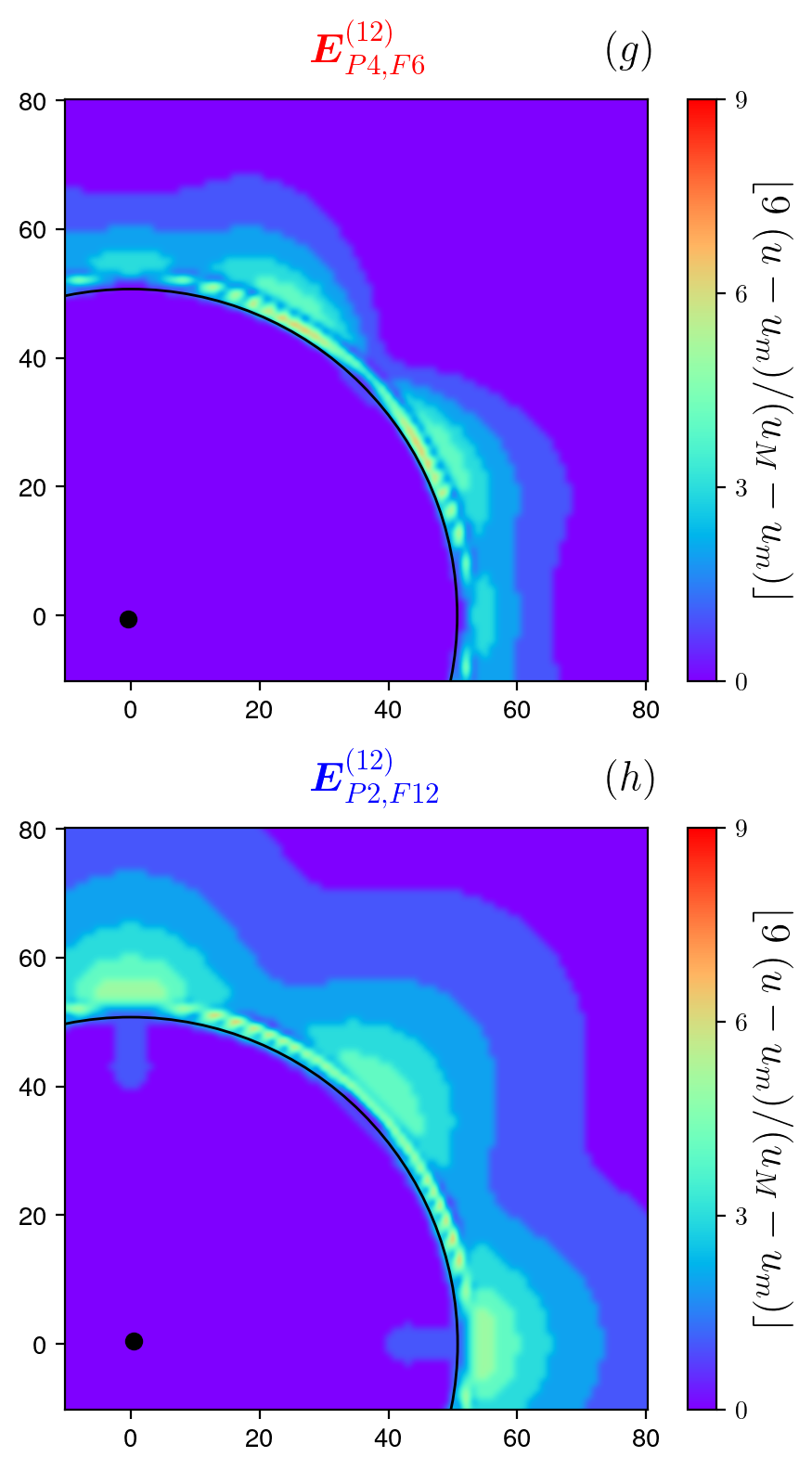}\\
  \caption{\label{fig:res_u_map} Maps of the normalized spurious currents intensity $u = |\textbf{u}|$ for different stencils with fixed $Gc_s^2=-3.6$, $L=255$ and $\psi = \exp(-1/n)$. In the first row we report the results for the new forcing schemes with a $4$-th order isotropic pressure tensor, while in the second row we report the standard ones. The normalization is performed with respect to $u_m = \min u(\textbf{x})$ and $u_M = \max u(\textbf{x})$. The values of the peak velocity are $u_M/c_s \simeq 0.019$ for $\boldsymbol{E}^{(6)}_{P4,F6}$ in $(a)$ and $u_M/c_s \simeq 0.022$ for $\boldsymbol{E}^{(6)}_{P2,F6}$ in $(b)$, and $u_M/c_s \simeq 0.012$ for $\boldsymbol{E}^{(8)}_{P4,F6}$ in $(c)$ and $u_M/c_s \simeq 0.014$ for $\boldsymbol{E}^{(8)}_{P2,F8}$ in $(d)$, $u_M/c_s \simeq 0.0088$ for $\boldsymbol{E}^{(10)}_{P4,F6}$ in $(e)$ and $u_M/c_s \simeq 0.0102$ $\boldsymbol{E}^{(10)}_{P2,F10}$ in $(f)$, $u_M/c_s \simeq 0.0071$ for $\boldsymbol{E}^{(12)}_{P4,F6}$ in $(g)$ and $u_M/c_s \simeq 0.0076$ for $\boldsymbol{E}^{(12)}_{P2,F12}$ in $(h)$. Coordinates have the origin in the center of the droplet. The relative intensities have been scaled and ``quantized'' by multiplying for an integer $N=9$ and taking the (\texttt{floor}) integer part $\lfloor \cdot \rfloor$. Hence, the color map has discrete changes, thus easing the area comparison in a given range. The spatial extension of the currents in the upper row (higher pressure tensor isotropy) is smaller than in the lower row.}
\end{figure*}

Let us begin by showing that the forcing schemes presented in Table~\ref{tab:W_vals} yield the same macroscopic properties, i.e. surface tension $\sigma$ and flat interface profiles $n(\textbf{x})$. Results on surface tension are reported in Fig.~\ref{fig:res_laplace}. For the evaluation of $\sigma$, we resorted to the Laplace test: we simulate various radii, measuring the pressure difference between the inside and the outside of the droplet, $\Delta p = p_{\text{in}} - p_{\text{out}}$. These values have been computed according to Eq.~\eqref{eq:bulk_p} since the gradients in the bulk regions of the two phases are negligible. The Young-Laplace law relates the pressure difference to the surface tension and the radius of the droplet through the well-known expression $\Delta p = \sigma/R$, in two dimensions. Hence, in order to estimate $\sigma$, given the values of $\Delta p$, one needs to measure the radius of the droplet $R$, which we obtain by means of the Gibbs criterion~\cite{RowlinsonWidom82}, i.e. by inverting the relation~\footnote{The relation for the radius can be obtained by computing the position of the interface yielding a vanishing adsorbance $\Gamma = \int_0^R \text{d}r (n(r) - n_l) + \int_R^\infty \text{d}r (n(r) - n_g) = 0$~\cite{RowlinsonWidom82}.} $L^2 \langle n \rangle = \pi R^2 \, n_{\text{in}} + (L^2 - \pi R^2)\,n_{\text{out}}$, where we used the average density $\langle n \rangle = L^{-2}\sum_{\textbf{x}}n(\textbf{x})$. The points $(R^{-1}, \Delta p)$ are reported in Fig.~\ref{fig:res_laplace} for different values of $G c_s^2$ and different choices of $\psi$. Red `x' points are associated to the new $4$-th order pressure tensor isotropy schemes, while blue `+' are those associated to the higher forcing isotropy schemes~\cite{Shan06,Sbragaglia07}. Finally, the slope of the lines represents the values of $\sigma$ obtained from the numerical integration of Eqs.~\eqref{eq:n_prof},~\eqref{eq:Maxwell} and~\eqref{eq:sigma},~\footnote{The surface tension values are $\sigma \simeq 0.00458, 0.04376, 0.09998$ (lbu) for $Gc_s^2 = -2.6, -3,1, -3.6$ and $\psi = \exp(-1/n)$, $\sigma \simeq 0.00387, 0.02904, 0.05442$ (lbu) for $Gc_s^2 = -1.4, -1.6, -1.75$ and $\psi = 1 - \exp(-n)$, where ``lbu'' stands for Lattice Boltzmann units.}. We first notice that blue and red points $(R^{-1}, \Delta p)$ superpose in good agreement with the slope given by $\sigma$ for all forcing values and choices of $\psi$, demonstrating that the newly proposed forcing schemes $\boldsymbol{E}^{(6)}_{P4,F6}$, $\boldsymbol{E}^{(8)}_{P4,F6}$, $\boldsymbol{E}^{(10)}_{P4,F6}$ and $\boldsymbol{E}^{(12)}_{P4,F6}$ yield the same surface tension as $\boldsymbol{E}^{(6)}_{P2,F6}$, $\boldsymbol{E}^{(8)}_{P2,F8}$, $\boldsymbol{E}^{(10)}_{P2,F10}$ and $\boldsymbol{E}^{(12)}_{P2,F12}$, respectively.

We continue with the analysis of the flat interface profiles, reported in Fig.~\ref{fig:res_prof}. We analyze the relative variation of the density profiles related to the new $4$-th order pressure tensor isotropy schemes $\boldsymbol{E}^{(6)}_{P4,F6}$, $\boldsymbol{E}^{(8)}_{P4,F6}$, $\boldsymbol{E}^{(10)}_{P4,F6}$ and $\boldsymbol{E}^{(12)}_{P4,F6}$, that we indicate for brevity as $n_{P4}$, with respect to the density profiles obtained using the standard schemes $\boldsymbol{E}^{(6)}_{P2,F6}$, $\boldsymbol{E}^{(8)}_{P2,F8}$, $\boldsymbol{E}^{(10)}_{P2,F10}$ and $\boldsymbol{E}^{(12)}_{P2,F12}$, labeled as $n_{P2}$. In the insets we report the profiles $n_{P4}$ for the same values of $Gc_s^2$. The data highlight that for $\boldsymbol{E}^{(6)}_{P4,F6}$ and $\boldsymbol{E}^{(8)}_{P4,F6}$ the magnitude of the largest deviation is of order $10^{-11}$ (compatibly with floating point rounding~\footnote{We remark that for $\boldsymbol{E}^{(6)}_{P4,F6}$ and $\boldsymbol{E}^{(8)}_{P4,F6}$ the exact value of the relative deviation is compatible with the double-precision floating point rounding, hence the details of these results can vary according to the implementation details, compiler and optimization options.}), changing for different values of the coupling constant and $\psi$. For $\boldsymbol{E}^{(10)}_{P4,F6}$ and $\boldsymbol{E}^{(12)}_{P4,F6}$, the deviation grows reaching a maximum value of the order $10^{-2}$ in the case of $\boldsymbol{E}^{(12)}_{P4,F6}$. However, such a discrepancy seems reasonable, as we are using only 5 weights to reproduce the bulk densities and flat interface profile of $\boldsymbol{E}^{(10)}_{P2,F10}$, defined using 7 weights, and of $\boldsymbol{E}^{(12)}_{P2,F12}$, defined using 10 weights.

Now that we have numerically verified that the macroscopic properties are consistent across the different schemes in a wide range of coupling values and for different choices of $\psi$, we continue with the analysis of the spurious currents. In Fig.~\ref{fig:res_u_map} we report the plots for the spatial distribution of the scaled velocity magnitude $u(\textbf{x})=|\textbf{u}(\textbf{x})|$. Each row refers to a different degree of isotropy of the pressure tensor, $4$-th and $2$-nd order for first and second row respectively. Starting from the leftmost column we consider the cases $\boldsymbol{E}^{(6)}_{P\#,Fk}$, $\boldsymbol{E}^{(8)}_{P\#,Fk}$, $\boldsymbol{E}^{(10)}_{P\#,Fk}$ and $\boldsymbol{E}^{(12)}_{P\#,Fk}$. The normalization is performed by means of the minimum $u_m = \min u(\textbf{x})$ and maximum $u_M = \max u(\textbf{x})$ in the whole domain, for each case. We multiply the normalized quantities by an arbitrary integer $N$ and then we take the integer part $\lfloor \cdot \rfloor$ so that only $N$ colors appear, with $N = 9$. To guide the eye, we report the center of the droplet, which is used as the origin of the coordinates, and the radius obtained with the Gibbs criterion. As apparent from Fig.~\ref{fig:res_u_map}, for $Gc_s^2 = -3.6$ and $\psi = \exp(-1/n)$, the extension of the spurious currents is always smaller for the new schemes. In particular $\boldsymbol{E}^{(8)}_{P4,F6}$, $\boldsymbol{E}^{(10)}_{P4,F6}$ and $\boldsymbol{E}^{(12)}_{P4,F6}$ have a lower isotropy degree than the target forcing. With respect to the previous literature~\cite{Shan06,Sbragaglia07,Falcucci07,Falcucci10}, this is a non-trivial result, that displays the role of the pressure tensor as a new ``dimension'' to be exploited for the imposition of the isotropy properties. Hence, the degree of isotropy of the pressure tensor tunes the spatial extension of the spurious currents, for the same values of the surface tension $\sigma$ and the reference (i.e. flat) interface profile.

\begin{figure*}[!ht]
  \includegraphics[scale=0.4]{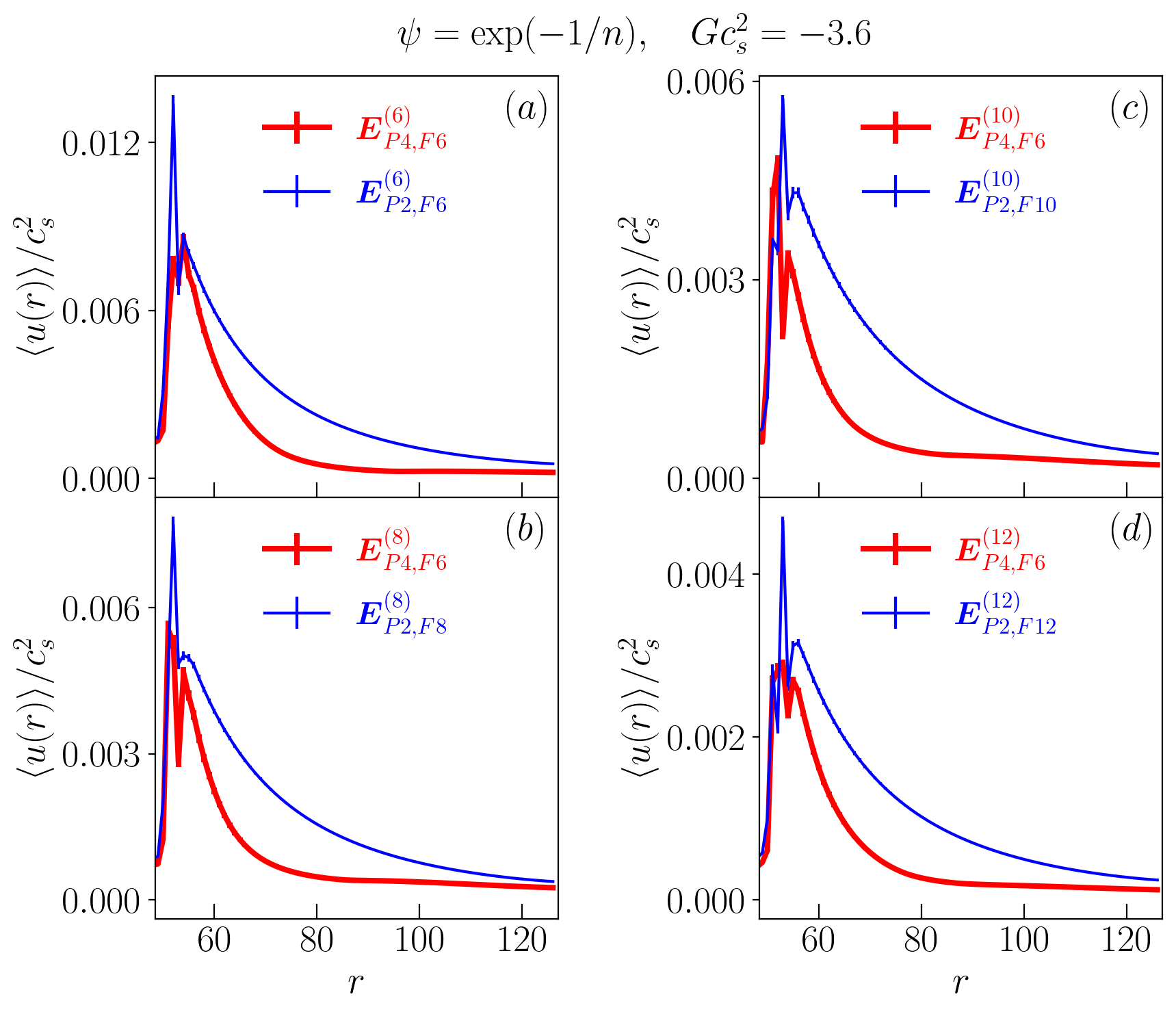}\quad  
  \includegraphics[scale=0.4]{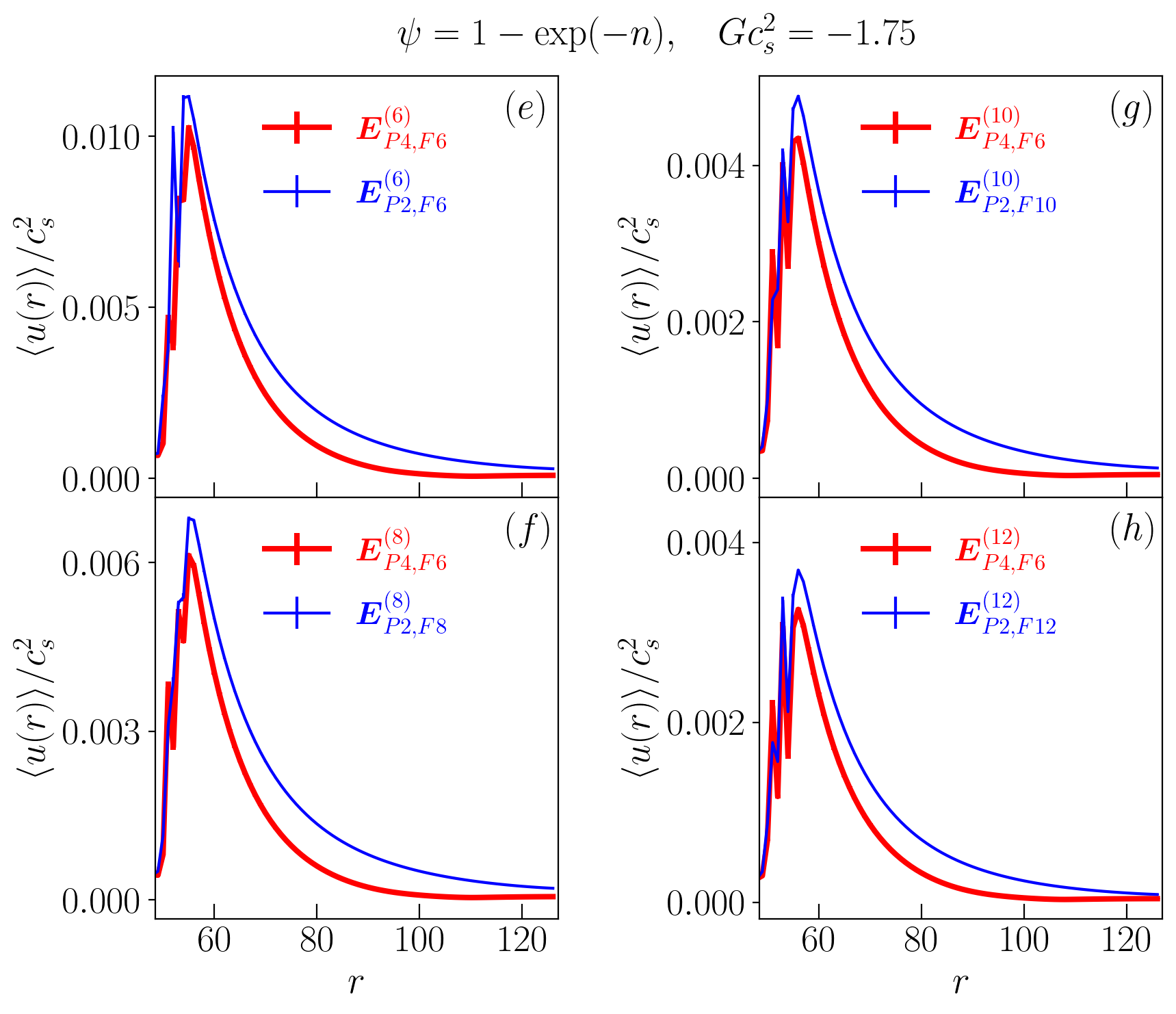}\\
  \caption{Average velocity profiles along the radial direction for different $\psi$ and $Gc_s^2$. The average is taken over an angle $\Delta\theta = \pi/4$. Thick red and thin blue lines represent the results for the new and old schemes, respectively. The average profiles of the new schemes have a faster convergence going away from the droplet surface. \label{fig:res_u_map_2}}
\end{figure*}

In Fig.~\ref{fig:res_u_map_2} we provide further evidence of the reduction of the currents, by displaying the average velocity profile along the radial direction for two different choices of $\psi$. Considering the symmetry of the velocity field, the average is taken over an angle $\Delta \theta = \pi/4$. Red thick lines are used for the new schemes $\boldsymbol{E}^{(6)}_{P4,F6}$, $\boldsymbol{E}^{(8)}_{P4,F6}$, $\boldsymbol{E}^{(10)}_{P4,F6}$ and $\boldsymbol{E}^{(12)}_{P4,F6}$, while blue thin ones for the old schemes $\boldsymbol{E}^{(6)}_{P2,F6}$, $\boldsymbol{E}^{(8)}_{P2,F8}$, $\boldsymbol{E}^{(10)}_{P2,F10}$ and $\boldsymbol{E}^{(12)}_{P2,F12}$. The profiles of the new schemes stay consistently below those of the older schemes, and especially for the case $\psi = \exp(-1/n)$, the new schemes yield the same velocity as the old ones a few tens of lattice sites closer to the surface of the droplet, thus demonstrating a sizable improvement. Furthermore, we can make a direct comparison of $\boldsymbol{E}^{(8)}_{P2,F8}$ and $\boldsymbol{E}^{(12)}_{P4,F6}$, since they are both defined on 5 weights: it is clear that the new set of weights allows to obtain \emph{far} weaker spurious currents (see the caption of Fig. \ref{fig:res_u_map}) without the need to use an even higher order scheme (as it was done previously with $\boldsymbol{E}^{(12)}_{P2,F12}$), with a much higher computational efficiency.

In previous studies~\cite{Shan06,Sbragaglia07,Falcucci07}, the intensity of the spurious currents has been mainly characterized by the maximum Mach number $u_M/c_s$. However, Fig.~\ref{fig:res_u_map} shows that only a very small fraction of the system area displays the strongest currents. In order to have a more informative characterization, we report in Fig.~\ref{fig:u_histo} the histograms of the logarithm of the normalized velocity magnitude $u/c_s$, i.e. $p(\log(u/c_s))$, for different values of $Gc_s^2$ and different $\psi$, as well as the complementary cumulative distribution $\tilde{F}(\log(u/c_s)) = 1 - F(\log(u/c_s))$ (starting from $1$ on the left side of the insets). This latter quantity represents the fraction of the area of the system where the currents are larger than a given value of $u/c_s$. The parameters used in Fig.~\ref{fig:res_u_map} are analyzed in Fig.~\ref{fig:u_histo} $(a)$, $(b)$, $(c)$ and $(d)$: thicker red lines refer to the new schemes $\boldsymbol{E}^{(6)}_{P4,F6}$, $\boldsymbol{E}^{(8)}_{P4,F6}$, $\boldsymbol{E}^{(10)}_{P4,F6}$ and $\boldsymbol{E}^{(12)}_{P4,F6}$ while the thinner blue curves refer to the standard ones~\cite{Shan06,Sbragaglia07,Falcucci07} $\boldsymbol{E}^{(6)}_{P2,F6}$, $\boldsymbol{E}^{(8)}_{P2,F8}$, $\boldsymbol{E}^{(10)}_{P2,F10}$ and $\boldsymbol{E}^{(12)}_{P2,F12}$ (see Table~\ref{tab:W_vals}). We can observe that the new schemes always yield the smallest peak value for the histograms, i.e. the majority of the system area is affected by smaller spurious currents with respect to the standard case. This automatically implies a smaller spatial extension of the currents. The insets in Fig.~\ref{fig:u_histo} show that the complementary cumulative distribution $\tilde{F} = 1 - F$ decreases faster for the new schemes, i.e. for a given value of $u/c_s$ the area of the system containing larger currents is sizeably smaller for the new schemes than for the standard ones. We verified that the histograms of the spurious currents eventually converge, independently on the pressure tensor isotropy order, for smaller coupling constants $Gc_s^2$, near the critical point. Finally, we verified that by changing the size of the system to $L=351$, while keeping fixed the ratio between $L$ and the initial droplet radius $L/R=5$, the histogram of $\log(u/c_s)$ does not change for $Gc_s^2 \leq -3.1$, for both choices of $\psi$.

In summary, with this series of numerical tests we showed that a higher order isotropy of the pressure tensor yields spurious currents that are both weaker and less spatially extended than those emerging from the standard multi-range approach~\cite{Shan06,Sbragaglia07,Falcucci07}. Such a result has been obtained comparing forcing schemes that share the same lattice force continuum expansion up to a given order, same surface tension and flat interface profile, for different values of the coupling constant $Gc_s^2$ and different choices of the pseudo-potential $\psi$, thus establishing the robustness of the findings.

\subsection{Computational Advantage}
In the light of the above discussion, we want to stress that the new scheme $\boldsymbol{E}^{(12)}_{P4,F6}$ has a significant numerical advantage over the so-far widely adopted $\boldsymbol{E}^{(8)}_{P2,F8}$, as well as over $\boldsymbol{E}^{(12)}_{P2,F12}$, since it basically brings all the benefits of $\boldsymbol{E}^{(12)}_{P2,F12}$ defined with 10 weights (and 56 lattice vectors), while using only 5 weights (24 lattice vectors). First of all, the number of memory reads and algebraic operations needed by $\boldsymbol{E}^{(12)}_{P4,F6}$ for computing the total force is roughly half of those necessary for $\boldsymbol{E}^{(12)}_{P2,F12}$. Furthermore, the handling of boundary conditions is drastically simplified, needing to deal only with a two-node thick boundary rather than four, as in the case of $\boldsymbol{E}^{(12)}_{P2,F12}$, which is extremely important for parallel implementations, where the boundaries need to be constantly exchanged. \\
With respect to $\boldsymbol{E}^{(8)}_{P2,F8}$, while keeping the same computational complexity, the new scheme $\boldsymbol{E}^{(12)}_{P4,F6}$ yields a better gain for the spurious currents than the one obtained by using the higher order stencil $\boldsymbol{E}^{(12)}_{P2,F12}$. Indeed, all the new stencils presented in this work can easily be used in any existing code where the forcing is implemented using 5 weights, simply by substituting the new proposed values. Hence, the advantages of the present analysis are readily accessible.

%%%%%%%%%%%%%%%%%%%%%%%%%%%%%%%%%%%%%%%%%%%%%%%%%%%%%%%%%%%%%%%
\section{Conclusions}\label{sec:CONCLUSIONS}
%%%%%%%%%%%%%%%%%%%%%%%%%%%%%%%%%%%%%%%%%%%%%%%%%%%%%%%%%%%%%%%
\begin{figure*}[t!]
%\begin{center}
%[scale=0.415]
%[scale=0.545]
\includegraphics[scale=0.57]{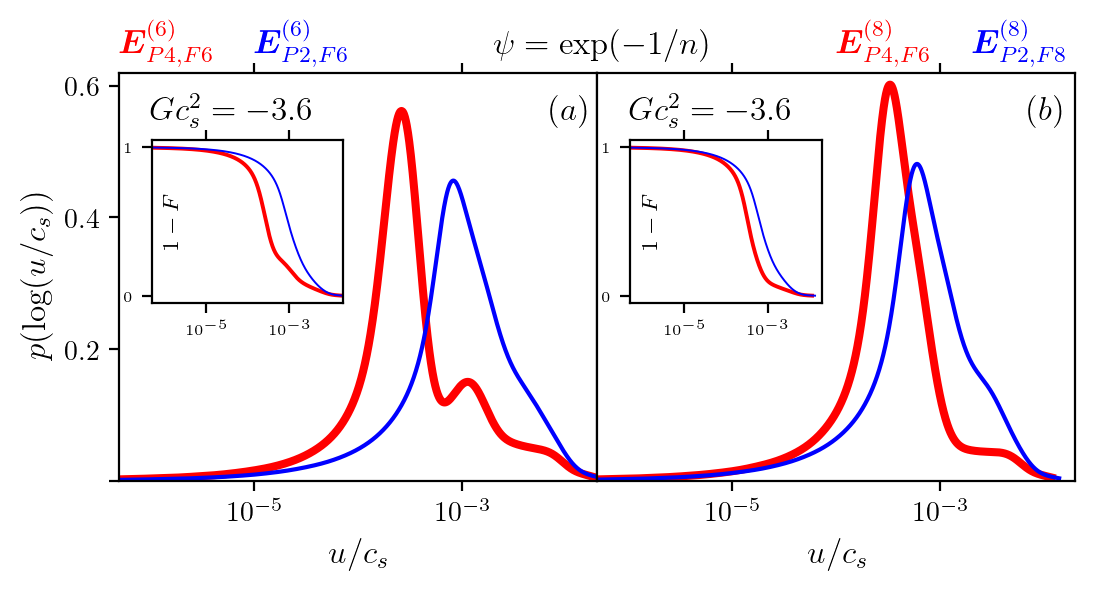}\qquad
\includegraphics[scale=0.57]{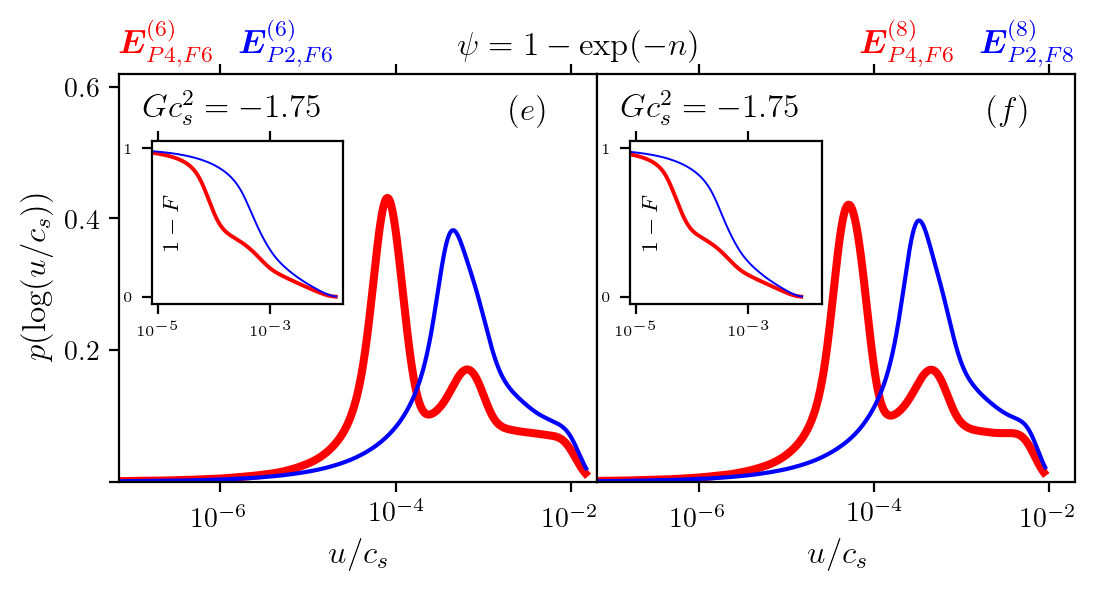}\\
\includegraphics[scale=0.57]{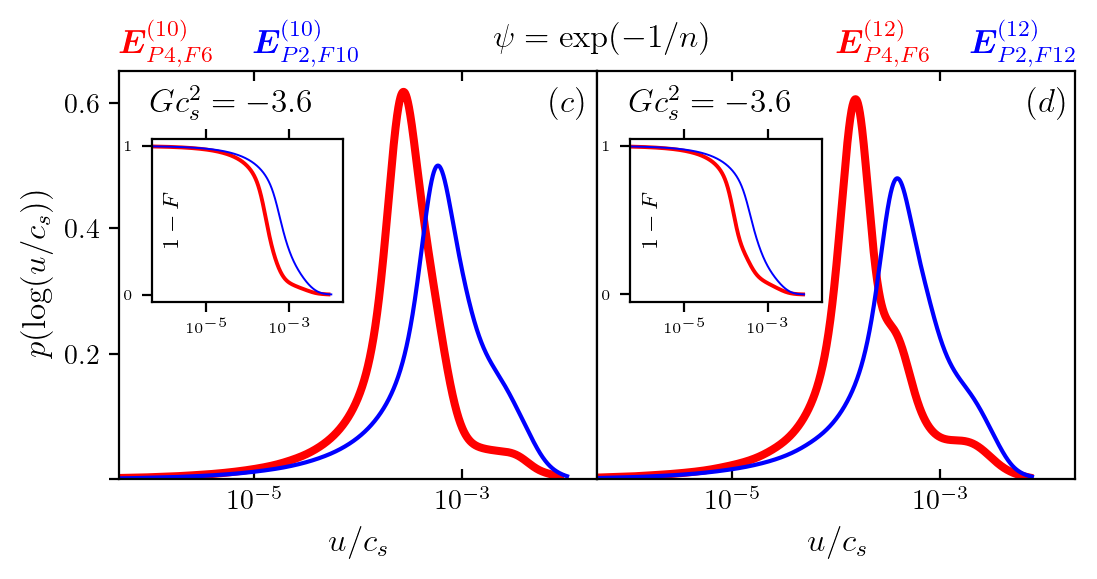}\qquad
\includegraphics[scale=0.57]{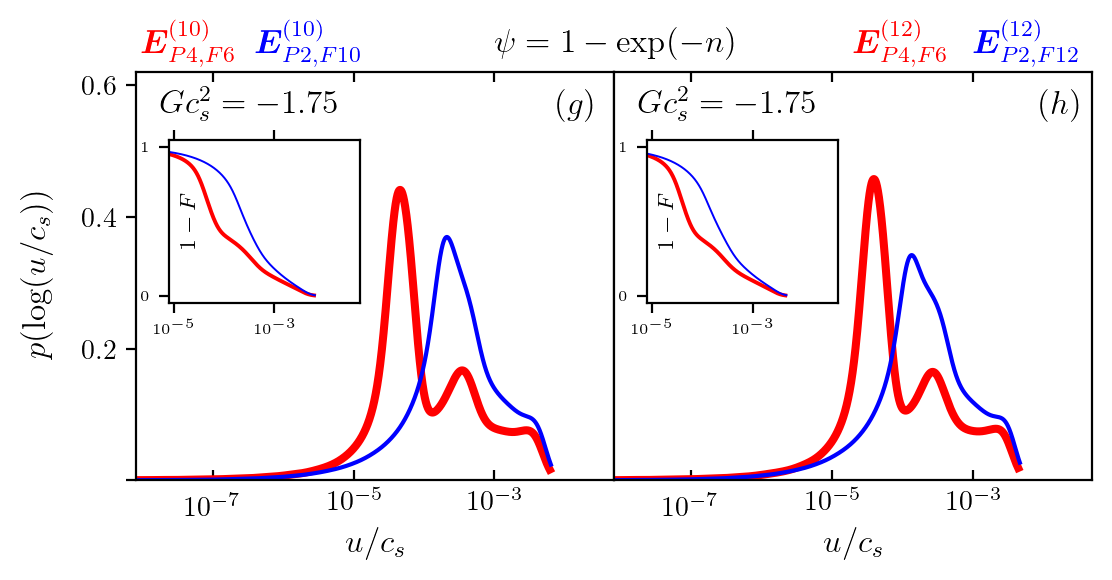}\\
%\end{center}
\caption{\label{fig:u_histo} Normalized velocity $u/c_s = |\textbf{u}|/c_s$ histogram $p(\log(u/c_s))$ for $L=255$ and complementary cumulative distribution $\tilde{F} = 1 - F(\log(u/c_s))$ for different values of $G$, choices of $\psi$ and forcing schemes. Data for the new $4$-th order isotropic pressure tensor schemes, i.e. $\boldsymbol{E}^{(6)}_{P4,F6}$, $\boldsymbol{E}^{(8)}_{P4,F6}$, $\boldsymbol{E}^{(10)}_{P4,F6}$ and $\boldsymbol{E}^{(12)}_{P4,F6}$ (see Tab.~\ref{tab:W_vals}), are reported in thick red lines while those related to the standard higher order schemes~\cite{Shan06,Sbragaglia07,Falcucci07}, i.e. $\boldsymbol{E}^{(6)}_{P2,F6}$, $\boldsymbol{E}^{(8)}_{P2,F8}$, $\boldsymbol{E}^{(10)}_{P2,F10}$ and $\boldsymbol{E}^{(12)}_{P2,F12}$ (see Tab.~\ref{tab:W_vals}), are reported in thin blue.}
\end{figure*}

In the present paper, we have reviewed the isotropy analysis of the Shan-Chen forcing scheme~\cite{ShanChen93,ShanChen94,Shan06,Sbragaglia07,Falcucci07} and generalized it to the lattice pressure tensor defined in~\cite{Shan08}. As a first step, we fine-grained the isotropy analysis to the single group of the forcing vectors used in the multi-range models by extending the parametrization of the relevant tensorial structures introduced in~\cite{Wolfram1986} (see Appendix~\ref{app:IsoDetails}). Such fine-grained approach, together with the treatment of mixed vectorial structures (see Appendix~\ref{app:PTDetails}), allowed us to write the general form of the fourth-order expansion of the lattice pressure tensor for the multi-range schemes $\boldsymbol{E}^{(4)}$, $\boldsymbol{E}^{(6)}$ and $\boldsymbol{E}^{(8)}$ [see Eq.~\eqref{eq:P4E8}]. Such general expression highlights the anisotropic contributions, allowing to define the new isotropy conditions for the lattice pressure tensor expansion, namely $\chi_I = \Lambda_I = 0$ [see Eq.~\eqref{eq:p4_niso}]. In particular, we noticed that the 4-th order isotropy condition for the forcing can be obtained by a linear combination of the pressure tensor conditions, i.e. $I_{4,0} = 4(\chi_I + \Lambda_I)$ [see Eq.~\eqref{eq:I4_LC}]. This result has the important consequence of making the value of the surface tension of the flat interface independent from the anisotropic coefficients $\chi_I$ and $\Lambda_I$ [see Eq.~\eqref{eq:I4_LC} and~\eqref{eq:sigma}], thus securing its physical meaning. Finally, we designed a numerical setup capable of keeping fixed the forcing expansion (up to the 4-th order) and the macroscopic flat interface properties (i.e. flat interface profile and surface tension), thus isolating the role of the pressure tensor isotropy. Hence, starting from the previously proposed $\boldsymbol{E}^{(6)}_{P2,F6}$, $\boldsymbol{E}^{(8)}_{P2,F8}$, $\boldsymbol{E}^{(10)}_{P2,F10}$ and $\boldsymbol{E}^{(12)}_{P2,F12}$ multi-range schemes~\cite{Shan06,Sbragaglia07,Falcucci07}, where we indicate with $P\#$ and $Fk$ the isotropy order of the pressure tensor and forcing respectively, we obtained the new schemes $\boldsymbol{E}^{(6)}_{P4,F6}$, $\boldsymbol{E}^{(8)}_{P4,F6}$, $\boldsymbol{E}^{(10)}_{P4,F6}$ and $\boldsymbol{E}^{(12)}_{P4,F6}$ (see Table~\ref{tab:W_vals}). We showed in Figures~\ref{fig:res_u_map} and~\ref{fig:u_histo} that the higher isotropy degree for the pressure tensor yields weaker and less spatially extended spurious currents, even when the forcing isotropy of the new schemes is lower than that of the old ones. The source code for the simulations can be found on the github repository \href{https://github.com/lullimat/idea.deploy}{https://github.com/lullimat/idea.deploy}~\cite{sympy, scipy, numpy0, numpy1, scikit-learn, matplotlib, ipython, pycuda_opencl}, where a Jupyter notebook~\cite{ipython} is available to reproduce the results reported in this paper.

On a more general perspective, the difference between the isotropy conditions of the lattice forcing and the lattice pressure tensor can be traced back to the different algebraic structure of their Taylor expansions: while the forcing expansion only involves higher order derivatives, the pressure tensor introduces products of lower order ones~\cite{SbragagliaBelardinelli13}. The possibility to express the $4$-th order isotropy condition of the forcing as a linear combination of the two new conditions, $\Lambda_I = \chi_I = 0$, for the pressure tensor, is striking and pointing at a more fundamental structure underlying both lattice quantities. It would be interesting to extend the present analysis to further orders and check whether the new isotropy conditions for the lattice pressure tensor would still be compatible with the forcing ones. Indeed, the analysis of the isotropy of the lattice pressure tensor opens up yet another ``dimension'' to study and control the spurious currents, yielding a more effective reduction of the latter at a fixed forcing isotropy order.

Finally, the possibility to isolate the anisotropic parts of the pressure tensor lays the foundation for a systematic treatment, in the multi-range case, of the remaining isotropic components. This is of utmost importance when bridging the Lattice Boltzmann method to other thermodynamic and mesoscopic descriptions of the physics of multi-phase interfaces~\cite{RowlinsonWidom82}. Future work will be focusing on the three-dimensional generalization of the present procedure, possibly considering a higher isotropy order for the lattice pressure tensor, as well as the extension to the multi-component case.

\begin{acknowledgements}
Luca Biferale thankfully acknowledges the hospitality from the Department of Mechanics and Aerospace Engineering of Southern University of Science and Technology. This work was supported by the National Natural Science Foundation of China Grants No. 12050410244, and No. 91752204, Science and Technology Innovation Committee Foundation of Shenzhen Grants No. JCYJ20170817105533245 and No. KQTD20180411143441009, Department of Science and Technology of Guangdong Province Grants No. 2019B121203001 and No. 2020B1212030001, and from the European Research Council (ERC) under the European Union’s Horizon 2020 research and innovation programme (grant agreement No. 882340).
\end{acknowledgements}

%%%%%%%%%%%%%%%%%%%%%%%%%%%%%%%%%%%%%%%%%%%%%%%%%%%%%%%%%%%%%%%
\appendix
%%%%%%%%%%%%%%%%%%%%%%%%%%%%%%%%%%%%%%%%%%%%%%%%%%%%%%%%%%%%%%%
\section{Lattice Pressure Tensor Definition}\label{app:LPT_def}
\begin{figure*}[!t]
\includegraphics[scale=0.87]{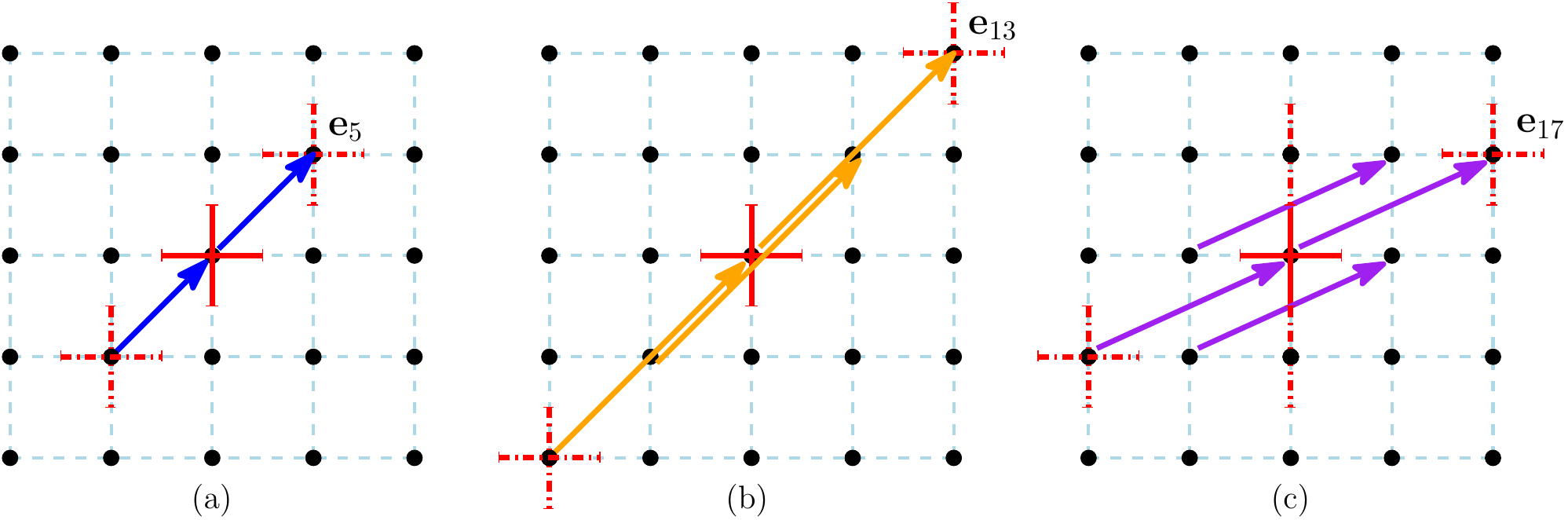}
\caption{\label{fig:pt_areas} Sketch for the computation of the number of contributions $N_{a,\left(k\right)}$ of forcing vectors crossing the area elements. The three examples in panels (a), (b) and (c), correspond to $a=5$, $13$ and $17$ (from left to right), respectively. In the middle we sketch in solid red lines the two area elements $\textbf{A}_{(x)} = \textbf{e}_2$ and $\textbf{A}_{(y)} = \textbf{e}_1$, while we report in dashed those parallel area element ``sharing'' a given forcing vector.}
\end{figure*}
In this Section, we provide a detailed review for the derivation of the lattice pressure tensor as described in~\cite{Shan08} and summarized in Section~\ref{sec:PT}. We write the total force crossing a given unit area element~\cite{Shan08,SbragagliaBelardinelli13} as the pressure flux through the same element, which for each group $\mathcal{G}_\ell$ reads
\begin{equation}\label{eq:discrete_p}
F_{\ell,\left(k\right)}^{\mu}\left(\mathbf{x}\right)=\sum_{\mathbf{e}_{a}\in\mathcal{G}_{\ell}}F_{a,\left(k\right)}^{\mu}\left(\mathbf{x}\right)=-\sum_{\mathbf{e}_{a}\in\mathcal{G}_{\ell}}P_{a}^{\mu\alpha}\left(\mathbf{x}\right)\;A_{\left(k\right)}^{\alpha},
\end{equation}
where $\mathbf{A}_{(y)} = \mathbf{e}_1$ and $\mathbf{A}_{(x)} = \mathbf{e}_2$ are the unit areas and $F_{\ell,\left(k\right)}^{\mu}\left(\mathbf{x}\right)$ is the group total force crossing the area element $\mathbf{A}_{(k)}$, while $F_{a,\left(k\right)}^{\mu}\left(\mathbf{x}\right)$ is the specific contribution along direction $\textbf{e}_a$. Let us come to the details of the calculation. A possible way to write $F_{a,\left(k\right)}^{\mu}$ is given by computing the contributions $N_{a,(k)}$ of the vectors $\mathbf{e}_a$ crossing $\mathbf{A}_{(k)}$ multiplied by the norm of an average force $\bar{F}_{a}(\textbf{x})$, i.e.
\begin{equation}\label{eq:force_sum}
F_{a,\left(k\right)}^{\mu}\left(\mathbf{x}\right)=\bar{F}_{a}\left(\mathbf{x}\right)N_{a,\left(k\right)}\,e_{a}^{\mu}.
\end{equation}
Hence, we need to specify both $N_{a,\left(k\right)}$ and $\bar{F}_{a}\left(\mathbf{x}\right)$. Let us start from the former. We draw in Fig.~\ref{fig:pt_areas} (a), (b) and (c) the force vectors intersecting the two unit area elements $\mathbf{A}_{(x)}$ (horizontal red line) and $\mathbf{A}_{(y)}$ (vertical red line), choosing, as an example, one direction for each of the three groups $\mathcal{G}_2$, $\mathcal{G}_8$ and $\mathcal{G}_5$ respectively. We determine $N_{a,\left(k\right)}$ using the following rules: i) if a vector $\mathbf{e}_a$, starting either at $\mathbf{x}+c_b\mathbf{e}_{b}$ or $\mathbf{x}-\mathbf{e}_{a}+c_b\mathbf{e}_{b}$ (with $c_b$ and $\mathbf{e}_b$ chosen in order to guarantee the intersection), crosses the area element  anywhere along its surface, excluding its boundary, then it contributes with weight $N_{\left(k\right)}\left(\mathbf{x}+c_{b}\mathbf{e}_{b},\mathbf{x}+\mathbf{e}_{a}+c_{b}\mathbf{e}_{b}\right)=N_{\left(k\right)}\left(\mathbf{x}-\mathbf{e}_{a}+c_{b}\mathbf{e}_{b},\mathbf{x}+c_{b}\mathbf{e}_{b}\right)=1$, ii) if a vector $\mathbf{e}_a$ starts or ends at the position where $\mathbf{A}_{(k)}$ is centered or it only superpose along the boundary, then it counts with weight $N_{\left(k\right)}\left(\mathbf{x}+c_{b}\mathbf{e}_{b},\mathbf{x}+\mathbf{e}_{a}+c_{b}\mathbf{e}_{b}\right)=N_{\left(k\right)}\left(\mathbf{x}-\mathbf{e}_{a}+c_{b}\mathbf{e}_{b},\mathbf{x}+c_{b}\mathbf{e}_{b}\right)=1/2$. The second rule is needed to avoid double counting the contribution of those vectors along the same direction that are ``shared'' by distinct parallel area elements (see Fig.~\ref{fig:pt_areas}). A supplementary rationalization of the last result for the ``shared'' forcing vectors~\cite{Shan08} has been given in~\cite{SbragagliaBelardinelli13}, following the pressure tensor construction of Irving \& Kirkwood~\cite{Irving1950}: the factor $1/2$ follows from choosing, on the basis of isotropy considerations, the normalization of the Dirac delta on half of the real line as $\int_{-\infty}^0 \delta(x) = 1/2$. In summary: each vector parallel to a given $\textbf{e}_a$ and crossing the area element $\textbf{A}_{(k)}$, contributes to the total sum $N_{a,(k)}$ by a weight that equals $1$, if the vector crosses the area element, or $1/2$ if the vector is shared by parallel area elements, i.e. if the vector starts or ends in the middle of the area element or simply touches the boundary of the area. Now we can determine the values of $N_{a,(k)}$ for the examples reported in Fig.~\ref{fig:pt_areas}. Let us begin with $\textbf{e}_5$ and $\textbf{e}_{13}$ for which the expression does not depend on the choice $k$ of the direction of the unit area element
\begin{equation}
  \begin{split}N_{5,\left(k\right)} & =N_{\left(k\right)}\left(\mathbf{x},\mathbf{x}+\mathbf{e}_{5}\right)+N_{\left(k\right)}\left(\mathbf{x}-\mathbf{e}_{5},\mathbf{x}\right)\\
 & =\frac{1}{2}+\frac{1}{2}=1,\\
N_{13,\left(k\right)} & =N_{\left(k\right)}\left(\mathbf{x}-\mathbf{e}_{13},\mathbf{x}\right)+N_{\left(k\right)}\left(\mathbf{x}-\frac{\mathbf{e}_{13}}{2},\mathbf{x}+\frac{\mathbf{e}_{13}}{2}\right)\\
 & +N_{\left(k\right)}\left(\mathbf{x},\mathbf{x}+\mathbf{e}_{13}\right)\\
 & =\frac{1}{2}+1+\frac{1}{2}=2,
\end{split}
\end{equation}
whereas in the case of $\textbf{e}_{17}$ we need to distinguish the area element directions
\begin{equation}
  \begin{split}
  N_{17,\left(y\right)} & =N_{\left(y\right)}\left(\mathbf{x}-\mathbf{e}_{17},\mathbf{x}\right)+N_{\left(y\right)}\left(\mathbf{x},\mathbf{x}+\mathbf{e}_{17}\right)\\
 & +N_{\left(y\right)}\left(\mathbf{x}-\mathbf{e}_{1},\mathbf{x}-\mathbf{e}_{1}+\mathbf{e}_{17}\right)\\
 & +N_{\left(y\right)}\left(\mathbf{x}-\mathbf{e}_{1}-\mathbf{e}_{2},\mathbf{x}-\mathbf{e}_{1}-\mathbf{e}_{2}+\mathbf{e}_{17}\right)\\
 & =4\times\frac{1}{2}=2,\\
  \end{split}
  \end{equation}
\begin{equation}
  \begin{split}
N_{17,\left(x\right)} & =N_{\left(x\right)}\left(\mathbf{x}-\mathbf{e}_{17},\mathbf{x}\right)+N_{\left(x\right)}\left(\mathbf{x},\mathbf{x}+\mathbf{e}_{17}\right)\\
 & =2\times\frac{1}{2}=1.
\end{split}
\end{equation}
As it was noticed in~\cite{Shan08}, the sum of these values coincides with the absolute value of the scalar product of the force direction and the area element
\begin{equation}
  N_{a,(k)} = e_a^\alpha A_{(k)}^\alpha,
\end{equation}
i.e. equal to $e_a^x$ and $e_a^y$ when crossing $\mathbf{A}_{(y)}$ and $\mathbf{A}_{(x)}$, respectively. Note that possible sign changes in $N_{a,(k)}$ reflect the possible choices of orientation of the area elements. We can rewrite Eq~\eqref{eq:force_sum} as
\begin{equation}\label{eq:single_farea}
F_{a,\left(k\right)}^{\mu}\left(\mathbf{x}\right)=\bar{F}_{a}\left(\mathbf{x}\right)\,e_{a}^{\mu}e_{a}^{\alpha}\,A_{\left(k\right)}^{\alpha}=-P_{a}^{\mu\alpha}\left(\mathbf{x}\right)\,A_{\left(k\right)}^{\alpha},
\end{equation}
from which one can read the definition of the lattice pressure tensor~\cite{Shan08}
\begin{equation}\label{eq:LP_def}
P_{a}^{\mu\nu}\left(\mathbf{x}\right)=-\bar{F}_{a}\left(\mathbf{x}\right)\,e_{a}^{\mu}e_{a}^{\nu}.
\end{equation}
We remark that the above definition of $N_{a,(k)}$ carries a sign of the relative orientation of the forcing vectors and the area element. While the vectorial nature of this sign is relevant for the definition of the pressure tensor, the contribution of a specific forcing vector is always assumed positive, i.e. the sign of $N_{a, (k)}$ is the same for a specific forcing vector $\textbf{e}_a$ and its opposite $\textbf{e}_{\bar{a}} = -\textbf{e}_a$, in agreement with the construction presented in~\cite{SbragagliaBelardinelli13}.

We now make some remarks about the symmetries of the terms in Eq.~\eqref{eq:LP_def}. We notice that the product of the stencil vectors on the right-hand side is invariant under axis reversal, or parity, transformations. Hence, opposite vectors, e.g. $\mathbf{e}_1$ and $\mathbf{e}_3 = - \mathbf{e}_1$, yield exactly the same contribution to the pressure tensor. On top of this we also notice that every time a $-\mathbf{e}_a$ appears in the pseudo-potential $\psi$ space dependence, it can be substituted with the opposite vector $\mathbf{e}_{\bar{a}} = -\mathbf{e}_a$ belonging to the same group. Hence, when considering all the vectors of the stencil, we need to multiply the total sum by 1/2.

Let us now define the average force $\bar{F}_a$. In order to take into account the variation of the force vectors crossing the area elements, we need to use an average force $\bar{F}_{a}\left(\mathbf{x}\right)$. In the multi-range case, one can immediately notice that the number of contributions for a given $\mathbf{e}_a$ may vary according to the direction of the area element. Let us use as a starting point the weighted sum of the crossing forces through $\textbf{A}_{(k)}$ along the direction $\textbf{e}_a$, i.e. the sum of the products between the weights $N_{(k)}\left(\mathbf{x}+c_{b}\mathbf{e}_{b},\mathbf{x}+\mathbf{e}_{a}+c_{b}\mathbf{e}_{b}\right)$, and the magnitude of the force defined between the same couple of points. For example, in the case of $\mathbf{e}_{17}$ one would obtain
\begin{equation}
  \begin{split}\bar{F}_{17,\left(y\right)}= & -Gc_{s}^{2}W\left(5\right)\psi\left(\mathbf{x}\right)\left[\frac{1}{2}\psi\left(\mathbf{x}-\mathbf{e}_{17}\right)+\frac{1}{2}\psi\left(\mathbf{x}+\mathbf{e}_{17}\right)\right]\\
-\frac{1}{2}G & c_{s}^{2}W\left(5\right)\psi\left(\mathbf{x}-\mathbf{e}_{1}\right)\psi\left(\mathbf{x}-\mathbf{e}_{1}+\mathbf{e}_{17}\right)\\
-\frac{1}{2}G & c_{s}^{2}W\left(5\right)\psi\left(\mathbf{x}-\mathbf{e}_{1}-\mathbf{e}_{2}\right)\psi\left(\mathbf{x}-\mathbf{e}_{1}-\mathbf{e}_{2}+\mathbf{e}_{17}\right),\\
\bar{F}_{17,\left(x\right)}= & -Gc_{s}^{2}W\left(5\right)\psi\left(\mathbf{x}\right)\left[\frac{1}{2}\psi\left(\mathbf{x}-\mathbf{e}_{17}\right)+\frac{1}{2}\psi\left(\mathbf{x}+\mathbf{e}_{17}\right)\right],
\end{split}
\end{equation}
A possible way to define a unique average force is to use the weighted sum with the largest total contribution and normalize it to the total sum of the weights. For the present case, we select $\bar{F}_{17}\left(\mathbf{x}\right)$ normalizing it by $|N_{17,(y)}| = |e_{17}^x|$, i.e.
\begin{equation}
\bar{F}_{17}\left(\mathbf{x}\right)=\frac{1}{|N_{17,\left(y\right)}|}\bar{F}_{17,\left(y\right)}\left(\mathbf{x}\right).
\end{equation}
Such a choice implies that, when considering the contribution of the forcing direction $\mathbf{e}_a$ crossing the surface area $\mathbf{A}_{(k)}$ (i.e.,  $F_{a,\left(k\right)}^{\mu}\left(\mathbf{x}\right)$, with $k=x,y$), one would obtain
\begin{equation}
\begin{split}F_{17,\left(y\right)}^{\mu}(\textbf{x}) & =\bar{F}_{17}\left(\mathbf{x}\right)e_{17}^{\alpha}e_{17}^{\mu}\,A_{\left(y\right)}^{\alpha}=\frac{A_{\left(y\right)}^{\alpha}e_{17}^{\alpha}}{|N_{17,\left(y\right)}|}\bar{F}_{17,\left(y\right)}\left(\mathbf{x}\right)e_{17}^{\mu}\\
=\frac{N_{17,\left(y\right)}}{|N_{17,\left(y\right)}|} & \bar{F}_{17,\left(y\right)}\left(\mathbf{x}\right)e_{17}^{\mu}=\bar{F}_{17,\left(y\right)}\left(\mathbf{x}\right)e_{17}^{\mu}\,\text{sign}\left(e_{17}^{x}\right),\\
F_{17,\left(x\right)}^{\mu}(\textbf{x}) & =\bar{F}_{17}\left(\mathbf{x}\right)e_{17}^{\alpha}e_{17}^{\mu}\,A_{\left(x\right)}^{\alpha}=\frac{A_{\left(x\right)}^{\alpha}e_{17}^{\alpha}}{|N_{17,\left(y\right)}|}\bar{F}_{17,\left(y\right)}\left(\mathbf{x}\right)e_{17}^{\mu}\\
=\frac{N_{17,\left(x\right)}}{|N_{17,\left(y\right)}|} & \bar{F}_{17,\left(y\right)}\left(\mathbf{x}\right)e_{17}^{\mu}=\frac{1}{2}\bar{F}_{17,\left(y\right)}\left(\mathbf{x}\right)e_{17}^{\mu}\,\text{sign}\left(e_{17}^{y}\right),
\end{split}
\end{equation}
hence, $|\mathbf{F}_{17,(y)}\left(\mathbf{x}\right)|=|\mathbf{F}_{17,(x)}\left(\mathbf{x}\right)|/2$ which is consistent with the ratio of the number of contributing vectors for the two area elements.

The above discussion has focused on those force vectors whose components do not have equal magnitude, or are not proportional to the coordinate basis. However, the above construction naturally applies to those vectors whose components have the same magnitude, i.e. $\{\mathcal{G}_2,\mathcal{G}_8\}$, since the intersecting vectors yield the same contribution for both area elements $|\mathbf{F}_{a,(y)}\left(\mathbf{x}\right)|=|\mathbf{F}_{a,(x)}\left(\mathbf{x}\right)|$, and also to the vectors proportional to the coordinate basis, i.e. $\{\mathcal{G}_1,\mathcal{G}_4\}$, for which the number of crossings alternatively equals zero according to $N_{a,(k)} = e_a^\alpha A_{(k)}^\alpha$.

Now, we can write the contribution to the lattice pressure tensor $P_a^{\mu\nu}$ for a specific vector belonging to each group, ordered according to squared norm of the group vectors $\ell = |\textbf{e}_a|^2$
\begin{equation}\label{eq:SingleEPT}
  \begin{split}P_{1}^{\mu\nu}= & Gc_{s}^{2}W\left(1\right)\psi\left(\mathbf{x}\right)\left[\frac{1}{2}\psi\left(\mathbf{x}+\mathbf{e}_{1}\right)+\frac{1}{2}\psi\left(\mathbf{x}-\mathbf{e}_{1}\right)\right]\,e_{1}^{\mu}e_{1}^{\nu},\\
P_{5}^{\mu\nu}= & Gc_{s}^{2}W\left(2\right)\psi\left(\mathbf{x}\right)\left[\frac{1}{2}\psi\left(\mathbf{x}+\mathbf{e}_{5}\right)+\frac{1}{2}\psi\left(\mathbf{x}-\mathbf{e}_{5}\right)\right]\,e_{5}^{\mu}e_{5}^{\nu},\\
P_{9}^{\mu\nu}= & \frac{Gc_{s}^{2}}{2}W\left(4\right)\psi\left(\mathbf{x}\right)\left[\frac{1}{2}\psi\left(\mathbf{x}+\mathbf{e}_{9}\right)+\frac{1}{2}\psi\left(\mathbf{x}-\mathbf{e}_{9}\right)\right]\,e_{9}^{\mu}e_{9}^{\nu}\\
+ & \frac{Gc_{s}^{2}}{2}W\left(4\right)\psi\left(\mathbf{x}-\frac{\mathbf{e}_{9}}{2}\right)\psi\left(\mathbf{x}+\frac{\mathbf{e}_{9}}{2}\right)\,e_{9}^{\mu}e_{9}^{\nu},\\
P_{17}^{\mu\nu}= & \frac{Gc_{s}^{2}}{4}W\left(5\right)\psi\left(\mathbf{x}\right)\psi\left(\mathbf{x}-\mathbf{e}_{17}\right)\,e_{17}^{\mu}e_{17}^{\nu}\\
+ & \frac{Gc_{s}^{2}}{4}W\left(5\right)\psi\left(\mathbf{x}\right)\psi\left(\mathbf{x}+\mathbf{e}_{17}\right)\,e_{17}^{\mu}e_{17}^{\nu}\\
+ & \frac{Gc_{s}^{2}}{4}W\left(5\right)\psi\left(\mathbf{x}-\mathbf{e}_{1}\right)\psi\left(\mathbf{x}-\mathbf{e}_{1}+\mathbf{e}_{17}\right)\,e_{17}^{\mu}e_{17}^{\nu}\\
+ & \frac{Gc_{s}^{2}}{4}W\left(5\right)\psi\left(\mathbf{x}-\mathbf{e}_{1}-\mathbf{e}_{2}\right)\\
& \times\psi\left(\mathbf{x}-\mathbf{e}_{1}-\mathbf{e}_{2}+\mathbf{e}_{17}\right)\,e_{17}^{\mu}e_{17}^{\nu},\\
P_{13}^{\mu\nu}= & \frac{Gc_{s}^{2}}{4}W\left(8\right)\psi\left(\mathbf{x}\right)\psi\left(\mathbf{x}+\mathbf{e}_{13}\right)\,e_{13}^{\mu}e_{13}^{\nu}\\
+ & \frac{Gc_{s}^{2}}{4}W\left(8\right)\psi\left(\mathbf{x}\right)\psi\left(\mathbf{x}-\mathbf{e}_{13}\right)\,e_{13}^{\mu}e_{13}^{\nu}\\
+ & \frac{Gc_{s}^{2}}{2}W\left(8\right)\psi\left(\mathbf{x}-\frac{\mathbf{e}_{13}}{2}\right)\psi\left(\mathbf{x}+\frac{\mathbf{e}_{13}}{2}\right)\,e_{13}^{\mu}e_{13}^{\nu}.
\end{split}
\end{equation}
The latter quantities can be used to define the different contributions to the full lattice pressure tensor reported in Eqs.~\eqref{eq:p_12},~\eqref{eq:p_48},~\eqref{eq:p_5a} and~\eqref{eq:p_5b} of Section~\ref{sec:PT}.

\section{Lattice Pressure Tensor Continuum Expansion}\label{app:PTDetails}
In this Section we provide some detailed calculations for the $4$-th order continuum expansion of the pressure tensor.
Let us start from the leading order $P_{\left[0\right]}^{\mu\nu}$: one can check that the contributions from all groups sum up to yield the second order isotropy constant $e_2$ (cf. Eq.~\eqref{eq:e_w})
\begin{equation}\label{eq:zeroth}
P_{\left[0\right]}^{\mu\nu}=\frac{Gc_{s}^{2}}{2}\psi^{2}\sum_{\mathbf{e}_{a}\in \mathcal{G} }W\left(|\mathbf{e}_{a}|^{2}\right)e_{a}^{\mu}e_{a}^{\nu}=\frac{Ge_{2}c_{s}^{2}}{2}\psi^{2}\delta^{\mu\nu}.
\end{equation}
If we sum this expression to the kinetic ideal gas contribution $P^{\mu\nu}_{\text{kin}}(\textbf{x})=n(\textbf{x})c_{s}^{2}\delta^{\mu\nu}$, we obtain the well-known expression for the bulk pressure~\cite{ShanChen93,ShanChen94,Shan06}:
\begin{equation}\label{eq:bulk_p}
P_{b}^{\mu\nu}=\left(nc_{s}^{2}+\frac{Gc_{s}^{2}e_{2}}{2}\psi^{2}\right)\delta^{\mu\nu}.
\end{equation}
Let us now analyze the second order derivatives terms (indicated with subscript $\left[\partial^{2}\right]$) from the groups $\mathcal{G}_1, \mathcal{G}_2, \mathcal{G}_4$ and $\mathcal{G}_{5a}$
\begin{equation}\label{eq:second12485a}
\begin{split}P_{\text{\ensuremath{\left(1,2\right)}}\left[\partial^{2}\right]}^{\mu\nu} & =Gc_{s}^{2}W(2)\Delta^{\alpha\beta\mu\nu}\psi\partial_{\alpha}\partial_{\beta}\psi\\
 & +\frac{Gc_{s}^{2}}{2}[W(1)-4W(2)]\delta^{\alpha\beta\mu\nu}\psi\partial_{\alpha}\partial_{\beta}\psi,\\
P_{\text{\ensuremath{\left(4,8\right)\left[\partial^{2}\right]}}}^{\mu\nu} & =12Gc_{s}^{2}W(8)\Delta^{\alpha\beta\mu\nu}\psi\partial_{\alpha}\partial_{\beta}\psi\\
 & +\frac{3Gc_{s}^{2}}{2}[4W(4)-16W(8)]\delta^{\alpha\beta\mu\nu}\psi\partial_{\alpha}\partial_{\beta}\psi,\\
P_{5a\left[\partial^{2}\right]}^{\mu\nu} & =4Gc_{s}^{2}W(5)\Delta^{\alpha\beta\mu\nu}\psi\partial_{\alpha}\partial_{\beta}\psi\\
 & -\frac{7Gc_{s}^{2}}{2}W(5)\delta^{\alpha\beta\mu\nu}\psi\partial_{\alpha}\partial_{\beta}\psi.
\end{split}
\end{equation}
These terms can be obtained by applying the results of Section~\ref{sec:F} and Appendix~\ref{app:IsoDetails} and computing the coefficients multiplying the isotropic $\Delta^{\alpha\beta\mu\nu}=\delta^{\alpha\beta}\delta^{\mu\nu}+\delta^{\alpha\mu}\delta^{\beta\nu}+\delta^{\alpha\nu}\delta^{\beta\mu}$ and anisotropic $\delta^{\alpha\beta\mu\nu}$ tensors according to $\sum_{\mathbf{e}_{a}\in\mathcal{G}_{\ell}}e_{a}^{\alpha}e_{a}^{\beta}e_{a}^{\mu}e_{a}^{\nu}=\mathcal{A}^{\left(4\right)}\left(\ell\right)\Delta^{\alpha\beta\mu\nu}+\mathcal{B}_{4}^{\left(4\right)}\left(\ell\right)\delta^{\alpha\beta\mu\nu}$. The expressions for the coefficients read
\begin{equation}
\begin{split}\mathcal{A}^{\left(4\right)}\left(\ell\right)= & \sum_{\mathbf{e}_{a}\in\mathcal{G}_{\ell}}\left(e_{a}^{x}\right)^{2}\left(e_{a}^{y}\right)^{2},\\
\mathcal{B}_{4}^{\left(4\right)}\left(\ell\right)= & \sum_{\mathbf{e}_{a}\in\mathcal{G}_{\ell}}\left(e_{a}^{x}\right)^{4}-3\mathcal{A}^{\left(4\right)}\left(\ell\right).
\end{split}
\end{equation}
Similarly, one finds the terms containing the product of first order derivatives (indicated with subscript $\left[\partial\partial\right]$) yielded by $\mathcal{G}_8$ and $\mathcal{G}_4$
\begin{equation}\label{eq:first8}
\begin{split}P_{\text{\ensuremath{\left(4,8\right)}}\left[\partial\partial\right]}^{\mu\nu}= & -4Gc_{s}^{2}W\left(8\right)\Delta^{\alpha\beta\mu\nu}\partial_{\alpha}\psi\partial_{\beta}\psi\\
  & -\frac{Gc_{s}^{2}}{2}[4W(4)-16W(8)]\delta^{\alpha\beta\mu\nu}\partial_{\alpha}\psi\partial_{\beta}\psi,\\
  \end{split}
\end{equation}
The only contributions to the expansion of the lattice pressure tensor that require further attention are the ones related to the shifted vectors of $\mathcal{G}_5$ reported in Eq.~\eqref{eq:p_5b}. Differently from the other contributions, Eq.~\eqref{eq:p_5b} yields an expansion where the product of two pairs of different vectors appears, namely terms of the type $e_1^\alpha e_1^\beta e_2^\mu e_2^\nu$. In order to extract from the latter terms the same tensorial structures appearing in $E_\ell^{\alpha\beta\mu\nu}$, namely $\Delta^{\alpha\beta\mu\nu}$ and $\delta^{\alpha\beta\mu\nu}$, we first need to define some basic quantities in terms of the Cartesian basis vectors, i.e. $\textbf{e}_1$ and $\textbf{e}_2$. As a first step we express the Kronecker delta as
\begin{equation}
\delta^{\alpha\beta}=\delta_{\mu\nu}\delta^{\alpha\mu}\delta^{\beta\nu}=\delta_{x}^{\alpha}\delta_{x}^{\beta}+\delta_{y}^{\alpha}\delta_{y}^{\beta}=e_{1}^{\alpha}e_{1}^{\beta}+e_{2}^{\alpha}e_{2}^{\beta}.
\end{equation}
Hence, by the same token, we write the rank-4 Kronecker delta as
\begin{equation}
\delta^{\alpha\beta\mu\nu}=e_{1}^{\alpha}e_{1}^{\beta}e_{1}^{\mu}e_{1}^{\nu}+e_{2}^{\alpha}e_{2}^{\beta}e_{2}^{\mu}e_{2}^{\nu}.
\end{equation}
In order to compute the 4-th order expansion of Eq.~\eqref{eq:p_5b} we need to manipulate the quantity $8e_{1}^{(\mu}e_{2}^{\nu)}e_{1}^{(\rho}e_{2}^{\sigma)}$, where we indicate the symmetric part of the vectors product as $e_{1}^{(\mu}e_{2}^{\nu)} = (e_{1}^{\mu}e_{2}^{\nu} + e_{1}^{\nu}e_{2}^{\mu})/2$. Since we want to retrieve terms related to $\Delta^{\alpha\beta\mu\nu}$ and $\delta^{\alpha\beta\mu\nu}$ we sum and subtract a few terms as follows
\begin{equation}
\begin{split}8e_{1}^{(\alpha}e_{2}^{\beta)}e_{1}^{(\mu}e_{2}^{\nu)}&=2\left(e_{1}^{\alpha}e_{2}^{\beta}+e_{2}^{\alpha}e_{1}^{\beta}\right)\left(e_{1}^{\mu}e_{2}^{\nu}+e_{2}^{\mu}e_{1}^{\nu}\right)\\=2(e_{1}^{\alpha}e_{1}^{\mu}&(e_{1}^{\beta}e_{1}^{\nu}+e_{2}^{\beta}e_{2}^{\nu})-e_{1}^{\alpha}e_{1}^{\mu}e_{1}^{\beta}e_{1}^{\nu})\\+2(e_{1}^{\beta}e_{1}^{\mu}&(e_{1}^{\alpha}e_{1}^{\nu}+e_{2}^{\alpha}e_{2}^{\nu})-e_{1}^{\beta}e_{1}^{\mu}e_{1}^{\alpha}e_{1}^{\nu})\\+2(e_{1}^{\alpha}e_{1}^{\nu}&(e_{1}^{\beta}e_{1}^{\mu}+e_{2}^{\beta}e_{2}^{\mu})-e_{1}^{\alpha}e_{1}^{\nu}e_{1}^{\beta}e_{1}^{\mu})\\+2(e_{1}^{\beta}e_{1}^{\nu}&(e_{1}^{\alpha}e_{1}^{\mu}+e_{2}^{\alpha}e_{2}^{\mu})-e_{1}^{\beta}e_{1}^{\nu}e_{1}^{\alpha}e_{1}^{\mu}).\end{split}
\end{equation}
It is still possible to perform a similar manipulation that would finally yield the desired tensorial structure and the very same term we started with but with opposite sign
\begin{equation}
\begin{split}8e_{1}^{(\alpha}e_{2}^{\beta)}e_{1}^{(\mu}e_{2}^{\nu)}=&+2[\left(e_{1}^{\alpha}e_{1}^{\mu}+e_{2}^{\alpha}e_{2}^{\mu}\right)\left(e_{1}^{\beta}e_{1}^{\nu}+e_{2}^{\beta}e_{2}^{\nu}\right)\\&\quad-e_{2}^{\alpha}e_{1}^{\beta}e_{2}^{\mu}e_{1}^{\nu}-\left(e_{1}^{\alpha}e_{1}^{\mu}e_{1}^{\beta}e_{1}^{\nu}+e_{2}^{\alpha}e_{2}^{\mu}e_{2}^{\beta}e_{2}^{\nu}\right)]\\&+2[\left(e_{1}^{\beta}e_{1}^{\mu}+e_{2}^{\beta}e_{2}^{\mu}\right)\left(e_{1}^{\alpha}e_{1}^{\nu}+e_{2}^{\alpha}e_{2}^{\nu}\right)\\&\quad-e_{1}^{\alpha}e_{2}^{\beta}e_{2}^{\mu}e_{1}^{\nu}-\left(e_{1}^{\beta}e_{1}^{\mu}e_{1}^{\alpha}e_{1}^{\nu}+e_{2}^{\alpha}e_{2}^{\nu}e_{2}^{\beta}e_{2}^{\mu}\right)]\\&+2[\left(e_{1}^{\alpha}e_{1}^{\nu}+e_{2}^{\alpha}e_{2}^{\nu}\right)\left(e_{1}^{\beta}e_{1}^{\mu}+e_{2}^{\beta}e_{2}^{\mu}\right)\\&\quad-e_{1}^{\beta}e_{2}^{\alpha}e_{2}^{\nu}e_{1}^{\mu}-\left(e_{1}^{\beta}e_{1}^{\mu}e_{1}^{\alpha}e_{1}^{\nu}+e_{2}^{\beta}e_{2}^{\mu}e_{2}^{\alpha}e_{2}^{\nu}\right)]\\&+2[\left(e_{1}^{\beta}e_{1}^{\nu}+e_{2}^{\beta}e_{2}^{\nu}\right)\left(e_{1}^{\alpha}e_{1}^{\mu}+e_{2}^{\alpha}e_{2}^{\mu}\right)\\&\quad-e_{1}^{\alpha}e_{2}^{\beta}e_{2}^{\nu}e_{1}^{\mu}-\left(e_{1}^{\alpha}e_{1}^{\mu}e_{1}^{\beta}e_{1}^{\nu}+e_{2}^{\alpha}e_{2}^{\mu}e_{2}^{\beta}e_{2}^{\nu}\right)]\\=2(2\delta^{\alpha\mu}\delta^{\beta\nu}+&2\delta^{\beta\mu}\delta^{\alpha\nu}-4\delta^{\alpha\beta\mu\nu})-8e_{1}^{(\alpha}e_{2}^{\beta)}e_{1}^{(\mu}e_{2}^{\nu)},\end{split}
\end{equation}
thus, we can write the following relation
\begin{equation}\label{eq:finale2e1}
8e_{1}^{(\alpha}e_{2}^{\beta)}e_{1}^{(\mu}e_{2}^{\nu)}=2\delta^{\alpha\mu}\delta^{\beta\nu}+2\delta^{\beta\mu}\delta^{\alpha\nu}-4\delta^{\alpha\beta\mu\nu}.
\end{equation}

Now, we examine the derivative expansion. Starting from Eq.~\eqref{eq:p_5b}, we begin by selecting the terms that are proportional to the second order derivative, bearing in mind to decompose the vectors $\textbf{e}_{17}$, $\textbf{e}_{18}$, $\textbf{e}_{19}$ and $\textbf{e}_{20}$ as a sum of $\textbf{e}_{1}$ and $\textbf{e}_{2}$. Hence, we obtain
\begin{equation}\label{eq:second5b}
\begin{split}
P_{5b\left[\partial^{2}\right]}^{\mu\nu}=&\frac{Gc_{s}^{2}}{4}W\left(5\right)e_{2}^{\alpha}e_{2}^{\beta}(\left[e_{18}^{\mu}e_{18}^{\nu}\right]+\left[e_{19}^{\mu}e_{19}^{\nu}\right])\psi\partial_{\alpha}\partial_{\beta}\psi\\+\frac{Gc_{s}^{2}}{4}W\left(5\right)&(e_{1}^{\alpha}e_{1}^{\beta}\left[e_{17}^{\mu}e_{17}^{\nu}\right]+e_{3}^{\alpha}e_{3}^{\beta}\left[e_{20}^{\mu}e_{20}^{\nu}\right])\psi\partial_{\alpha}\partial_{\beta}\psi\\+\frac{Gc_{s}^{2}}{4}W\left(5\right)&e_{5}^{\alpha}e_{5}^{\beta}(\left[e_{17}^{\mu}e_{17}^{\nu}\right]+\left[e_{18}^{\mu}e_{18}^{\nu}\right])\psi\partial_{\alpha}\partial_{\beta}\psi\\+\frac{Gc_{s}^{2}}{4}W\left(5\right)&e_{6}^{\alpha}e_{6}^{\beta}(\left[e_{19}^{\mu}e_{19}^{\nu}\right]+\left[e_{20}^{\mu}e_{20}^{\nu}\right])\psi\partial_{\alpha}\partial_{\beta}\psi\\=+\frac{3Gc_{s}^{2}}{2}W&\left(5\right)(e_{1}^{\mu}e_{1}^{\nu}e_{1}^{\alpha}e_{1}^{\beta}+e_{2}^{\alpha}e_{2}^{\beta}e_{2}^{\mu}e_{2}^{\nu})\psi\partial_{\alpha}\partial_{\beta}\psi\\+\frac{Gc_{s}^{2}}{2}W\left(5\right)&(e_{1}^{\alpha}e_{1}^{\beta}+e_{2}^{\alpha}e_{2}^{\beta})(e_{1}^{\mu}e_{1}^{\nu}+e_{2}^{\mu}e_{2}^{\nu})\psi\partial_{\alpha}\partial_{\beta}\psi\\+\frac{5Gc_{s}^{2}}{2}W\left(5\right)&(e_{1}^{\alpha}e_{1}^{\beta}+e_{2}^{\alpha}e_{2}^{\beta})(e_{1}^{\mu}e_{1}^{\nu}+e_{2}^{\mu}e_{2}^{\nu})\psi\partial_{\alpha}\partial_{\beta}\psi\\+Gc_{s}^{2}W\left(5\right)&(8e_{1}^{(\alpha}e_{2}^{\beta)}e_{1}^{(\mu}e_{2}^{\nu)})\psi\partial_{\alpha}\partial_{\beta}\psi\\=\frac{Gc_{s}^{2}}{4}W&\left(5\right)(8\Delta^{\alpha\beta\mu\nu}+4\delta^{\alpha\beta}\delta^{\mu\nu}-10\delta^{\alpha\beta\mu\nu})\psi\partial_{\alpha}\partial_{\beta}\psi
\end{split}
\end{equation}
Similarly, we consider the terms proportional to the product of two first derivatives from the expansion of Eq.~\eqref{eq:p_5b}, and finally obtain
\begin{equation}\label{eq:first5b}
\begin{split}P_{5b\left[\partial\partial\right]}^{\mu\nu}&=-\frac{Gc_{s}^{2}}{2}W\left(5\right)e_{5}^{\alpha}\left(e_{1}^{\beta}\left[e_{17}^{\mu}e_{17}^{\nu}\right]+e_{2}^{\beta}\left[e_{18}^{\mu}e_{18}^{\nu}\right]\right)\partial_{\alpha}\psi\partial_{\beta}\psi\\&-\frac{Gc_{s}^{2}}{2}W\left(5\right)e_{6}^{\alpha}\left(e_{2}^{\beta}\left[e_{19}^{\mu}e_{19}^{\nu}\right]+e_{3}^{\beta}\left[e_{20}^{\mu}e_{20}^{\nu}\right]\right)\partial_{\alpha}\psi\partial_{\beta}\psi\\&=-3Gc_{s}^{2}W\left(5\right)\left(e_{1}^{\alpha}e_{1}^{\beta}e_{1}^{\mu}e_{1}^{\nu}+e_{2}^{\alpha}e_{2}^{\beta}e_{2}^{\mu}e_{2}^{\nu}\right)\partial_{\alpha}\psi\partial_{\beta}\psi\\&-Gc_{s}^{2}W\left(5\right)\left(e_{1}^{\mu}e_{1}^{\nu}+e_{2}^{\mu}e_{2}^{\nu}\right)\left(e_{1}^{\alpha}e_{1}^{\beta}+e_{2}^{\alpha}e_{2}^{\beta}\right)\partial_{\alpha}\psi\partial_{\beta}\psi\\&-Gc_{s}^{2}W\left(5\right)(8e_{2}^{(\alpha}e_{1}^{\beta)}e_{1}^{(\mu}e_{2}^{\nu)})\partial_{\alpha}\psi\partial_{\beta}\psi\\&=-Gc_{s}^{2}W\left(5\right)\left(2\Delta^{\alpha\beta\mu\nu}-\delta^{\mu\nu}\delta^{\alpha\beta}-\delta^{\alpha\beta\mu\nu}\right)\partial_{\alpha}\psi\partial_{\beta}\psi
\end{split}
\end{equation}
It is now possible to sum up all the contributions, i.e. Eqs~\eqref{eq:zeroth},~\eqref{eq:second12485a},~\eqref{eq:first8},~\eqref{eq:second5b} and~\eqref{eq:first5b}, and recover the full expansion reported in Eq.~\eqref{eq:P4E8}.

%%%%%%%%%%%%%%%%%%%%%%%%%%%%%%%%%%%%%%%%%%%%%%%%%%%%%%%%%%%%%%%%%%%%%%%%%%%%%%%%%%%%%%%%%%%

\section{Forcing weights as a function of $\{e_{2n}\}$ and $\varepsilon$}\label{app:w2e}
By treating the forcing weights $\{W(|\textbf{e}_a|^2)\}$ as degrees of freedom, we can write them as functions of the first four isotropy constants $\{e_{2n}\}$ and the parameter $\varepsilon$. We do so in order to gain insight on the definition of the new forcing schemes, $\boldsymbol{E}^{(6)}_{P4,F6}$, $\boldsymbol{E}^{(8)}_{P4,F6}$ $\boldsymbol{E}^{(10)}_{P4,F6}$ and $\boldsymbol{E}^{(12)}_{P4,F6}$, yielding a higher order pressure tensor isotropy. The advantage results in a better understanding of the implications on the isotropy conditions when fixing the force expansion coefficients $e_{2n}$ and the macroscopic flat interface properties by means of $\varepsilon$.

We start by explicitly writing the expressions of $\{e_{2n}\}$ and $\varepsilon$ according to the new parametrization reported in Eq.~\eqref{eq:E_2n} (see Appendix~\ref{app:IsoDetails} for details)
\begin{equation}\label{eq:e_w}
  \begin{split}e_{2} & =2W\left(1\right)+4W\left(2\right)+8W\left(4\right)+20W\left(5\right)+16W\left(8\right),\\
e_{4} & =4W\left(2\right)+32W\left(5\right)+64W\left(8\right),\\
e_{6} & =\frac{4}{3}W\left(2\right)+\frac{80}{3}W\left(5\right)+\frac{256}{3}W\left(8\right),\\
e_{8} & =\frac{4}{9}W\left(2\right)+\frac{128}{9}W\left(5\right)+\frac{1024}{9}W\left(8\right),\\
\varepsilon & =\frac{48W\left(4\right)+96W\left(5\right)+96W\left(8\right)}{6W\left(1\right)+12W\left(2\right)+72W\left(4\right)+156W\left(5\right)+144W\left(8\right)}.
\end{split}
\end{equation}
It is possible to invert this system of equations and obtain the five weights as functions of the four isotropy coefficients and $\varepsilon$
\begin{equation}\label{eq:W_E}
\begin{split}W\left(1\right)= & \frac{1}{24}\left[\frac{6e_{2}}{\left(\varepsilon-1\right)}+18e_{2}-20e_{4}+27e_{6}-9e_{8}\right],\\
W\left(2\right)= & \frac{1}{36}\left(16e_{4}-24e_{6}+9e_{8}\right),\\
W\left(4\right)= & -\frac{1}{96}\left[\frac{6e_{2}}{\left(\varepsilon-1\right)}+6e_{2}-5e_{4}+18e_{6}-9e_{8}\right],\\
W\left(5\right)= & -\frac{1}{144}\left(4e_{4}-15e_{6}+9e_{8}\right),\\
W\left(8\right)= & \frac{1}{576}\left(e_{4}-6e_{6}+9e_{8}\right).
\end{split}
\end{equation}
We can use the above transformation to rewrite in the new variables the forcing isotropy conditions
\begin{equation}\label{eq:I_E}
  \begin{split}I_{4,0} & =+2W\left(1\right)-8W\left(2\right)+32W\left(4\right)-28W\left(5\right)-128W\left(8\right)\\
 & =-\frac{3e_{2}}{2\left(\varepsilon-1\right)}-\frac{1}{2}\left(e_{2}+6e_{4}\right),\\
I_{6,0} & =+2W\left(1\right)-16W\left(2\right)+128W\left(4\right)\\
 & \quad-140W\left(5\right)-1024W\left(8\right)\\
 & =-\frac{15e_{2}}{2\left(\varepsilon-1\right)}-\frac{1}{2}\left(13e_{2}+30e_{6}\right),\\
I_{8,0} & =+2W\left(1\right)+32W\left(2\right)+512W\left(4\right)\\
 & \quad-2108W\left(5\right)+8192W\left(8\right)\\
 & =-\frac{63e_{2}}{2\left(\varepsilon-1\right)}-\frac{1}{2}\left(61e_{2}-224e_{4}+840e_{6}-630e_{8}\right),\\
I_{8,1} & =-\frac{8}{3}W\left(2\right)+\frac{176}{3}W\left(5\right)-\frac{2048}{3}W\left(8\right)\\
 & =-4e_{4}+15e_{6}-15e_{8},
\end{split}
\end{equation}
and the pressure tensor ones
\begin{equation}\label{eq:P_E}
\begin{split}\chi_{I}= & -\frac{1}{144}\left[\frac{18e_{2}}{\left(\varepsilon-1\right)}+18e_{2}-17e_{4}+57e_{6}-18e_{8}\right],\\
\Lambda_{I}= & -\frac{1}{144}\left[\frac{36e_{2}}{\left(\varepsilon-1\right)}+125e_{4}-57e_{6}+18e_{8}\right].
\end{split}
\end{equation}
Given the condition $I_{4,0} = 0$, and matching both $e_2$ and $\varepsilon$, it follows that, at least for $\boldsymbol{E}^{(6)}_{P4,F6}$ and $\boldsymbol{E}^{(8)}_{P4,F6}$ (for which the above equations are valid), also the value of $e_4$, i.e. the surface tension, is matched. Our strategy (cf. Section~\ref{subsec:match}) yields the same result also for $\boldsymbol{E}^{(10)}_{P4,F6}$, while for $\boldsymbol{E}^{(12)}_{P4,F6}$ the value of $e_4$ differs from the target one $e_4(\boldsymbol{E}^{(12)}_{P2,F12})$ by $10^{-4}$, as reported in Table~\ref{tab:FP_vals}. Such a discrepancy will be the subject of further studies, and it only appears when mimicking with 5 weights $W(\ell)$ the isotropy properties of a stencil defined using 10 different weights.

Let us conclude this section by proving the relation in Eq.~\eqref{eq:sigma_e4}. We only need to use the definition of the coefficients $\chi_T = 4W(5) + 8W(8)$ and $\Lambda_T = 2W(2) + 24W(8) + 12W(5)$, provided right after the general expansion of the pressure tensor in Eq.~\eqref{eq:p4_niso}, and compare with the definition of $e_4$ in Eq.~\eqref{eq:e_w} obtaining
\begin{equation}
  \chi_T + \Lambda_T = \frac{1}{2}\left[4W(2) + 32W(5) + 64W(8)\right] = \frac{e_4}{2}
\end{equation}

\section{One dimensional Lattice Pressure Tensor}\label{app:1dPT}
In this Section, we provide a few details that allow to quickly compute the lattice pressure tensor for a one-dimensional interface without starting from the two-dimensional expression. This is instrumental for computing, in the case of $\boldsymbol{E}^{(10)}_{P2,F10}$ and $\boldsymbol{E}^{(12)}_{P2,F12}$, the values of the different coefficients $\alpha$, $\beta$, $\gamma$ and $\eta$ that have been provided in Sec.~\ref{sec:THEORY} for the case of a stencil with five weights only. Thus, we determine the expression for $\varepsilon(\boldsymbol{E}^{(10)}_{P2,F10})$ and $\varepsilon(\boldsymbol{E}^{(12)}_{P2,F12})$. Let us start by considering a planar interface between gas and liquid phases whose normal is oriented along the $x$ axis. To illustrate the key steps, let's focus on the vectors of the group $\mathcal{G}_1$: given the arguments in Appendix~\ref{app:LPT_def}, we only need to consider half of the vectors of each group; moreover by symmetry, we already know that in this case $P^{xy}_{\ell} = 0$, for each group $\mathcal{G}_{\ell}$. Hence, we only need to consider the diagonal terms of the lattice pressure tensor. We focus on $P^{xx}$ first: all terms are multiplied by $e_a^xe_a^x$, hence only $\mathbf{e}_1$ contributes. Considering that the pseudopotential only depends on $x$, we can follow the construction presented in Appendix~\ref{app:LPT_def} and write the average force as
\begin{equation}
\bar{F}_{1}=-\frac{Gc_{s}^{2}}{2}\psi\left(x\right)\left[\psi\left(x+1\right)+\psi\left(x-1\right)\right],
\end{equation}
hence, the the contribution to $P^{xx}$ from the group $\mathcal{G}_1$ is
\begin{equation}
P_{(1)}^{xx}\left(x\right)=-\bar{F}_{1}\left(x\right)e_{1}^{x}e_{1}^{x}=\frac{Gc_{s}^{2}}{2}\psi\left(x\right)\left[\psi\left(x+1\right)+\psi\left(x-1\right)\right].
\end{equation}
Let us now consider $P^{yy}$: all terms will be multiplied by $e_a^y e_a^y$ so that only the direction $\mathbf{e}_2$ contributes. However, along this direction, the pseudopotential keeps the constant value $\psi(x)$ so that we can immediately find
\begin{equation}
P_{(1)}^{yy}\left(x\right)=-\bar{F}_{2}\left(x\right)e_{2}^{y}e_{2}^{y}=Gc_{s}^{2}\psi^{2}\left(x\right).
\end{equation}
This construction is straightforward, and by making use of the results in Appendix~\ref{app:LPT_def}, we can write the two diagonal components of the lattice pressure tensor for the stencil $\boldsymbol{E}^{(12)}$. Let us begin with $P^{xx}$
\begin{equation}\label{eq:ptflatxx}
  \begin{split}
    P^{xx}\left(x\right)&=Gc_{s}^{2}a_{\left[-4,0,4\right]}^{\left(xx\right)}\psi\left(x\right)\left(\psi\left(x+4\right)+\psi\left(x-4\right)\right)\\+&Gc_{s}^{2}a_{\left[-3,0,3\right]}^{\left(xx\right)}\psi\left(x\right)\left(\psi\left(x+3\right)+\psi\left(x-3\right)\right)\\+&Gc_{s}^{2}a_{\left[-2,0,2\right]}^{\left(xx\right)}\psi\left(x\right)\left(\psi\left(x+2\right)+\psi\left(x-2\right)\right)\\+&Gc_{s}^{2}a_{\left[-1,0,1\right]}^{\left(xx\right)}\psi\left(x\right)\left(\psi\left(x+1\right)+\psi\left(x-1\right)\right)\\+&Gc_{s}^{2}b_{\left[2,2\right]}^{\left(xx\right)}\psi\left(x+2\right)\psi\left(x-2\right)\\+&Gc_{s}^{2}b_{\left[1,1\right]}^{\left(xx\right)}\psi\left(x+1\right)\psi\left(x-1\right)\\+&Gc_{s}^{2}b_{\left[1,3\right]}^{\left(xx\right)}\left[\psi\left(x+3\right)\psi\left(x-1\right)+\psi\left(x+1\right)\psi\left(x-3\right)\right]\\+&Gc_{s}^{2}b_{\left[1,2\right]}^{\left(xx\right)}\left[\psi\left(x+2\right)\psi\left(x-1\right)+\psi\left(x+1\right)\psi\left(x-2\right)\right],
  \end{split}
\end{equation}
with the coefficients given by
\begin{equation}
  \begin{split}
    a_{\left[-4,0,4\right]}^{\left(xx\right)}=&2W\left(16\right)+4W\left(17\right),\\a_{\left[-3,0,3\right]}^{\left(xx\right)}=&\frac{3}{2}W\left(9\right)+3W\left(10\right)+3W\left(13\right),\\a_{\left[-2,0,2\right]}^{\left(xx\right)}=&W\left(4\right)+2W\left(5\right)+2W\left(8\right)+\frac{4}{3}W\left(13\right),\\a_{\left[-1,0,1\right]}^{\left(xx\right)}=&\frac{1}{2}W\left(1\right)+W\left(2\right)+W\left(5\right)+W\left(10\right)+W\left(17\right),\\b_{\left[2,2\right]}^{\left(xx\right)}=&4W\left(16\right)+8W\left(17\right),\\b_{\left[1,1\right]}^{\left(xx\right)}=&2W\left(4\right)+4W\left(5\right)+4W\left(8\right)+\frac{16}{3}W\left(13\right),\\b_{\left[1,3\right]}^{\left(xx\right)}=&4W\left(16\right)+8W\left(17\right),\\b_{\left[1,2\right]}^{\left(xx\right)}=&3W\left(9\right)+6W\left(10\right)+6W\left(13\right).
  \end{split}
\end{equation}
Finally we write $P^{yy}$
\begin{equation}\label{eq:ptflatyy}
  \begin{split}
    P^{yy}\left(x\right)&=Gc_{s}^{2}a_{\left[-4,0,4\right]}^{\left(yy\right)}\psi\left(x\right)\left(\psi\left(x+4\right)+\psi\left(x-4\right)\right)\\+&Gc_{s}^{2}a_{\left[-3,0,3\right]}^{\left(yy\right)}\psi\left(x\right)\left(\psi\left(x+3\right)+\psi\left(x-3\right)\right)\\+&Gc_{s}^{2}a_{\left[-2,0,2\right]}^{\left(yy\right)}\psi\left(x\right)\left(\psi\left(x+2\right)+\psi\left(x-2\right)\right)\\+&Gc_{s}^{2}a_{\left[-1,0,1\right]}^{\left(yy\right)}\psi\left(x\right)\left(\psi\left(x+1\right)+\psi\left(x-1\right)\right)\\+&Gc_{s}^{2}a_{\left[0\right]}^{\left(yy\right)}\psi^{2}\left(x\right)\\+&Gc_{s}^{2}b_{\left[2,2\right]}^{\left(yy\right)}\psi\left(x-2\right)\psi\left(x+2\right)\\+&Gc_{s}^{2}b_{\left[1,1\right]}^{\left(yy\right)}\psi\left(x-1\right)\psi\left(x+1\right)\\+&Gc_{s}^{2}b_{\left[1,3\right]}^{\left(yy\right)}\left[\psi\left(x+3\right)\psi\left(x-1\right)+\psi\left(x+1\right)\psi\left(x-3\right)\right]\\+&Gc_{s}^{2}b_{\left[1,2\right]}^{\left(yy\right)}\left[\psi\left(x+2\right)\psi\left(x-1\right)+\psi\left(x+1\right)\psi\left(x-2\right)\right],
  \end{split}
\end{equation}
and the related coefficients
\begin{equation}
  \begin{split}
    a_{\left[-4,0,4\right]}^{\left(yy\right)}=&\frac{1}{4}W\left(17\right),\\a_{\left[-3,0,3\right]}^{\left(yy\right)}=&\frac{1}{3}W\left(10\right)+\frac{4}{3}W\left(13\right),\\a_{\left[-2,0,2\right]}^{\left(yy\right)}=&\frac{1}{2}W\left(5\right)+2W\left(8\right)+3W\left(13\right),\\a_{\left[-1,0,1\right]}^{\left(yy\right)}=&W\left(2\right)+4W\left(5\right)+9W\left(10\right)+16W\left(17\right),\\a_{\left[0\right]}^{\left(yy\right)}=&W\left(1\right)+4W\left(4\right)+9W\left(9\right)+16W\left(16\right),\\b_{\left[2,2\right]}^{\left(yy\right)}=&\frac{1}{2}W\left(17\right),\\b_{\left[1,1\right]}^{\left(yy\right)}=&W\left(5\right)+4W\left(8\right)+12W\left(13\right),\\b_{\left[1,3\right]}^{\left(yy\right)}=&\frac{1}{2}W\left(17\right),\\b_{\left[1,2\right]}^{\left(yy\right)}=&\frac{2}{3}W\left(10\right)+\frac{8}{3}W\left(13\right).
  \end{split}
\end{equation}
The expressions in Eqs~\eqref{eq:ptflatxx} and~\eqref{eq:ptflatyy} include all the lower isotropy stencils as subcases. Let us now examine the Taylor expansion of $P^{xx}$ from which we can extract the expression for $\varepsilon$ for $\boldsymbol{E}^{(12)}$. Let us report once again the general expression (see Eq.~\eqref{eq:p1d})
\begin{equation}
  P^{xx}=nc_{s}^{2}+\frac{Gc_{s}^{2}e_{2}}{2}\psi^{2}+\frac{Gc_{s}^{2}}{12}\left[\beta\psi\frac{\text{d}^{2}\psi}{\text{d}x^{2}}+\alpha\left(\frac{\text{d}\psi}{\text{d}x}\right)^{2}\right]
\end{equation}
for which the coefficients are now given by
\begin{equation}
  \begin{split}
    \alpha=&-\left[24W\left(4\right)+48W\left(5\right)+48W\left(8\right)+144W\left(9\right)\right]\\&-\left[288W\left(10\right)+352W\left(13\right)+480W\left(16\right)+960W\left(17\right)\right]\\&\\\beta=&6W\left(1\right)+12W\left(2\right)+72W\left(4\right)+156W\left(5\right)+144W\left(8\right)\\+&342W\left(9\right)+696W\left(10\right)+812W\left(13\right)+1056W\left(16\right)\\+&2124W\left(17\right)
  \end{split}
\end{equation}
so that by following the definition $\varepsilon = -2\alpha/\beta$ one gets the extended expression for $\varepsilon$.

We also wish to check the surface tension coefficient. In order to do so, we first report the Taylor expansion for $P^{yy}$
\begin{equation}
P^{yy} = nc_{s}^{2}+\frac{Gc_{s}^{2}e_{2}}{2}\psi^{2}+\frac{Gc_{s}^{2}}{4}\left[\eta\psi\frac{\text{d}^{2}\psi}{\text{d}x^{2}}+\gamma\left(\frac{\text{d}\psi}{\text{d}x}\right)^{2}\right]
\end{equation}
and its coefficients
\begin{equation}
  \begin{split}
    \eta=&4\left[W\left(2\right)+7W\left(5\right)+12W\left(8\right)\right]\\+&4\left[\frac{46}{3}W\left(10\right)+\frac{148}{3}W\left(13\right)+27W\left(17\right)\right]\\&\\\gamma=&-4\left[W\left(5\right)+4W\left(8\right)\right]\\&-4\left[\frac{8}{3}W\left(10\right)+\frac{68}{3}W\left(13\right)+5W\left(17\right)\right]
  \end{split}
\end{equation}
We notice that it is only possible to translate these combinations of weights in terms of the isotropy coefficients $e_{2n}$ and $\varepsilon$ only for stencils up to $\boldsymbol{E}^{(8)}$: starting from $\boldsymbol{E}^{(10)}$, the number of weights outgrows the number of isotropy coefficients at which order the forcing is isotropic. Using the isotropy coefficients of the orders for which the isotropy conditions are not satisfied only brings in linearly dependent equations, so it is not a viable alternative.

Finally, we write the surface tension as
\begin{equation}
  \begin{split}
    \sigma&=\int_{-\infty}^{+\infty}\mbox{d}x\left[P^{xx}(x)-P^{yy}(x)\right]\\&=-\frac{Gc_{s}^{2}}{12}\left[\beta-\alpha+3\left(\gamma-\eta\right)\right]\int_{-\infty}^{+\infty}\mbox{d}x\left[\frac{\text{d}\psi\left(x\right)}{\text{d}x}\right]^{2}
  \end{split}
\end{equation}
from which we define the constant coefficient $\hat{\sigma} = -[\beta-\alpha+3\left(\gamma-\eta\right)]/12$
\begin{equation}
  \begin{split}
    \hat{\sigma}=&-\frac{1}{2}\left[W\left(1\right)+16W\left(2\right)+18W\left(5\right)+81W\left(9\right)\right]\\&-\frac{1}{2}\left[128W\left(10\right)+50W\left(13\right)+256W\left(16\right)+450W\left(17\right)\right].
  \end{split}
\end{equation}
We provide here the expressions of $e_4$ and $I_{4,0}$ for $\boldsymbol{E}^{(12)}$
\begin{equation}
  \begin{split}
    e_{4}=&4W(2)+32W(5)+64W(8)+72W(10)\\+&288W(13)+128W(17)\\
    I_{4,0}=&2W(1)-8W(2)+32W(4)-28W(5)-128W(8)\\+&162W(9)+112W(10)-476W(13)\\+&512W(16)+644W(17),
  \end{split}
\end{equation}
so that one can check that the same result as in Eq.~\eqref{eq:sigma_e4} still holds
\begin{equation}
\hat{\sigma}=-\frac{e_{4}}{2}-\frac{I_{4,0}}{4}=-\frac{e_{4}}{2},
\end{equation}
assuming that $I_{4,0} = 0$, i.e. the $4$-th order isotropy condition is satisfied.

The different expressions for the isotropy coefficients are reported in the Jupyter notebook~\cite{ipython} relative to this paper, accessible on the github repository \href{https://github.com/lullimat/idea.deploy}{https://github.com/lullimat/idea.deploy}~\cite{sympy, scipy, numpy0, numpy1, scikit-learn, matplotlib, ipython, pycuda_opencl}.

\section{Stencil Isotropy Details}\label{app:IsoDetails}
In this Section, we present the details of the derivation of the expressions for the isotropy constants, i.e. $e_{2n}$, and forcing isotropy conditions, i.e. $I_{2n,k}$, which have been introduced in Section~\ref{sec:F} as the isotropic and anisotropic contributions  to $E^{\mu_{1}\ldots\mu_{2n}}$ in Eq.~\eqref{eq:Eall} and further specified in Eqs.~\eqref{eq:iso_cond_short} and~\eqref{eq:E_aniso}. Expressing the isotropy constants $e_{2n}$ as functions of the weights, as in Appendix~\ref{app:w2e}, allows us to define the system of equations whose solution is the set of weights defining $\boldsymbol{E}^{(6)}_{P4,F6}$, $\boldsymbol{E}^{(8)}_{P4,F6}$, $\boldsymbol{E}^{(10)}_{P4,F6}$ and $\boldsymbol{E}^{(12)}_{P4,F6}$ (see Table~\ref{tab:W_vals}), yielding a 4-th order isotropic pressure tensor. The presentation below provides a basis for the generalization of the results presented in this paper at higher order and in three-dimensions, which must be complemented by a parallel development of the results obtained in Appendix~\ref{app:PTDetails} relative to the product of vectors belonging to different groups [see Eq.~\eqref{eq:finale2e1}]. Technically, we adopt a slightly different perspective with respect to earlier multi-range works~\cite{Shan06,Sbragaglia07}, by generalizing (to the best of our knowledge) the analysis reported in~\cite{Wolfram1986}, which was limited to the $6$-th isotropy order~\footnote{See Eq.~(3.5.5), (3.5.6) and (3.5.7) in~\cite{Wolfram1986}}.

Let us start from the definition of $E^{\mu_{1}\ldots\mu_{2n}}$ in Eq.~\eqref{eq:Eall}: we can see that a summation over all groups is used. However, we can split the definition for each group, i.e. keeping fixed the square norm $|\textbf{e}_a|^2 = \ell$, so that we can write the group-wise quantities as
\begin{equation}\label{eq:Esingle}
E_{\ell}^{\mu_{1}\ldots\mu_{2n}}=\sum_{\mathbf{e}_{a}\in\mathcal{G}_{\ell}}e_{a}^{\mu_{1}}e_{a}^{\mu_{2}}\cdots e_{a}^{\mu_{2n}},
\end{equation}
for which a possible parametrization for $n\geq 2$ can be written as
\begin{equation}\label{eq:EDef}
\begin{split}E_{\ell}^{\mu_{1}\ldots\mu_{2n}}= & \mathcal{A}^{\left(2n\right)}\left(\ell\right)\Delta^{\mu_{1}\ldots\mu_{2n}}\\
+ & \mathcal{B}_{2n}^{\left(2n\right)}\left(\ell\right)\delta^{\mu_{1}\ldots\mu_{2n}}\\
+ & \mathcal{B}_{2n-2}^{\left(2n\right)}\left(\ell\right)\left[\delta^{\mu_{1}\mu_{2}}\delta^{\mu_{3}\ldots\mu_{2n}}+\text{perms}\right]+\cdots\\
+ & \mathcal{B}_{2n-M\left(n\right)}^{\left(2n\right)}\left(\ell\right)\left[\delta^{\mu_{1}\ldots\mu_{M\left(n\right)}}\delta^{\mu_{M\left(n\right)+1}\ldots\mu_{2n}}+\text{perms}\right]
\end{split}
\end{equation}
or in a more compact form
\begin{equation}\label{eq:E_2n}
  \begin{split}E_{\ell}^{\mu_{1}\ldots\mu_{2n}} & =\mathcal{A}^{\left(2n\right)}\left(\ell\right)\Delta^{\mu_{1}\ldots\mu_{2n}}\\
+\sum_{k=0}^{M\left(n\right)/2} & \mathcal{B}_{2n-2k}^{\left(2n\right)}\left(\ell\right)\left[\delta^{\mu_{1}\ldots\mu_{2k}}\delta^{\mu_{2k+1}\ldots\mu_{2n}}+\text{perms}\right].
\end{split}
\end{equation}
In the above expressions $\Delta^{\mu_1\ldots\mu_{2n}}$ is the $2n$-rank isotropic tensor~\cite{Wolfram1986, Shan06, Sbragaglia07}, $\delta^{\mu_1\ldots\mu_{2n}}$ is the $2n$-rank Kronecker delta (which equals one only if all indices take the same value) and $M(n) = n - (2 + n\,\mbox{mod}\,2)$ (notice that we use both $n$ and $2n$ in the definitions). Finally, we set the convention $\delta^{\mu_n\mu_k} = 1$ for $n\geq 1$ and $k=0$, e.g. $\delta^{\mu_1\mu_0} = 1$, $\delta^{\mu_2\mu_0} = 1$ and so on. The constants $\mathcal{A}^{\left(2n\right)}\left(\ell\right)$ take on different values for each group of vectors of squared length $\ell=|\mathbf{e}_a|^2$ and {\it they all multiply isotropic tensors}. Similarly, the coefficients $\mathcal{B}_{2n-2k}^{\left(2n\right)}\left(\ell\right)$ depend on the specific group and {\it they all multiply the anisotropic contributions} given by the higher rank Kronecker deltas. Hence, given Eq.~\eqref{eq:E_2n}, it is clear that a single group of vectors cannot be used as a basis for $2n$-rank isotropic tensors, because it is not possible to eliminate the anisotropic contributions. The solution is to use more than a group as it is done in Eq.~\eqref{eq:sc_f}, so that the total sum of the $2n$-indices quantities can be made fully isotropic. By summing $E_\ell$ over the different groups, we single out the coefficients $e_{2n}$ (cf. Eq.\eqref{eq:iso_cond_short} and nearby discussion) multiplying the fully isotropy tensors of rank $2n$, i.e. the isotropy coefficients, and the isotropy conditions $I_{2n,k} = 0$ assuring the vanishing of the anisotropic contributions
\begin{equation}\label{eq:IsoCond}
  \begin{split} & e_{2n}=\sum_{\ell}\mathcal{A}^{\left(2n\right)}\left(\ell\right)W\left(\ell\right),\\
 & \left\{ I_{2n,k}=\sum_{\ell}\mathcal{B}_{2n-2k}^{\left(2n\right)}\left(\ell\right)W\left(\ell\right)=0\right\} .
\end{split}
\end{equation}
We remark that Eq.~\eqref{eq:E_2n} only represents a definition of the anisotropic contribution coefficients $\mathcal{B}_{2n-2k}^{\left(2n\right)}\left(\ell\right)$ allowing to set their combination to zero as in Eq.~\eqref{eq:IsoCond}.

Let us now discuss the combinatorial aspect of Eq. \eqref{eq:E_2n}. We remark that the present discussion assumes $n\geq 2$. The quantity $M(n) = n - (2 + n\,\mbox{mod}\,2)$ is related to the maximum of the sum. The limit $k = M(n)/2$ is imposed in order to avoid double counting the tensorial structures. This point can be better understood by some direct examples: choosing $2n=4$ we get $M(2) = 0$, i.e. the above sum only contains the $k=0$ element, which is indeed the case, since at 4-th order one can only have either the full isotropic tensor $\Delta^{(4)}$ or the higher rank Kronecker delta $\delta^{(4)}$, whose coefficients are going to be captured by the $k=0$ terms. If we consider $2n=6$, then $M(3)=0$ yielding only $\Delta^{(6)}$ and $\delta^{(6)}$ in agreement with the highest order explicitly treated in~\cite{Wolfram1986}. For $2n=8$ one would get $M(4) = 2$, so that the sum would end at $k = M(4)/2 = 1$. This result is compatible with the analysis reported in~\cite{Shan06,Sbragaglia07} yielding two isotropy conditions for the forcing at the $8$-th order.

Let us now look at the possible arrangements of an even number of the two variables $x$ and $y$ in a set of $2n$ elements. For $2n = 4$ it is clear that only two arrangements are possible, either $\{x,x,x,x\}$ or $\{x,x,y,y\}$ since the ones obtained from the exchange $x\leftrightarrow y$, namely $\{y,y,y,y\}$ and $\{y,y,x,x\}$, are expected to yield the same expressions, given the invariance of the vectors of the group under coordinate permutations. Hence, for the problem of finding the independent indices arrangements, one needs to consider all those permutations that are not trivially linked by coordinates exchange. In the case of $2n = 6$, one still has two possible arrangements $\{x,x,x,x,x,x\}$ and $\{x,x,x,x,y,y\}$, while for $2n = 8$ there are three, namely $\{x,x,x,x,x,x,x,x\}$, $\{x,x,x,x,x,x,y,y\}$ and $\{x,x,x,x,y,y,y,y\}$.

Furthermore, we notice that, at each order $2n$, all arrangements different from the homogeneous one $\{x,x,\ldots,x,x\}$, would allow at most two tensorial structures to yield a contribution. Let us analyze again the previous examples: for $2n=4$ the combination $\{x,x,x,x\}$ is such that both $\Delta^{xxxx} = 3$ (see~\cite{Wolfram1986,Sbragaglia07}) and $\delta^{xxxx} = 1$ differ from zero, while for $\{x,x,y,y\}$ the only non-zero contribution would be $\Delta^{xxyy} = 1$ since $\delta^{xxyy}=0$. Similar arguments hold for $2n=6$. For $2n=8$ one has three tensorial structures, namely $\Delta^{(8)}$, $\delta^{(8)}$ and $\delta^{(2)}\delta^{(6)}$, which in Eq.~\eqref{eq:E_2n} are multiplied by $\mathcal{A}^{(8)}(\ell)$, $\mathcal{B}^{(8)}_{8}(\ell)$ and $\mathcal{B}^{(8)}_{6}(\ell)$ respectively. For $\{x,x,x,x,x,x,x,x\}$ all three terms survive yielding $\mathcal{A}^{(8)}(\ell)\,\Delta^{xxxxxxxx} = \mathcal{A}^{(8)}(\ell)\,7!!$, $\mathcal{B}^{(8)}_{8}(\ell)\,\delta^{xxxxxxxx} = \mathcal{B}^{(8)}_{8}(\ell)$ and $\mathcal{B}^{(8)}_{6}(\ell)\,(\delta^{xx}\delta^{xxxxxx} + \text{perms.}) = \mathcal{B}^{(8)}_{6}(\ell)\,\binom{8}{2}$, while for $\{x,x,x,x,x,x,y,y\}$ one has $\mathcal{A}^{(8)}(\ell)\,\Delta^{xxxxxxyy} = \mathcal{A}^{(8)}(\ell)\,5!!$, $\mathcal{B}^{(8)}_{8}(\ell)\,\delta^{xxxxxxyy} = 0$ and $\mathcal{B}^{(8)}_{6}(\ell)\,(\delta^{yy}\delta^{xxxxxx} + \text{perms.}) = \mathcal{B}^{(8)}_{6}(\ell)$, where in the last term only one of the possible combinations survives. The last permutation $\{x,x,x,x,y,y,y,y\}$ yields only the term proportional to the fully isotropic tensor $\mathcal{A}^{(8)}(\ell)\,\Delta^{xxxxyyyy} = \mathcal{A}^{(8)}(\ell)\,3!!3!!$. Thus, we can define a system of equations to determine the coefficients $\mathcal{A}^{(8)}(\ell)$, $\mathcal{B}^{(8)}_{8}(\ell)$ and $\mathcal{B}^{(8)}_{6}(\ell)$ for any value of $\ell$, by means of Eq.~\eqref{eq:E_2n}: we enumerate all possible independent indices permutations and isolate the non-vanishing terms in
\begin{equation}
\begin{split}
  & \sum_{\mathbf{e}_{a}\in\mathcal{G}_{\ell}}e_{a}^{\mu_{1}}e_{a}^{\mu_{2}}\cdots e_{a}^{\mu_{8}} \\
  &= \mathcal{A}^{(8)}(\ell)\,\Delta^{\mu_1\ldots\mu_8} + \mathcal{B}^{(8)}_{8}(\ell)\,\delta^{\mu_1\ldots\mu_8} \\
  & + \mathcal{B}^{(8)}_{6}(\ell)\,(\delta^{\mu_1\mu_2}\delta^{\mu_3\ldots\mu_8} + \text{perms}),\\
\end{split}
\end{equation}
yielding, for each permutation, a linear equation. The system can then be solved for the coefficients $\mathcal{A}^{(8)}(\ell)$, $\mathcal{B}^{(8)}_{8}(\ell)$ and $\mathcal{B}^{(8)}_{6}(\ell)$.

Let us now analyze the general case in which we select the first $2n_x$ indices to be equal to $x$ and the remaining $2n - 2n_x = 2n_y$ to be equal to $y$, so that one would get
\begin{equation}
  \begin{split} & \quad\quad\sum_{\mathbf{e}_{a}\in\mathcal{G}_{\ell}}\left(e_{a}^{x}\right)^{2n_{x}}\left(e_{a}^{y}\right)^{2n_{y}}=\\
 & =\mathcal{A}^{\left(2n\right)}\left(\ell\right)\left(2n_{x}-1\right)!!\left(2n_{y}-1\right)!!+\mathcal{B}_{2n}^{\left(2n\right)}\left(\ell\right)\delta_{\text{k}}\left(2n_{y}\right)\\
 & +\sum_{k=1}^{M\left(n\right)/2}\mathcal{B}_{2n-2k}^{\left(2n\right)}\left(\ell\right)\;\left[Z_{2n-2k}^{\left(2n\right)}\delta_{\text{k}}\left(2n_{y}\right)+\delta_{\text{k}}\left(2n_{y}-2k\right)\right],
\end{split}
\end{equation}
where $\delta_{\text{k}}(a)$ is the Kronecker delta being equal to 1 when $a=0$, and $Z^{(2n)}_{2n - 2k} = \binom{2n}{2n - 2k} = \binom{2n}{2k}$ indicates the number of possible independent permutations of the indices in the terms $\delta^{\mu_{1}\ldots\mu_{2k}}\delta^{\mu_{2k+1}\ldots\mu_{2n}}$. 

%%  which allows for a total range of variation $m(n) - M(n) = 2(n\;\mbox{mod}\;2 + 1)$
The above arguments of symmetry under coordinate exchange $x\leftrightarrow y$ impose a lower limit $2n_x \geq m(n) = n + n\;\mbox{mod}\;2$: all indices permutations below this value, i.e.  $2n_x < m(n)$, coincide, under coordinates exchange $x\leftrightarrow y$, with those such that $2n_x \geq m(n)$. At the lower bound, for $2n_x = m(n)$, remembering the upper limit of the summation $M(n) = n - (2 + n\,\mbox{mod}\,2)$, one has $2n_y = 2n - 2n_x = n - n\;\mbox{mod}\;2 > M(n)$ so that all the $\mathcal{B}^{(2n)}_{2q}$ terms disappear allowing to compute the coefficient $\mathcal{A}^{(2n)}$ as
\begin{equation}\label{eq:A2n}
  \mathcal{A}^{\left(2n\right)}\left(\ell\right)=\frac{\sum_{\mathbf{e}_{a}\in\mathcal{G}_{\ell}}\left(e_{a}^{x}\right)^{n+n\;\text{mod}\;2}\left(e_{a}^{y}\right)^{n-n\;\text{mod}\;2}}{\left(n+n\;\text{mod}\;2-1\right)!!\left(n-n\;\text{mod}\;2-1\right)!!}.
\end{equation}
For $2n - M(n) \leq 2q < 2n$ the coefficients $\mathcal{B}^{(2n)}_{2q}$ can be computed as
\begin{equation}\label{eq:B2q}
  \begin{split}\mathcal{B}_{2q}^{\left(2n\right)}\left(\ell\right) & =\sum_{\mathbf{e}_{a}\in\mathcal{G}_{\ell}}\left(e_{a}^{x}\right)^{2q}\left(e_{a}^{y}\right)^{2n-2q}\\
 & -\mathcal{A}^{\left(2n\right)}\left(\ell\right)\left(2q-1\right)!!\left(2n-2q-1\right)!!,
\end{split}
\end{equation}
while in the limiting case $2q=2n$ one has
\begin{equation}\label{eq:B2n}
  \begin{split}\mathcal{B}_{2n}^{\left(2n\right)}\left(\ell\right) & =\sum_{\mathbf{e}_{a}\in\mathcal{G}_{\ell}}\left(e_{a}^{x}\right)^{2n}-\mathcal{A}^{\left(2n\right)}\left(\ell\right)\left(2n-1\right)!!\\
 & -\sum_{k=1}^{M\left(n\right)/2}\mathcal{B}_{2n-2k}^{\left(2n\right)}\left(\ell\right)\;Z_{2n-2k}^{\left(2n\right)}.
\end{split}
\end{equation}
The above equations can be solved by first computing the value of the coefficient $\mathcal{A}^{(2n)}(\ell)$ in Eq.~\eqref{eq:A2n}, which in turn allows to compute any of the coefficients $\mathcal{B}^{(2n)}_{2q}(\ell)$ as in Eq.~\eqref{eq:B2q}. Once computed the above values one can finally evaluate the remaining $\mathcal{B}^{(2n)}_{2n}(\ell)$ as in Eq.~\eqref{eq:B2n}. 

\section{Forcing Isotropy Comparison}\label{app:IsoCondCompare}
Let us now connect the results in Appendix~\ref{app:IsoDetails} to the previous literature on the forcing isotropy~\cite{Shan06,Sbragaglia07,Falcucci07}. Indeed, we defined the forcing isotropy conditions $\{I_{2n,k} = 0\}$, according to our new parametrization, in Eq.~\eqref{eq:IsoCond} as
\begin{equation}\label{eq:IOnly}
  \left\{I_{2n,k}=\sum_{\ell}\mathcal{B}_{2n-2k}^{\left(2n\right)}\left(\ell\right)W\left(\ell\right)=0\right\},
\end{equation}
which can be explicitly written once all the coefficients $\mathcal{B}_{2n-2k}^{\left(2n\right)}\left(\ell\right)$ are computed according to Eqs.~\eqref{eq:A2n},~\eqref{eq:B2q} and~\eqref{eq:B2n}. However, the above conditions do not have the same form as those reported in~\cite{Sbragaglia07}, where the isotropy is obtained by requiring that the sum, over all groups, $\sum_{\mathbf{e}_{a}\in\mathcal{G}}W\left(|\mathbf{e}_{a}|^{2}\right)\left(e_{a}^{x}\right)^{2n_x}\left(e_{a}^{y}\right)^{2n_y}$, only yield isotropic contributions. Such request is expressed by the following sequence of ratios~\cite{Sbragaglia07}
\begin{widetext}
\begin{equation}\label{eq:mauro_sys}
\begin{cases}
\begin{split}\sum_{\mathbf{e}_{a}\in\mathcal{G}}W\left(|\mathbf{e}_{a}|^{2}\right)\left(e_{a}^{x}\right)^{m\left(n\right)+2}\left(e_{a}^{y}\right)^{2n-m\left(n\right)-2}/\sum_{\mathbf{e}_{a}\in\mathcal{G}}W\left(|\mathbf{e}_{a}|^{2}\right)\left(e_{a}^{x}\right)^{m\left(n\right)}\left(e_{a}^{y}\right)^{2n-m\left(n\right)}=\frac{\left[m\left(n\right)+1\right]!!\left[2n-m\left(n\right)-3\right]!!}{\left[m\left(n\right)-1\right]!!\left[2n-m\left(n\right)-1\right]!!}\\
\sum_{\mathbf{e}_{a}\in\mathcal{G}}W\left(|\mathbf{e}_{a}|^{2}\right)\left(e_{a}^{x}\right)^{m\left(n\right)+4}\left(e_{a}^{y}\right)^{2n-m\left(n\right)-4}/\sum_{\mathbf{e}_{a}\in\mathcal{G}}W\left(|\mathbf{e}_{a}|^{2}\right)\left(e_{a}^{x}\right)^{m\left(n\right)+2}\left(e_{a}^{y}\right)^{2n-m\left(n\right)-2}=\frac{\left[m\left(n\right)+3\right]!!\left[2n-m\left(n\right)-5\right]!!}{\left[m\left(n\right)+1\right]!!\left[2n-m\left(n\right)-3\right]!!}\\
\vdots\\
\sum_{\mathbf{e}_{a}\in\mathcal{G}}W\left(|\mathbf{e}_{a}|^{2}\right)\left(e_{a}^{x}\right)^{2n}/\sum_{\mathbf{e}_{a}\in\mathcal{G}}W\left(|\mathbf{e}_{a}|^{2}\right)\left(e_{a}^{x}\right)^{2n-2}\left(e_{a}^{y}\right)^{2}=\frac{\left(2n-1\right)!!}{\left(2n-3\right)!!}
\end{split}
\end{cases}
\end{equation}
Equations in~\eqref{eq:mauro_sys} must then be linear combinations of those in~\eqref{eq:IOnly}. Such combinations can be computed by straightforward (although tedious) manipulations. We report now, in the same order, the isotropy conditions in Eq.~\eqref{eq:mauro_sys}, expressed in terms of the coefficients $\mathcal{B}^{(2n)}_{2n - 2k}$ of the new parametrization
\begin{equation}\label{eq:mauro_sys_b}
\begin{cases}
\begin{split} & \sum_{\ell}W\left(\ell\right)\mathcal{B}_{2n-M\left(n\right)}^{\left(2n\right)}\left(\ell\right)=0,\;\text{for}\;M\left(n\right)>0\\
 & \sum_{\ell}W\left(\ell\right)\left[\mathcal{B}_{2q+2}^{\left(2n\right)}\left(\ell\right)-\frac{\left(2q+1\right)}{\left(2n-2q-1\right)}\mathcal{B}_{2q}^{\left(2n\right)}\left(\ell\right)\right]=0,\;\text{for}\;M\left(n\right)>0\;\text{and}\;2n-M\left(n\right)\leq2q<2n\\
 & \sum_{\ell}W\left(\ell\right)\left[\mathcal{B}_{2n}^{\left(2n\right)}\left(\ell\right)+\theta\left(M\left(n\right)\right)\left(Z_{2n-2}^{\left(2n\right)}-2n+1\right)\mathcal{B}_{2n-2}^{\left(2n\right)}\left(\ell\right)+\theta\left(M\left(n\right)-2\right)\sum_{k=2}^{M\left(n\right)/2}Z_{2n-2k}^{\left(2n\right)}\mathcal{B}_{2n-2k}^{\left(2n\right)}\left(\ell\right)\right]=0
\end{split}
\end{cases}
\end{equation}
\end{widetext}
Each equation involves a combination of our new isotropy conditions $I_{2n,k} = \sum_{\ell} \mathcal{B}^{(2n)}_{2n - 2k} W(\ell) = 0$, proving the linear dependence of Eq.~\eqref{eq:IOnly} and Eq.~\eqref{eq:mauro_sys}. In the Jupyter notebook~\cite{ipython} relative to this paper, accessible on the github repository \href{https://github.com/lullimat/idea.deploy}{https://github.com/lullimat/idea.deploy}~\cite{sympy, scipy, numpy0, numpy1, scikit-learn, matplotlib, ipython, pycuda_opencl}, it is possible to find the comparison of Eq.~\eqref{eq:mauro_sys_b} against Eq.~\eqref{eq:mauro_sys} for multi-range forcing schemes up to the 14-th isotropy order.

As an aside, the above analysis allows to compute the number of equations $N_{\mbox{\tiny{eq}}}$ needed to satisfy the isotropy conditions at the $2n$-th order, which is simply given by $N_{\mbox{\tiny{eq}}}(2n) = (2n - m(n))/2 = (n - n\;\mbox{mod}\;2)/2$, i.e. by the difference between the maximum values of $2n_x = 2n$ and the minimum $2n_x = m(n)$, divided by 2 since only even changes in $2n_x$ would yield a non-zero result. Hence, the total number of weights $N_{\text{w}}$ required to obtain isotropy at the $2n$-th order is given by the following equation
\begin{equation}
  N_{\text{w}} - 1 = \frac{1}{2} \sum_{k=2}^n \;[2k - m(k)] = \frac{1}{2} \sum_{k=2}^n \;(k - k\;\mbox{mod}\;2),
\end{equation}
where with $-1$ we are indicating that one of the equations is typically used to set the value of the second order isotropy constant $e_2$. This is the common practice, even though this is not necessary from the mathematical point of view.

%%%%%%%%%%%%%%%%%%%%%%%%%%%%%%%%%%%%%%%%%%%%%%%%%%%%%%%%%%%%%%%%%%%%%%%%%%%%%%%%%%%%%%%%%%%

%\bibliographystyle{spphys}
\bibliography{prex}
\end{document}